\documentclass[structabstract]{aa}  
\usepackage{graphicx}
\usepackage{txfonts}
\usepackage{multirow}
\usepackage{longtable,lscape}
\usepackage[]{natbib}
\bibpunct{(}{)}{;}{a}{}{,}
\usepackage{stfloats}
\usepackage{fixltx2e}
\begin{document}
   \title{The distribution of equivalent widths in long GRB afterglow spectra}

   \author{A. de Ugarte Postigo
          \inst{1,2}
          \and
          J.P.U.~Fynbo\inst{2}
          \and
          C.C.~Th\"one \inst{1}
          \and
          L.~Christensen \inst{2}
          \and
          J.~Gorosabel \inst{1}
	\and
	B. Milvang-Jensen \inst{2}
          \and
          S.~Schulze \inst{3}
          \and
          P.~Jakobsson \inst{3}
          \and
          K.~Wiersema \inst{4}
          \and
          R.~S\'anchez-Ram\'irez \inst{1}
	\and
	G. Leloudas \inst{5,2}
          \and
          T.~Zafar \inst{6}
          \and
          D.~Malesani \inst{2}
          \and
          J.~Hjorth \inst{2}
          }

   \institute{Instituto de Astrof\' isica de Andaluc\' ia (IAA-CSIC), Glorieta de la Astronom\' ia s/n, E-18008, Granada, Spain\\
                     \email{adeugartepostigo@gmail.com}
         \and
             {Dark Cosmology Centre, Niels Bohr Institute, Juliane Maries Vej 30, Copenhagen \O, D-2100, Denmark}
         \and
             {Centre for Astrophysics and Cosmology, Science Institute, University of Iceland, Dunhagi 5, IS-107 Reykjavik, Iceland}
         \and
             {University of Leicester, University Road, Leicester LE1 7RH, UK}
         \and
             {The Oskar Klein Centre, Department of Physics, Stockholm University, 106 91 Stockholm, Sweden}
         \and
             {Laboratoire d'Astrophysique de Marseille - LAM, Universit\'e Aix-Marseille \& CNRS, UMR7326, 38 rue F. Joliot-Curie, 13388 Marseille Cedex 13, France}
             }

   \date{Received ; accepted }

  \abstract
   {The extreme brightness of gamma-ray burst (GRB) afterglows and their simple spectral shape make them ideal beacons to study the interstellar medium of their host galaxies through absorption line spectroscopy at almost any redshift.}
   {We describe the distribution of rest-frame equivalent widths (EWs) of the most prominent absorption features in GRB afterglow spectra, providing the means to compare individual spectra to the sample and identify its peculiarities.}
  {Using 69 low-resolution GRB afterglow spectra, we conduct a study of the rest-frame EWs distribution of features with an average rest-frame EW larger than 0.5 \AA. 
  To compare an individual GRB with the sample, we develop \textit{EW diagrams} as a graphical tool, and we give a catalogue with diagrams for the 69 spectra. We introduce a \textit{line strength parameter} ($LSP$) that allows us to quantify the strength of the absorption features in a GRB spectrum as compared to the sample by a single number. Using the distributions of EWs of single-species features, we derive the distribution of their column densities by a curve of growth (CoG) fit.}
   {We find correlations between the $LSP$ and the extinction of the GRB, the UV brightness of the host galaxies and the neutral  hydrogen column density. However, we see no significant evolution of the $LSP$ with the redshift. 
There is a weak correlation between the ionisation of the absorbers and the energy of the GRB, indicating that, either the GRB event is responsible for part of the ionisation, or that galaxies with high-ionisation media produce more energetic GRBs.   
Spectral features in GRB spectra are, on average, 2.5 times stronger than those seen in QSO intervening damped Lyman-$\alpha$ (DLA) systems and slightly more ionised. In particular we find a larger excess in the EW of \ion{C}{iv}$\lambda$$\lambda$1549 relative to QSO DLAs, which could be related to an excess of Wolf-Rayet stars in the environments of GRBs. From the CoG fitting we obtain an average number of components in the absorption features of GRBs of $6.00_{-1.25}^{+1.00}$. The most extreme ionisation ratios in our sample are found for GRBs with low neutral hydrogen column density, which could be related to ionisation by the GRB emission. }
   {}


   \maketitle
%

\section{Introduction}

Long gamma-ray burst afterglows (GRBs) shine, during the first hours, as the most luminous objects that can be detected in the Universe \citep{kan07,rac08}. They have been observed at almost any redshift, from nearby (GRB\,980425 was the nearest one, at $z=0.008$, \citealt{gal98}) to the very distant Universe (GRB\,090429B is the current record-holder with a photometric redshift of $z=9.4$, \citealt{cuc11}, GRB\,090423 being the furthest spectroscopically confirmed at $z=8.2$, \citealt{tan09,sal09}). Their optical spectra are normally well reproduced by a clean, simple power-law, making them ideal beacons to probe the material in the line of sight of the GRBs. Using optical spectroscopy it is possible to study the dust extinction, that introduces a curvature in the spectra \citep[see e.g. ][]{kan10,sch10,zaf11}, we can learn about intervening systems, located between the host galaxy of the GRB and the observer \citep{pro06,tej07,ver09} or about the material located within the host galaxy itself \citep{vre07,del09}.

Most of the knowledge that  we currently have about the interstellar medium (ISM) of GRBs comes from the study of individual events. However, as the samples of GRB afterglows become larger, much can be gained from statistical studies. This however, does not come without complications: Due to the very wide range of redshifts, very different rest-frame wavelength ranges are probed and tools need to be developed for their common study. Furthermore, the large samples of GRB afterglows are dominated by low resolution spectra. In these spectra, the line profiles cannot be fitted to derive column densities and the different components that form each feature cannot be resolved. This means that we can only measure equivalent widths (EW) of the lines, which limits the information that can be extracted from a single spectrum. However, we can attempt to derive some of the information contained in these spectra through the use of statistical analysis of large samples.

In this paper, we use 69 low resolution optical spectra to study the absorption features created by material within the GRB host galaxies. In particular we look at the distribution of the rest-frame EWs of the most prominent lines found in GRB afterglow spectra. Based on this, we devise observational tools to easily evaluate the different types of environments that host GRBs and understand their properties and diversity.

The paper is structured as follows: Section 2 presents the sample, the methods used  and the biases and limitations of our study, section 3 presents the results of the analysis, section 4 discusses the results and finally, section 5 gives our conclusions.


\section{Sample and method}

This work is based on the sample presented by \citet{fyn09} selecting those spectra with known redshifts and a resolution $300 < R <  2000$, to have data of similar characteristics. The selection criteria of this sample, first proposed by \citet{jak06} and later refined by \citet{fyn09} were chosen to achieve a high degree of completeness and to be as independent as possible of the optical properties of the afterglow: (1) burst duration longer than 2 s, (2) precise localisation, normally provided in X-rays by \textit{Swift}/XRT, and distributed within 12 hr of the burst, to allow a rapid follow-up, (3) small Galactic extinction ($A_{V} < 0.5$ mag), (4) favourable declination for optical follow-up observations (--70$^{\circ}< \delta <  +70^{\circ}$), and a distance between the sun and the field of more than $55^{\circ}$, to allow at least 1 hr of observations to secure spectroscopy.

We add 8 further low-resolution spectra to those presented by \citet{fyn09}, obtained after the publication but with similar criteria, as detailed in Appendix II of the online material. The total sample comprises 69 spectra. This includes 7 spectra that were part of the sample of \citet{fyn09} but that did not meet all the criteria outlined in the previous paragraph.

This study analyses the distribution of rest-frame EWs, which is only possible for the most prominent absorption features in GRB afterglow spectra (the number of detections is still too small for faint features). Our sample is limited to those absorption features that have a rest-frame EW of at least 0.5 {\AA} in the composite spectrum presented by \citet{chr10}, which add to a total of 22 features. Some lines produced by transitions from excited fine structure and metastable states are known to vary during the evolution of the GRB and would introduce a further uncertainty in the results, so we exclude them from our analysis. Using the same argument, we exclude from our dataset the line at 2186.96 {\AA} that \citet{chr10} identify as due to \ion{Mn}{i}, as we believe that it is actually not due to this species but most probably to a transition from a fine structure line or a metastable state. This is supported by the fact that it is only seen in  spectra with many other fine structure features and that it is detected only in early spectra. Furthermore, other \ion{Mn}{i} lines that should have been simultaneously detected are not present in the composite spectrum.

We use the EW measurements provided by \citet{fyn09}, together with the 8 other unpublished spectra, and complement them with detection limits that we calculate for each feature in the individual spectra. To obtain the detection limits we use the individual spectra with their corresponding error spectra and measure the uncertainty at the location of the features over a wavelength span equivalent to the instrumental resolution, which is determined by the width of the slit (with this we assume that the lines are unresolved with our resolution, which is generally the case). The limits presented in this work are 3-$\sigma$. We excluded from the analysis the regions of the spectra that were affected by strong telluric features. Tables~\ref{table:data1} and \ref{table:data2} display the rest-frame EWs and limits for the lines in each spectrum.

\addtocounter{table}{2}

\subsection{EW histogram fitting}

\begin{figure}[h]
   \centering
   \includegraphics[width=8cm]{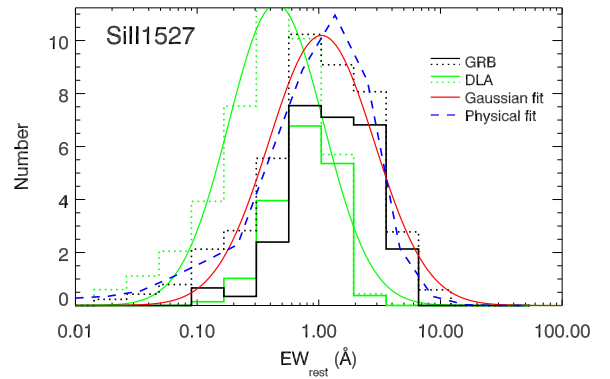}
      \caption{EW distribution of the SiII feature at $1527 \AA$. The thick black line histogram indicates only detections, whereas the dotted histogram includes also the limits. The red line is the best fit of the complete histogram with a lognormal distribution and the dashed blue line is a physical fit as described in Section~\ref{sec:fit}. In green we show a histogram of EWs in a sample of DLA systems (see Section~\ref{sec:dla}) and its lognormal fit. Similar histograms for all the lines in the sample can be found in Fig.~\ref{Fig:fits} of the online material.
              }
         \label{Fig:SiII1260}
   \end{figure}

Using the data described above we build histograms of EWs in a logarithmic scale (see Fig.~\ref{Fig:SiII1260} for an example). To consider the EW uncertainties, we do not give all the weight of the measurement to a single bin, but instead we use a gaussian distribution to distribute the weight over the different bins according to the uncertainty of the measurement. For the detection limits, we distribute the weight evenly over all the possible values below the 3-$\sigma$ limit in linear space.

    \begin{figure*}[h]
   \centering
   \includegraphics[width=17cm]{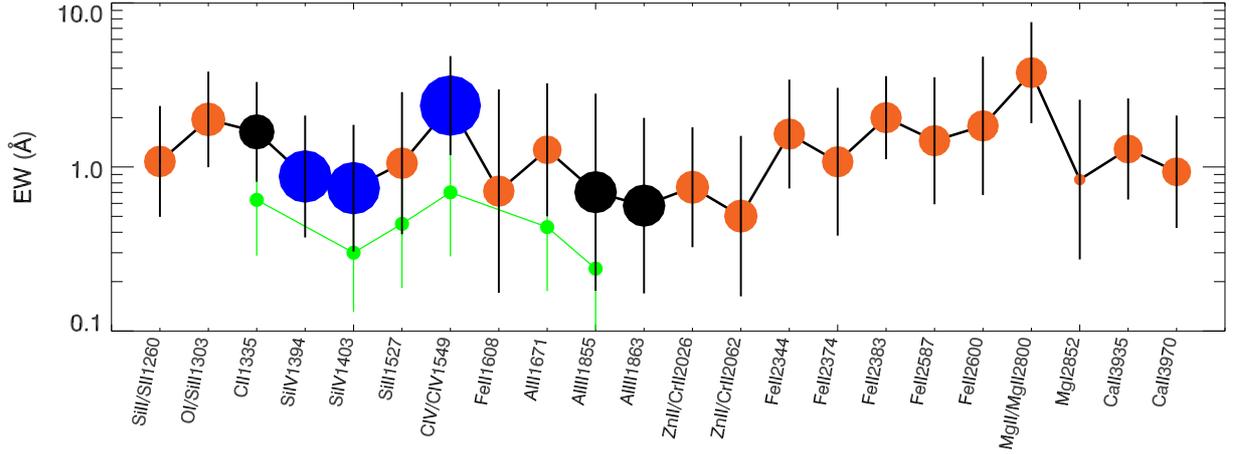}
      \caption{Average strength of the different lines together with their standard deviation, as displayed in Table~\ref{table:fits}. The area of the dots are proportional to the ionisation potential of each species. We mark in orange the low ionisation lines ($<10$ eV), in blue the high ionisation lines ($>30$ eV) and in black the intermediate lines and the blend of \ion{O}{i}/\ion{Si}{ii}$\lambda$1302,1304. In green we plot the average EWs for quasar intervening DLAs for a subsample of lines (see Sect.~\ref{sec:dla}).
              }
         \label{Fig:strength}
   \end{figure*}

\vspace{-0.5cm} 
 
   \begin{table}[h!]
\begin{minipage}[t]{\columnwidth}
\caption{Fits to the distribution of lines, including the number of detections and detection limits for each species. We show $10^{\left\langle\log\,\rm{EW}\right\rangle}$ instead of $\left\langle\log\,\rm{EW}\right\rangle$ as it allows a more intuitive interpretation.}             
\label{table:fits}      
\centering                          
\begin{tabular}{c c c c c }        
\hline\hline                 
Feature                                &  Det. &  Lim. & $10^{\left\langle\log\,\rm{EW}\right\rangle}$ (\AA) & $\sigma_{\log\,(\rm{EW/\AA})}$ \\   
\hline                        
\ion{Si}{ii}/\ion{S}{ii}$\lambda$$\lambda$1260,1260                      & 15 & 20 & 1.08 & 0.34 \\
\ion{O}{i}/\ion{Si}{ii}$\lambda\lambda$1302,1304$^a$      & 22 & 15 & 1.95 & 0.29 \\
\ion{C}{ii}$\lambda$1335\footnote{In some cases this line is blended with a fine structure transition of the same species. This can be significant in early spectra, as the fine structure line is normally excited by the prompt GRB emission.}                        & 23 & 16 & 1.64 & 0.30 \\
\ion{Si}{iv}$\lambda$1394                      & 19 & 19 & 0.88 & 0.37 \\
\ion{Si}{iv}$\lambda$1403                      & 19 & 19 & 0.74 & 0.39 \\
\ion{Si}{ii}$\lambda$1527                       & 27 & 16 & 1.05 & 0.43 \\
\ion{C}{iv}$\lambda\lambda$1548,1551   & 35 &  7  & 2.37 & 0.30 \\
\ion{Fe}{ii}$\lambda$1608                      & 19 & 24 & 0.71 & 0.61 \\
\ion{Al}{ii}$\lambda$1671                       & 26 & 13 & 1.27 & 0.41 \\
\ion{Al}{iii}$\lambda$1855                      & 15 & 27 & 0.70 & 0.60 \\
\ion{Al}{iii}$\lambda$1863                      & 14 & 29 & 0.58 & 0.53 \\
\ion{Zn}{ii}/\ion{Cr}{ii}$\lambda\lambda$2026,2026                      & 16 & 27 & 0.75 & 0.37 \\
\ion{Zn}{ii}/\ion{Cr}{ii}$\lambda\lambda$2063,2062                      & 11 & 26 & 0.50 & 0.49 \\
\ion{Fe}{ii}$\lambda$2344$^a$                      & 23 & 14 & 1.59 & 0.33 \\
\ion{Fe}{ii}$\lambda$2374                      & 17 & 20 & 1.08 & 0.45 \\
\ion{Fe}{ii}$\lambda$2383$^a$                      & 23 & 15 & 2.00 & 0.25 \\
\ion{Fe}{ii}$\lambda$2587                      & 20 & 16 & 1.45 & 0.39 \\
\ion{Fe}{ii}$\lambda$2600$^a$                      & 18 & 16 & 1.78 & 0.42 \\
\ion{Mg}{ii}$\lambda\lambda$2796,2803 & 27 &  4  & 3.76 & 0.30 \\
\ion{Mg}{i}$\lambda$2852                      & 12 &  18 & 0.84 & 0.49 \\
\ion{Ca}{ii}$\lambda$3935                     & 7    &  8 & 1.29 & 0.31 \\
\ion{Ca}{ii}$\lambda$3970                     & 7    &   8  & 0.93 & 0.34 \\
\hline                                   
\end{tabular}
\end{minipage}
\end{table}

In contrast with the work published by \citet{chr10}, where the EWs are weighted with the signal to noise ratio of the individual spectra to get a typical GRB spectrum, our aim is to study the actual distribution, independently of the signal to noise ratio of the spectra. To give a statistical value to the distributions that we obtain, as a first order approximation, we fit the histograms (detections plus limits) to a log-normal distribution. This gives us a typical EW value for the line and a standard deviation in the logarithmic space. The results of these fits are displayed in Table~\ref{table:fits}. The table includes also the number of detections and limits for each of the features in our sample. Figure~\ref{Fig:strength} shows these results graphically, including information on the ionisation potential of each transition (see Table~\ref{table:ion}).

\begin{table}[h]
\caption{Ionisation potential of the ionised species in our sample, from \citet{mor03}.}             
\label{table:ion}      
\centering                          
\begin{tabular}{c c}        
\hline\hline                 
Species		& Ionisation 				\\ 
			& potential (eV) 			\\
\hline                        
\ion{Al}{ii}		& 5.99					\\
\ion{Ca}{ii}	& 6.11					\\
\ion{Mg}{ii}	& 7.65					\\
\ion{Fe}{ii}		& 7.90					\\
\ion{Si}{ii}		& 8.15					\\
\ion{Zn}{ii}		& 9.39					\\
\ion{C}{ii}		& 11.26					\\
\ion{Al}{iii}		& 18.83					\\
\ion{Si}{iv}	& 33.49					\\
\ion{C}{iv}		& 47.89					\\
\hline                                   
\end{tabular}
\end{table}

\subsection{On the fitting function}
\label{sec:fitfun}

In our study we assume that the column density of each individual species is described by a log-normal distribution. However, this does not directly imply that the distribution of EWs will also have a similar distribution. The column densities will be convolved with the curve of growth (CoG), distorting the original distribution. In this section we simulate the effect of a log-normal distribution convolved with typical CoG using Monte Carlo simulations.

\begin{figure}[h]
   \centering
   \includegraphics[width=8cm]{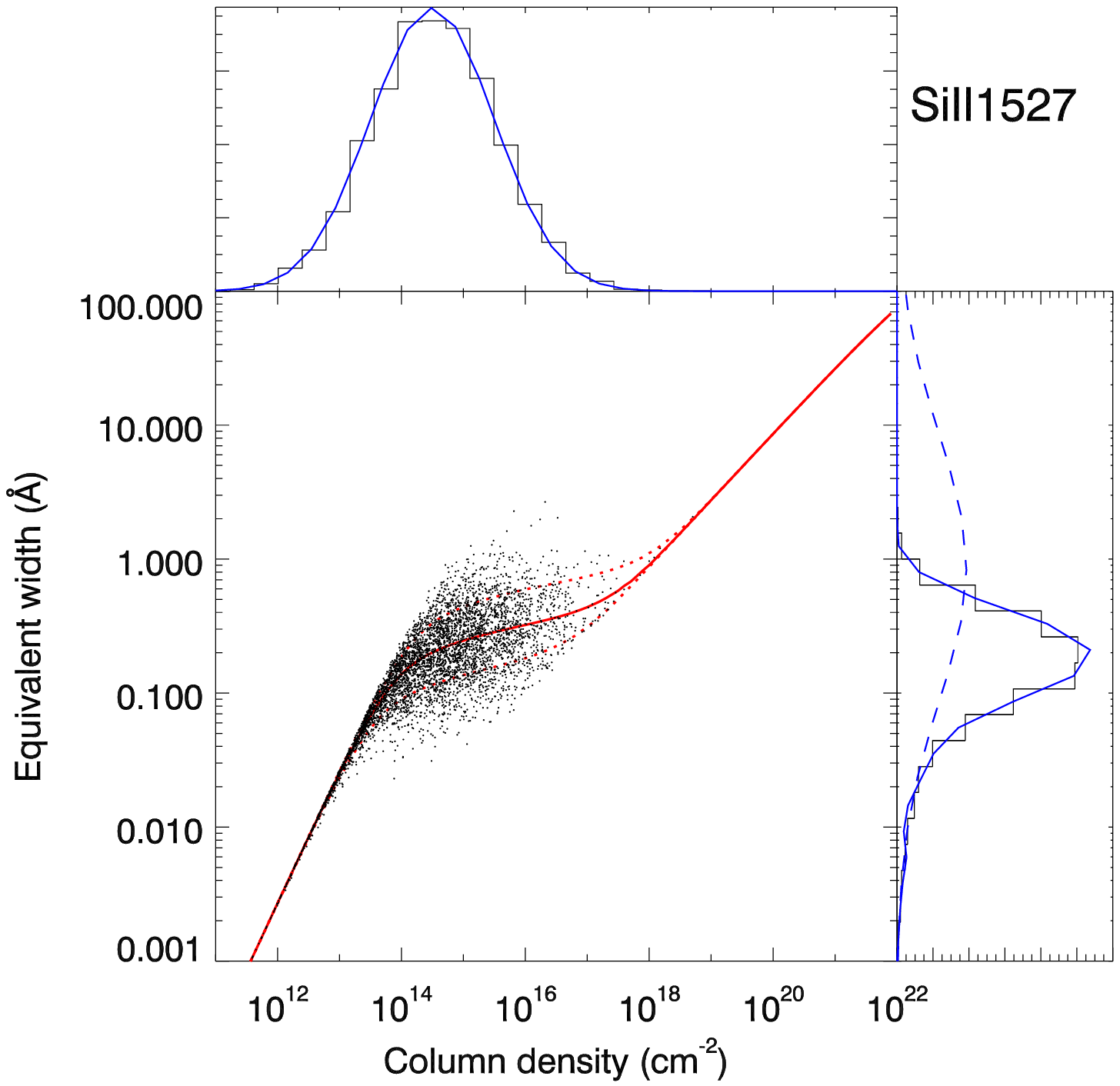}
      \caption{Convolution of a log-normal distribution of column densities of \ion{Si}{ii}$\lambda$1527, randomly generated, with a CoG to a distribution of EWs. We have assumed a distribution with $\log\,(N/\rm{cm}^{-2}) = 14.5\pm1.0$ and also a log-normal distribution of particle velocities with $log\,(b/km s^{-1}) = 1.1\pm0.3$ (equivalent to $12.6\,km\,s^{-1}$). The red continuous line indicates the CoG using the central b parameter, while the dotted lines are for b values 1-$\sigma$ above or below. Continuous blue lines mark the distribution of column densities and its corresponding distribution of EWs once convolved with the CoG. The blue dashed line is the EW distribution that would be produced if the CoG would only have a linear regime.
              }
         \label{Fig:CoG_SiII1260}
   \end{figure}

Figure \ref{Fig:CoG_SiII1260} shows how a log-normal histogram of column densities for \ion{Si}{ii}$\lambda$1527 with $\log\,(N/\rm{cm}^{-2}) = 14.5\pm1.0$ randomly generated (5000 tests) is transformed into a distribution of EWs. To do this, we also consider a random distribution of particle velocities following a log-normal distribution with $\log\,(b/\rm{km s}^{-1}) = 1.1\pm0.3$ (equivalent to $12.6\,km\,s^{-1}$). We can see how, in the border between the linear and the saturated regime, where we find the features studied in this paper, the histogram has a tail to lower values of EWs and is truncated for high EWs due to the effect of saturation. This is similar to what we see in Fig.~\ref{Fig:SiII1260} and Fig.~\ref{Fig:fits} for most of the lines in our sample.

This implies that using log-normal distributions to fit the EW histograms is not an accurate physical description. However, due to the limited size of the sample, and the fact that the tail of the distribution towards low EWs is not well sampled in most of the cases (it is mainly populated by detection limits), we use the log-normal distribution as a first degree approximation. Furthermore, the use of such a simple fitting function allows us to perform an efficient and model independent observational analysis. In section~\ref{sec:fit} we attempt to use a more complex fit of the data to derive statistical distributions of column densities.

\begin{figure*}[h]
\centering
\includegraphics[width=14cm,clip]{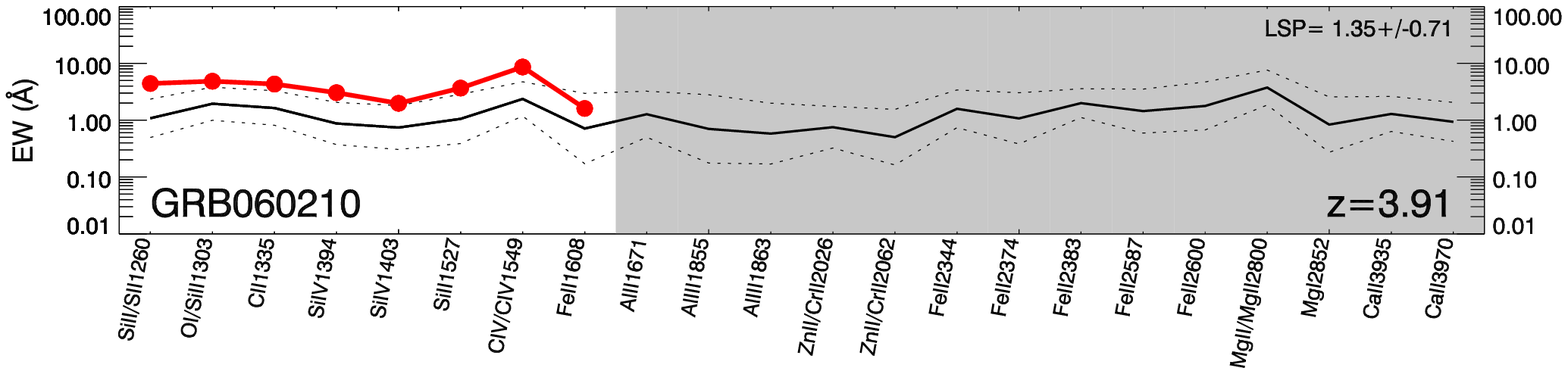}
\includegraphics[width=14cm,clip]{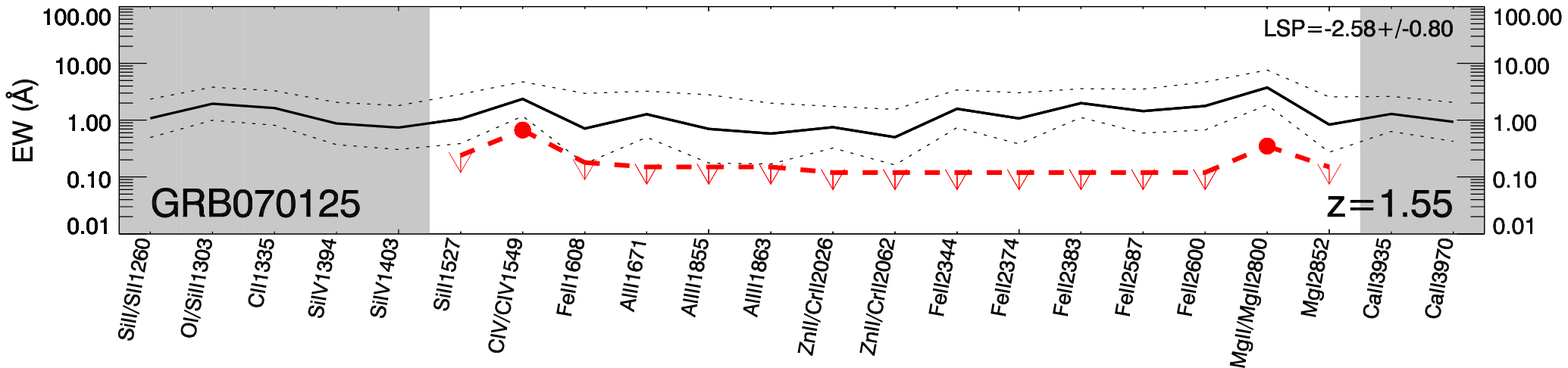}
\includegraphics[width=14cm,clip]{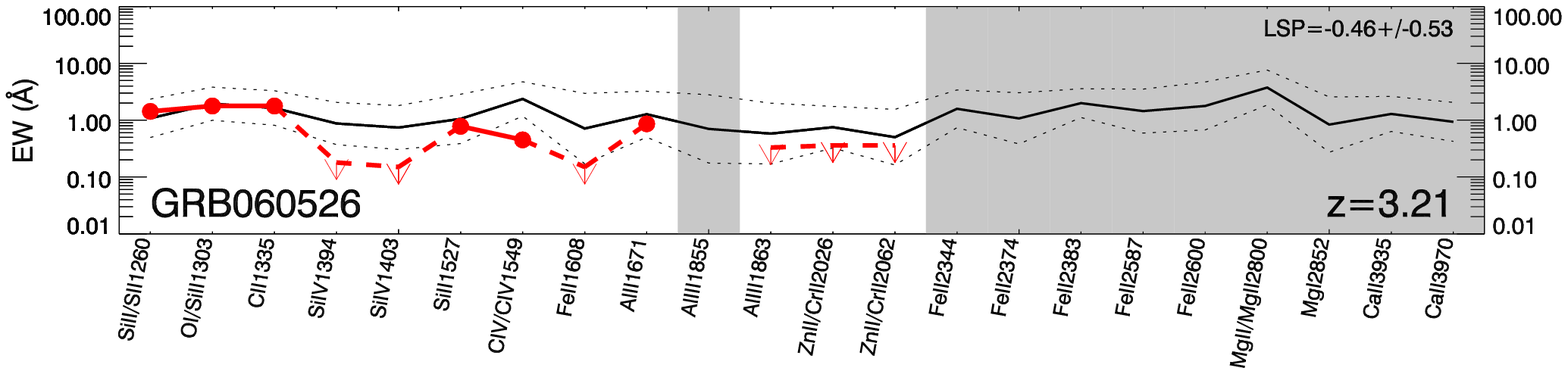}
\includegraphics[width=14cm,clip]{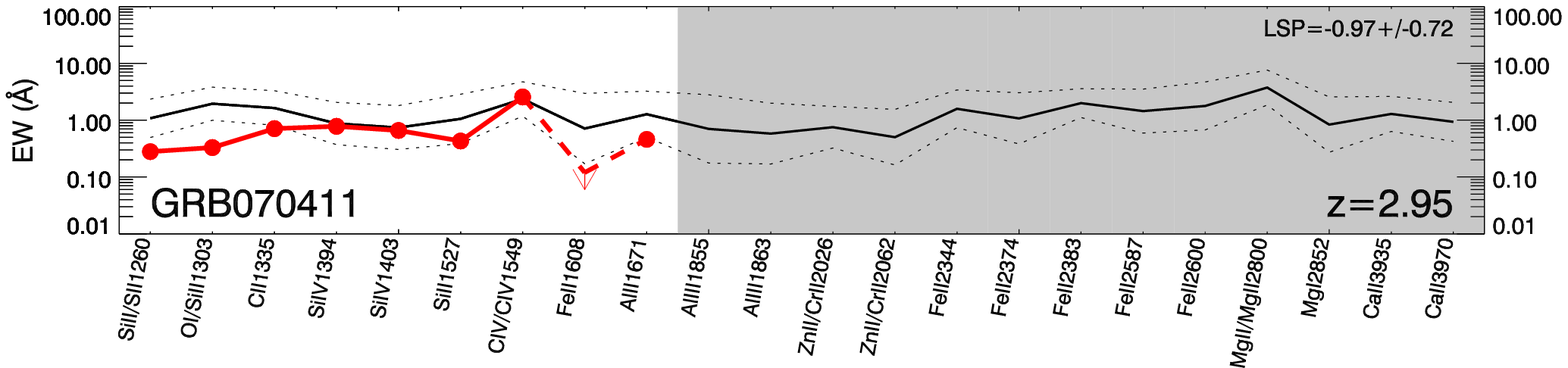}
      \caption{EW diagrams:  In red we represent the EWs of the spectral features if each afterglow spectrum. The continuous black line is the average of the sample of GRBs (see Fig.~\ref{Fig:strength}), whereas the dotted lines indicate the standard deviations over and under this value. Gray regions mark the range where there is no spectral coverage or it is affected by strong telluric features. Each panel shows examples of individual bursts: GRB\,060210, an afterglow with strong lines; GRB\,070125, with weak features; GRB\,060526, with a very weakly ionised environment; GRB\,070411, with very high ionisation.
              }
         \label{Fig:EW_diag}
\end{figure*}

\subsection{Biases and limitations}

When using this sample, we must be aware that there are observational biases that limit the statistical value of the study, even if the data were obtained using criteria designed to minimise them. Within the sample criteria, there were cases for which a spectrum could not be obtained and/or the redshift was not constrained. These missing data will be dominated by faint afterglows, probably due to high extinctions, or by low density environments, which produce weaker features. The biases of this sample, which includes $\sim$1/3 of the bursts meeting the observational criteria during the period of the study, are explained in detail by \citet{fyn09}.

For the spectra with a large amount of limits as compared to line detections, the way in which we treat the limits will affect our fits. We have chosen to give equal probability in the linear space to all the possible values below the 3-$\sigma$ limits as this is the less model-dependent method.

On the other hand, there are limitations on the study of EWs in low resolution spectra as compared to determining the column densities of material with Voigt profile fitting in high resolution spectra. In most of the cases, the lines that we are studying are in the saturated regime, where the strength of the lines are not linearly related with the amount of material, so line ratios must be handled with care. In this sense, the study of the EWs of individual bursts can give us an idea of the environment within GRB host galaxies, but a quantitative description can be more challenging (see section~\ref{sec:fit}).

Finally, one must be cautious when studying some of the individual features  due to blending of several species. The most prominent case in our sample is the feature at 1303 {\AA} produced by the blend of \ion{O}{i}$\lambda$1302.17, \ion{Si}{ii}$\lambda$1304.37 and \ion{O}{i}*$\lambda$1304.86. Both \ion{O}{i}$\lambda$1302.17 and \ion{Si}{ii}$\lambda$1304.37 are intense features that cannot be disentangled at our resolution and that add similar contributions, whereas \ion{O}{i}*$\lambda$1304.86 is only important in some very early spectra. In a similar way the two \ion{Zn}{ii} features are blended with \ion{Cr}{ii}. Although in most cases the \ion{Zn}{ii} is dominant, the contribution of \ion{Cr}{ii} cannot be neglected. In the case of early spectra, some \ion{Fe}{ii} features as well as \ion{C}{ii} can be blended with fine structure transitions of the same species. These fine structure transitions can be induced by the radiation from the prompt GRB emission and in consequence have an EW that can vary with time. These features are disregarded in the physical fit performed in Sect.~\ref{sec:fit}. In Table~\ref{table:fits} we indicate the lines that are affected by blends with other lines.

\section{Results}

\subsection{EW diagrams}

By comparing an individual spectrum with the sample of EW distributions, one can understand the general characteristics of the GRB environment. To do this in an effective way, we create comparative \textit{EW diagrams}, where the EWs of an individual burst are plotted together with the average and standard deviations of the sample, which are the same as shown in Fig.~\ref{Fig:strength} (See Fig.~\ref{Fig:EW_diag} for some examples and Fig.~\ref{Fig:GRBs} of the online material for the complete catalogue of EW diagrams).

These diagrams are useful to quickly estimate the strength of the absorption features as compared to the sample as well as to identify environments with unusual characteristics. Independently of the strength of the features, the relative line strengths should be similar to the sample average if the composition of GRB host galaxies is uniform. An EW diagram that does not follow the shape of the average diagram (see the 3 bottom panels of Fig.~\ref{Fig:EW_diag}), immediately tells us that the environment of the GRB has peculiarities. As an example, GRB\,060526, second from the bottom of Fig.~\ref{Fig:EW_diag},  shows very low EWs for highly ionised species, such as \ion{Si}{iv}, \ion{C}{iv} or \ion{Al}{iii} as compared to their lower ionised transitions \ion{Si}{ii}, \ion{C}{ii} and \ion{Al}{ii}. For GRB\,070411 the behaviour is the opposite. 

As EW diagrams can easily reveal anomalous feature strengths, they are also a useful tool to identify faulty measurements, features that are affected by contamination of intervening material, artefacts in the data, etc. In fact, while writing this paper we have identified several such cases, allowing us to revise the data and, when possible, correct the measurements.

\subsection{Line Strength Parameter}

\begin{table}[]
\caption{Neutral hydrogen column density, derived from Ly$\alpha$, $LSP$, low-ionisation $LSP$ and high-ionisation $LSP$ for the 69 spectra in our sample. }             
\label{table:LSP}      
\centering                          
\begin{scriptsize}
\begin{tabular}{l c c c c }       
\hline\hline                 
GRB & $\log\,N_{\rm \ion{H}{i}}$ & $LSP$ & $LSP_{L}$ & $LSP_{H}$ \\
\hline                        
050319   &20.90$\pm$0.20 &  0.69$\pm$0.30 &  0.67$\pm$0.14 &  1.20$\pm$0.25 \\
050401   &22.60$\pm$0.30 &  0.57$\pm$0.46 &  0.34$\pm$0.31 &  0.12$\pm$0.03 \\
050408   &          ---   &  0.49$\pm$0.37 &  0.34$\pm$0.22 &           ---   \\
050730   &22.10$\pm$0.10 & $-$0.46$\pm$0.27 & $-$0.48$\pm$0.22 & $-$0.31$\pm$0.12 \\
050802   &          ---   &  0.41$\pm$0.18 &  0.48$\pm$0.18 &  0.20$\pm$0.04 \\
050824   &          ---   & $-$0.68$\pm$0.24 & $-$0.68$\pm$0.24 &           ---   \\
050908   &17.60$\pm$0.10 & $-$1.12$\pm$0.96 & $<-3.19$        &  0.09$\pm$0.11 \\
050922C  &21.55$\pm$0.10 & $-$0.90$\pm$0.57 & $-$0.64$\pm$0.26 & $-$0.75$\pm$0.27 \\
060115   &21.50$\pm$0.10 &  0.01$\pm$0.22 &  0.28$\pm$0.06 & $-$0.84$\pm$0.18 \\
060124   &18.50$\pm$0.50 &  0.28$\pm$0.08 & $<-1.55$        &  0.28$\pm$0.08 \\
060206   &21.85$\pm$0.10 &  0.50$\pm$0.15 & $< 1.01$        &           ---   \\
060210   &21.55$\pm$0.15 &  1.35$\pm$0.71 &  1.21$\pm$0.47 &  1.48$\pm$0.53 \\
060502A  &          ---   & $-$0.16$\pm$0.21 & $-$0.16$\pm$0.21 &           ---   \\
060526   &20.00$\pm$0.15 & $-$0.46$\pm$0.53 & $-$0.12$\pm$0.14 & $-2.39\pm$0.51 \\
060614   &          ---   & $<-0.37$        & $<-0.37$        &           ---   \\
060707   &21.00$\pm$0.20 &  0.36$\pm$0.35 &  0.66$\pm$0.20 & $-0.87\pm$0.19 \\
060714   &21.80$\pm$0.10 &  0.62$\pm$0.33 &  0.52$\pm$0.25 &  0.64$\pm$0.23 \\
060729   &          ---   & $-0.84\pm$0.40 & $-0.84\pm$0.40 &           ---   \\
060904B  &          ---   &  0.04$\pm$0.13 &  0.04$\pm$0.13 &           ---   \\
060906   &21.85$\pm$0.10 & $-0.42\pm$0.19 & $-0.31\pm$0.07 & $-0.29\pm$0.06 \\
060908   &          ---   &  0.45$\pm$0.10 & $<-0.18$        &  0.45$\pm$0.10 \\
060926   &22.60$\pm$0.15 & $-0.25\pm$0.09 & $-0.10\pm$0.02 & $-0.32\pm$0.07 \\
060927   &          ---   & $< 0.23$        & $< 0.23$        &           ---   \\
061007   &          ---   &  0.65$\pm$0.34 &  0.65$\pm$0.34 &           ---   \\
061021   &          ---   & $-1.00\pm$0.21 & $-1.00\pm$0.21 &           ---   \\
061110A  &          ---   & $-0.05\pm$0.01 & $-0.05\pm$0.01 &           ---   \\
061110B  &          ---   &  0.14$\pm$0.24 & $-0.37\pm$0.11 &  0.64$\pm$0.19 \\
061121   &          ---   &  0.76$\pm$0.37 &  0.79$\pm$0.43 &  0.90$\pm$0.19 \\
070110   &21.70$\pm$0.10 & $-0.33\pm$0.41 & $-0.09\pm$0.20 & $-0.95\pm$0.35 \\
070125   &          ---   & $-2.58\pm$0.80 & $-3.35\pm$0.71 & $-1.82\pm$0.39 \\
070306   &          ---   & $< 0.36$        & $< 0.36$        & $< 1.69$        \\
070318   &          ---   & $-0.39\pm$0.37 & $-0.39\pm$0.37 &           ---   \\
070411   &19.30$\pm$0.30 & $-0.97\pm$0.72 & $-1.24\pm$0.46 & $-0.05\pm$0.05 \\
070419A  &          ---   & $-0.29\pm$0.12 & $-0.29\pm$0.12 &           ---   \\
070506   &22.00$\pm$0.30 &  0.50$\pm$0.39 &  0.54$\pm$0.28 &  0.70$\pm$0.28 \\
070508   &          ---   & $< 0.39$        & $< 0.39$        &           ---   \\
070611   &          ---   & $-0.81\pm$0.17 & $-0.81\pm$0.17 & $<-1.84$        \\
070721B  &21.50$\pm$0.20 &  0.27$\pm$0.24 &  0.17$\pm$0.04 &  0.56$\pm$0.21 \\
070802   &21.50$\pm$0.20 &  1.12$\pm$0.68 &  1.35$\pm$0.68 & $-0.16\pm$0.03 \\
071020   &          ---   &  0.50$\pm$0.28 &  0.50$\pm$0.28 & $<-0.13$        \\
071031   &22.15$\pm$0.05 & $-0.40\pm$0.32 & $-0.47\pm$0.30 &  0.02$\pm$0.04 \\
071112C  &          ---   & $-0.59\pm$0.32 & $-0.59\pm$0.32 &           ---   \\
080210   &21.90$\pm$0.10 &  0.22$\pm$0.32 &  0.02$\pm$0.20 &  0.62$\pm$0.22 \\
080319B  &          ---   & $-0.75\pm$0.23 & $-0.75\pm$0.23 &           ---   \\
080319C  &          ---   &  0.46$\pm$0.24 &  0.45$\pm$0.10 &  0.60$\pm$0.13 \\
080330   &          ---   & $-0.43\pm$0.42 & $-1.12\pm$0.34 & $-0.47\pm$0.10 \\
080411   &          ---   & $-0.08\pm$0.12 & $-0.15\pm$0.11 &           ---   \\
080413B  &          ---   &  0.19$\pm$0.28 &  0.44$\pm$0.22 &           ---   \\
080520   &          ---   & $<-0.60$        & $<-0.60$        & $< 2.11$        \\
080603B  &21.85$\pm$0.05 & $-1.11\pm$0.59 & $-1.57\pm$0.48 & $-0.83\pm$0.30 \\
080604   &          ---   &  0.38$\pm$0.21 &  0.38$\pm$0.21 &           ---   \\
080605   &          ---   &  0.54$\pm$0.45 &  0.61$\pm$0.36 &  1.11$\pm$0.24 \\
080607   &22.70$\pm$0.15 &  0.88$\pm$0.47 &  0.80$\pm$0.30 &  1.06$\pm$0.38 \\
080707   &          ---   &  0.49$\pm$0.28 &  0.49$\pm$0.28 &           ---   \\
080710   &          ---   & $-1.33\pm$0.70 & $-1.33\pm$0.70 &           ---   \\
080721   &21.60$\pm$0.10 &  0.41$\pm$0.31 &  0.52$\pm$0.24 &  0.52$\pm$0.22 \\
080810   &          ---   & $-0.19\pm$0.19 & $-0.79\pm$0.17 &  0.40$\pm$0.09 \\
080905B  &          ---   & $-0.02\pm$0.32 & $-0.06\pm$0.25 & $-0.51\pm$0.11 \\
080913   &          ---   & $<-0.42$        & $<-0.42$        &           ---   \\
080916A  &          ---   &  0.38$\pm$0.08 &  0.38$\pm$0.08 &           ---   \\
080928   &          ---   & $-0.88\pm$0.62 & $-0.98\pm$0.62 & $-0.16\pm$0.03 \\
081007   &          ---   & $< 3.43$        & $< 3.43$        &           ---   \\
081008   &21.59$\pm$0.10 & $-0.59\pm$0.40 & $-0.30\pm$0.15 & $-0.83\pm$0.31 \\
090102   &          ---   &  0.75$\pm$0.49 &  0.87$\pm$0.50 &  0.56$\pm$0.12 \\
090516   &21.73$\pm$0.10 &  1.31$\pm$0.64 &  1.25$\pm$0.44 &  1.21$\pm$0.43 \\
090519   &          ---   & $<-0.55$        & $< 0.50$        & $<-0.55$        \\
090529   &          ---   &  0.07$\pm$0.17 &  0.61$\pm$0.13 & -0.47$\pm$0.10 \\
090812   &22.30$\pm$0.10 & $-0.14\pm$0.35 & $-0.07\pm$0.21 & $-0.65\pm$0.26 \\
090814A  &          ---   &  0.81$\pm$0.34 &  0.81$\pm$0.34 &           ---   \\
\hline                                   
\end{tabular}
\end{scriptsize}
\end{table}

Comparative studies of spectra over a wide range of redshifts can be a difficult task, as the spectral features can fall at very different wavelengths in the observer frame. To study the line strengths of the different bursts, independently of the wavelength range covered, we define the \textit{line strength parameter} ($LSP$ hereafter) as:

   \begin{equation}
	LSP=\frac{1}{N}\,\sum_{i=1}^{N}{\frac{\log\,EW_i - \left\langle\log\,EW\right\rangle_i}{\sigma_{\log\,EW,\,i}}},
   \end{equation}

\noindent where $EW_i$ is the equivalent width for each of the individual detected lines in the spectrum, $\left\langle\log\,EW\right\rangle_i$ the central equivalent width obtained from the fits in Sect.~3, $\sigma_{\log\,EW,i}$ is the standard deviation (both shown in Table~\ref{table:fits}) and $N$ the total number of lines used to calculate the $LSP$.

This parameter measures the strength of the absorption features of a spectrum as compared to the average GRB spectrum. A value of zero would mean that the absorption features in the spectrum have the same strength as the average spectrum. Positive values indicate bursts with stronger than average lines, and equal to 1.0 if the deviation is equivalent to 1-$\sigma$. In the same way a negative value implies weak lines. In Table~\ref{table:LSP} we display the $LSP$ for each of the GRBs considered in the sample. We also calculate the standard deviation (of the summation values in Eq.~1) for each of the GRBs that indicates how different is the distribution of line strengths with respect to the average of GRBs. In the case of a spectrum where we only have detection limits, we use the strongest limit given by a single line to calculate a limit value for the $LSP$.

\begin{figure}[h]
   \centering
   \includegraphics[width=8cm]{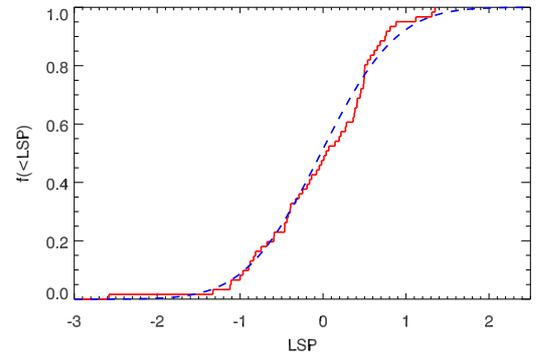}
      \caption{Cumulative distribution of the line strength parameter. The red line are the observations, whereas the the blue dashed line is the best fit to a gaussian distribution.}
         \label{Fig:LSP_cum}
   \end{figure}

The distribution of $LSPs$ can be fitted by a gaussian with an average value of $\left\langle{LSP}\right\rangle=-0.03$ and a standard deviation of $\sigma_{LSP}=0.72$. The average LSP is not centred on zero but slightly below, as the distribution of EWs is not fitted by a perfect gaussian but has a tail towards lower values (see Sect.~\ref{sec:fitfun}). These values can be used to calculate the percentile of line strength that corresponds to a specific afterglow (see Fig.~\ref{Fig:LSP_cum}) through the following formula:

\begin{equation}
P_{LSP,i}=50\left (1+{\rm ERF}\left (\frac{LSP_i-\left\langle{LSP}\right\rangle}{\sqrt{2}\sigma_{LSP}}\right ) \right )
\end{equation}

To extract further information, we calculate two additional $LSPs$, separately using just high-, or low-ionised species. In this way we can see if the behaviour is different for high ionisation and low ionised environments. The higher ionisation lines could be more closely associated with star forming regions, such as the ones that produce GRBs, whereas the lower ionisation species would be more representative of the overall galaxy. For the low-ionisation $LSP$ we use the lines of species with ionisation potential lower than 10 eV and for the high-ionisation $LSP$ those that have potentials higher than 30 eV, as shown in Table~\ref{table:ion}. Both low-ionisation $LSP$ ($LSP_{L}$) and high-ionisation $LSP$ ($LSP_{H}$) are displayed in Table~\ref{table:LSP}.

\subsection{Comparison with quasar DLAs}
\label{sec:dla}

In general, GRBs are produced in environments with large amounts of neutral gas, with neutral hydrogen column densities that lie in the damped Lyman-$\alpha$ (DLA) regime. In this section we measure the rest-frame EWs of a sample of intervening DLA systems in quasar spectra in a similar way to what we have done for GRBs. We have used the original QSO-DLA sample from \citet{not09} based on the Sloan Digital Sky Survey Data Release 7 (SDSS-DR7) database of QSO spectra. These authors automatically searched for DLA lines, refining their Ly$\alpha$ fits whenever metal lines are detected redward of the Ly$\alpha$ forest. We have selected all systems from their list with $\log\,(N_{\rm \ion{H}{i}}/cm^{-2}) > 20$ and redshifts in the range $2.2<z< 3.2$, located at least 5000 km s$^{-1}$ from a background QSO with $R < 21$. To simplify the analysis we have limited the comparison to the following spectral features: \ion{C}{ii}$\lambda$1335, \ion{Si}{iv}$\lambda$1403, \ion{Si}{ii}$\lambda$1527, \ion{C}{iv}$\lambda$1549, \ion{Al}{ii}$\lambda$1671 and \ion{Al}{iii}$\lambda$1855.

In Fig.~\ref{Fig:SiII1260} and Fig.~\ref{Fig:GRBs} of the Appendix we show the histograms of the EWs for GRBs and in the cases of the lines where we have DLA data, we also represent additional histograms with DLA data in green. The DLA histograms have been scaled down to match the smaller number of GRBs so that the histograms can be plotted together. Similar measurements to the ones displayed in Table~\ref{table:fits} are given for DLAs in Table~\ref{table:dla}. It is immediately clear that the features produced by quasar intervening DLAs are weaker. We find that the EWs of GRB absorption features are, on average, $2.5\pm0.6$ times larger than the DLAs. The excess of material in the line of sight of GRBs as compared to DLAs has been already described \citep{sav03,jak06b} and is interpreted as due to the different impact parameters: The material generating the DLA absorptions are produced as the line of sight to the quasar intercepts the intervening galaxy, which is statistically more probable to happen in the outskirts of the galaxy \citep{pro07,fyn08,pon10}. On the other hand, GRBs are produced inside the most luminous regions \citep{blo02,fru06,sve10} of star forming galaxies \citep{chr04}, implying that the light will have to interact with much more material before escaping the host galaxy. However, we note that all the lines in our DLA sample are between 2.3 and 2.9 times weaker than the ones in the GRB sample except for one. In the case of \ion{C}{iv}, the GRB lines are 3.4 times stronger than those found in DLAs. This excess of \ion{C}{iv} could be indicative of an excess of Wolf-Rayet stars \citep{ber06} in the GRB host galaxies, as compared to QSO DLAs.

For comparison, we have plotted, in Fig.~\ref{Fig:strength}, the average EWs measured for DLAs (in green) together with the values for GRBs.

\begin{table}[h]
\caption{Fits to the distribution of lines for quasar intervening DLA systems.}             
\label{table:dla}      
\centering                          
\begin{tabular}{c c c}        
\hline\hline                 
Feature                                & $10^{\left\langle\log\,\rm{EW}\right\rangle}$ (\AA)& $\sigma_{\log\,(\rm{EW}/\AA)}$ \\ 
\hline                        
\ion{C}{ii}$\lambda$1335                        & 0.63 & 0.34 \\
\ion{Si}{iv}$\lambda$1403                      & 0.30 & 0.36 \\
\ion{Si}{ii}$\lambda$1527                       & 0.45 & 0.39 \\
\ion{C}{iv}$\lambda\lambda$1548,1551  & 0.70  & 0.39 \\
\ion{Al}{ii}$\lambda$1671                       & 0.43 & 0.39 \\
\ion{Al}{iii}$\lambda$1855                      & 0.24 & 0.37 \\
\hline                                   
\end{tabular}
\end{table}

\subsection{Physical fit of the EW distributions}
\label{sec:fit}

\begin{table*}[h]
\begin{minipage}[t]{\textwidth}
\caption{Distribution of column densities derived from the EWs. From the fits we derive that the average number of line components is $6.00_{-1.00}^{+1.25}$.}             
\label{table:fitN}      
\centering                          
\begin{tabular}{c c c c c}        
\hline\hline                 
Species 		& Features (\AA)	& $\left\langle\log\,(N/cm^{-2})\right\rangle_{comp}$ 	& $\sigma_{\log\,(N/cm^{-2})}$ 		& $\left\langle\log\,(N/cm^{-2})\right\rangle_{total}$				\\    
\hline                        
\ion{Si}{ii}		& 1527		& $14.50_{-0.28}^{+0.28}$	& $1.10_{-0.30}^{+0.30}$	& $15.28_{-0.28}^{+0.28}$	\\
\ion{Si}{iv}	& 1394, 1403	& $13.80_{-0.21}^{+0.21}$	& $0.80_{-0.20}^{+0.30}$	& $14.58_{-0.21}^{+0.21}$	\\
\ion{C}{iv}		& 1549		& $14.50_{-0.21}^{+0.28}$	& $0.80_{-0.30}^{+0.30}$	& $15.28_{-0.21}^{+0.28}$	\\
\ion{Fe}{ii}	& 1608, 2374,2587	& $14.29_{-0.35}^{+0.35}$	& $1.00_{-0.30}^{+0.50}$	& $15.07_{-0.35}^{+0.35}$	\\
\ion{Al}{ii}		& 1671		& $13.38_{-0.35}^{+0.28}$	& $1.20_{-0.30}^{+0.30}$	& $14.16_{-0.35}^{+0.28}$	\\
\ion{Al}{iii}	$$	& 1855, 1863	& $13.17_{-0.28}^{+0.28}$	& $0.90_{-0.20}^{+0.40}$	& $13.95_{-0.28}^{+0.28}$	\\
\ion{Mg}{i}	$$	& 2852		& $12.19_{-0.21}^{+0.21}$	& $0.80_{-0.20}^{+0.30}$	& $12.97_{-0.21}^{+0.21}$	\\
\ion{Mg}{ii}	& 2800		& $13.59_{-0.28}^{+0.28}$	& $0.60_{-0.40}^{+0.40}$	& $14.37_{-0.28}^{+0.28}$	\\
\ion{Ca}{ii}$$    & 3935, 3970	& $12.54_{-0.21}^{+0.21}$	& $0.50_{-0.20}^{+0.40}$	& $13.32_{-0.21}^{+0.21}$	\\
\hline                                   
\end{tabular}
\end{minipage}
\end{table*}

In this section we attempt to fit the distributions of EWs that we have obtained, using CoGs to derive the actual distribution of column densities of each species. We use only features that are produced by single species (we exclude the \ion{O}{i}/\ion{Si}{ii}, \ion{Zn}{ii}/\ion{Cr}{ii} and \ion{Si}{ii}/\ion{S}{ii} line blends, as well as the \ion{C}{ii} and \ion{Fe}{ii} that can be blended with fine structure lines, but we include doublets of the same species such as \ion{C}{iv}, \ion{Mg}{ii}). 

The fit is determined from the distribution of EWs of all the features simultaneously. To do this, we must make some assumptions: (1) The distribution of column densities is described by a log-normal distribution (characterised by $\left\langle\log\,N\right\rangle$ and $\sigma_{\log\,N}$). (2) The distribution of particle velocities is described by a log-normal distribution (characterised by $\left\langle\log\,b\right\rangle$ and $\sigma_{\log\, b}$). (3) The material producing the features is located in a series of clumps, the same for all the species, this will create a specific number of components.

\begin{figure}[h]
      \centering
   \includegraphics[width=8cm]{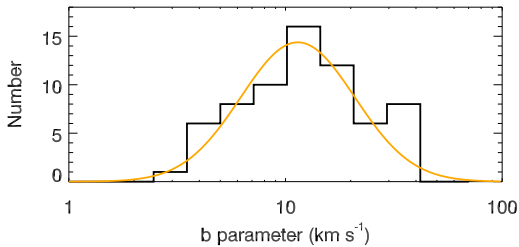}
      \caption{Distribution of the $b$ parameter for absorption components, measured by fitting voigt profiles in high-resolution spectra of GRB afterglows. The orange line is the log-normal fit.}
         \label{Fig:bpar}
   \end{figure}

To limit the amount of free parameters we will assume that the distribution of particle velocities has $\left\langle\log\,(b/km\,s^{-1})\right\rangle =1.1$ (equivalent to $12.6\,km\,s^{-1}$) and $\sigma_{\log\,(b/km\,s^{-1})}=0.3$. This is estimated from the values derived from line profile fitting in 10 GRB afterglows (with a total of 66 velocity components) observed with high spectral resolution using UVES at the Very Large Telescope \citep[see Fig.~\ref{Fig:bpar},][]{del07,pir08,fox08,del09}.

The fitting code that we have created compares the EW histograms to models created in a similar way to what was shown in section~\ref{sec:fitfun}. The models are computed for varying column density distributions and varying number of components. The method searches for minimum $\chi^2$ values within an allowed range of parameters.

To simultaneously use as much information as possible, we fit all the features that we have in our sample together. The results of the fits are displayed in Table~\ref{table:fitN}. The distributions giving the best fits, obtained by this method are shown in Fig.~\ref{Fig:SiII1260} and Fig.~\ref{Fig:GRBs} as dashed blue lines. From these results we obtain that the typical number of components generating absorption lines is $6.00_{-1.00}^{+1.25}$. Within our sample, the most abundant species that are detected in the host galaxies of GRBs are \ion{C}{iv} and \ion{Si}{ii}, with \ion{Ca}{ii} and \ion{Mg}{i} being the least common.

\section{Discussion}

\subsection{Ionisation ratios}

Using the elements of our sample that have different states of ionisation, we can derive equivalent width ratios that will give indicative measures of the typical degrees of ionisation of each element in GRB host galaxies. This can be used to compare with the environments probed in quasar intervening DLA systems or to identify peculiar GRB environments that depart from the typical values.

\begin{figure}[ht!]
   \centering
   \includegraphics[width=8cm]{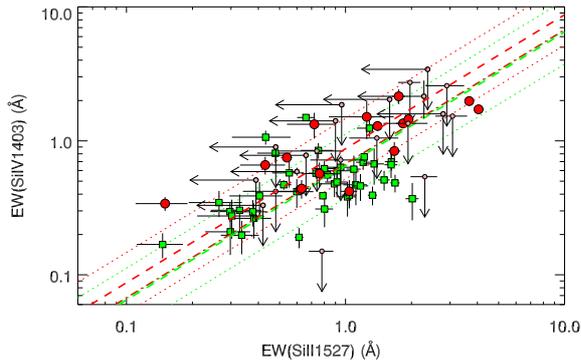}
      \caption{\ion{Si}{iv}$\lambda$1403 vs. \ion{Si}{ii}$\lambda$1527. Red dots are from GRB spectra whereas green boxes are from quasar intervening DLA systems. For clarity, detection limits are only plotted for GRBs. Dashed lines are the average values of the line ratios (detection limits are ignored), dotted are the standard deviation of these values and dashed-dotted lines are the ratios of the average values of the distributions. See Table~\ref{table:ratio} for a summary of the line ratios. 
              }
         \label{Fig:SiIV1402vsSiII1526}
   \end{figure}

We first compare \ion{Si}{iv}$\lambda$1403 with \ion{Si}{ii}$\lambda$1527. These two lines have the advantage of being separated by only 124 {\AA} in the rest-frame, so that they are normally covered in a single spectrum. Figure~\ref{Fig:SiIV1402vsSiII1526} shows the values of \ion{Si}{iv}$\lambda$1403 vs. \ion{Si}{ii}$\lambda$1527 for both GRBs (red dots) and QSO intervening DLAs (green squares). The average ratio of the detections (limits are not considered here) are shown with a dashed line (red for GRBs, green for DLAs), and dotted lines represent the standard deviations. At a first glance we see that GRBs seem to have higher ratios of \ion{Si}{iv}$\lambda$1403/\ion{Si}{ii}$\lambda$1527. However, when we plot the ratio of the distributions from Table~\ref{table:fits} (dashed-dotted line), where limits are considered, we see that it lies below the average value and that it is almost identical to the one from intervening DLAs. In the case of the DLAs the average value of the detections and the distribution are consistent. This implies that the previous average value is biased by non-detections of \ion{Si}{iv}$\lambda$1403, which is usually weaker than \ion{Si}{ii}$\lambda$1527. This does not seem to be so significant for the DLA systems, as the two ratios match.

\begin{figure}[h!]
   \centering
   \includegraphics[width=8cm]{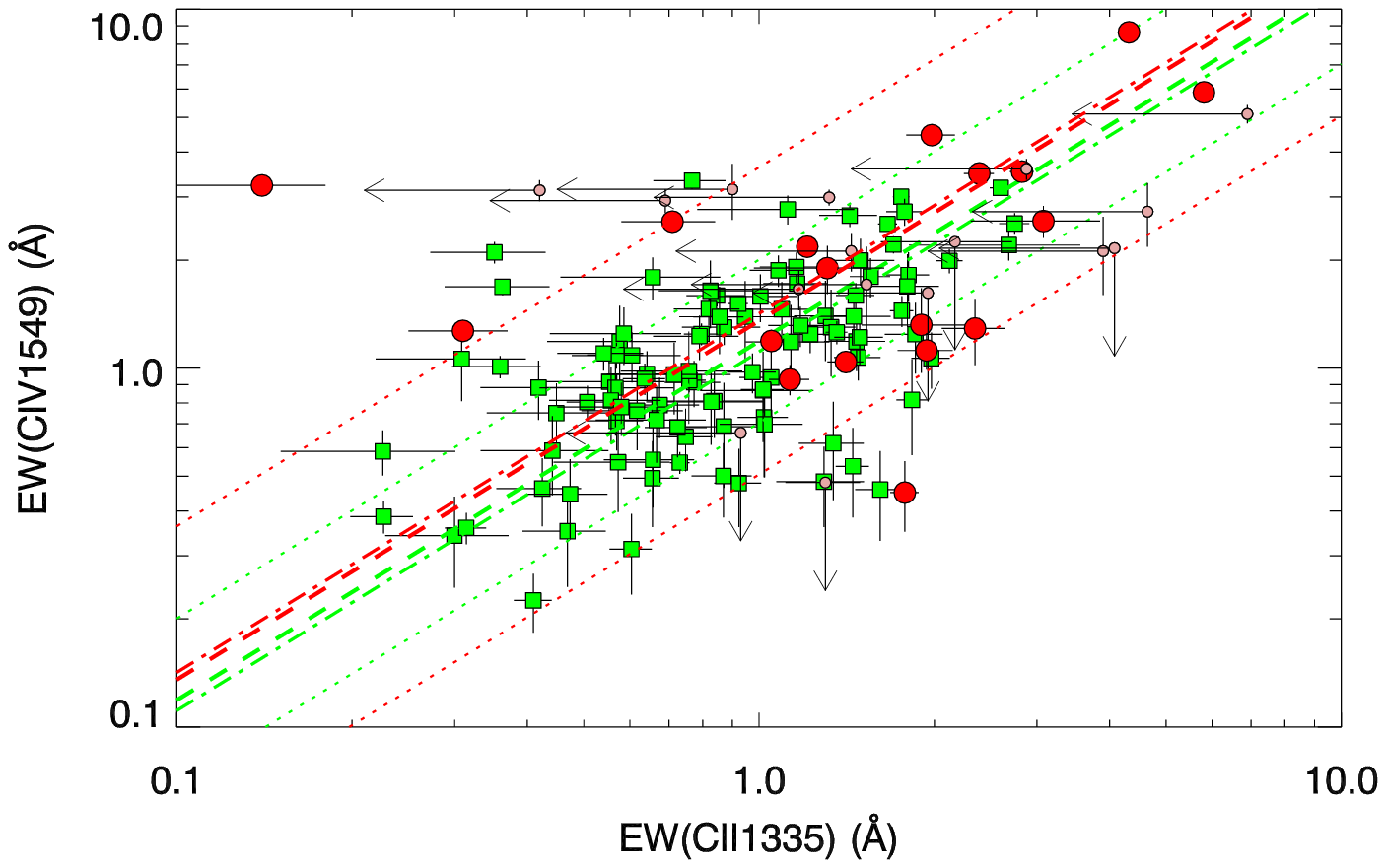}
      \caption{\ion{C}{iv}$\lambda$1549 vs. \ion{C}{ii}$\lambda$1335. The colour coding is the same as in Fig.~\ref{Fig:SiIV1402vsSiII1526}.
              }
         \label{Fig:CIV1548vsCII1334}
   \end{figure}

The ratio between \ion{C}{iv}$\lambda$1549 and \ion{C}{ii}$\lambda$1335 can be a better estimator of the ionisation as, being both very strong features, it provides a large number of detections (Fig.~\ref{Fig:CIV1548vsCII1334}). Furthermore, among the ionisation ratios given here, this one has the lowest dispersion in strength of the features, making the ionisation outliers stand out more clearly. Finally, the ionisation potential of CIV is the highest of our sample (see Table~\ref{table:ion}). In this case the average ratios of the detections and the ratio of the distribution average are consistent, meaning that the detection limits are not strongly affecting the results. 

\begin{figure}[h!]
   \centering
   \includegraphics[width=8cm]{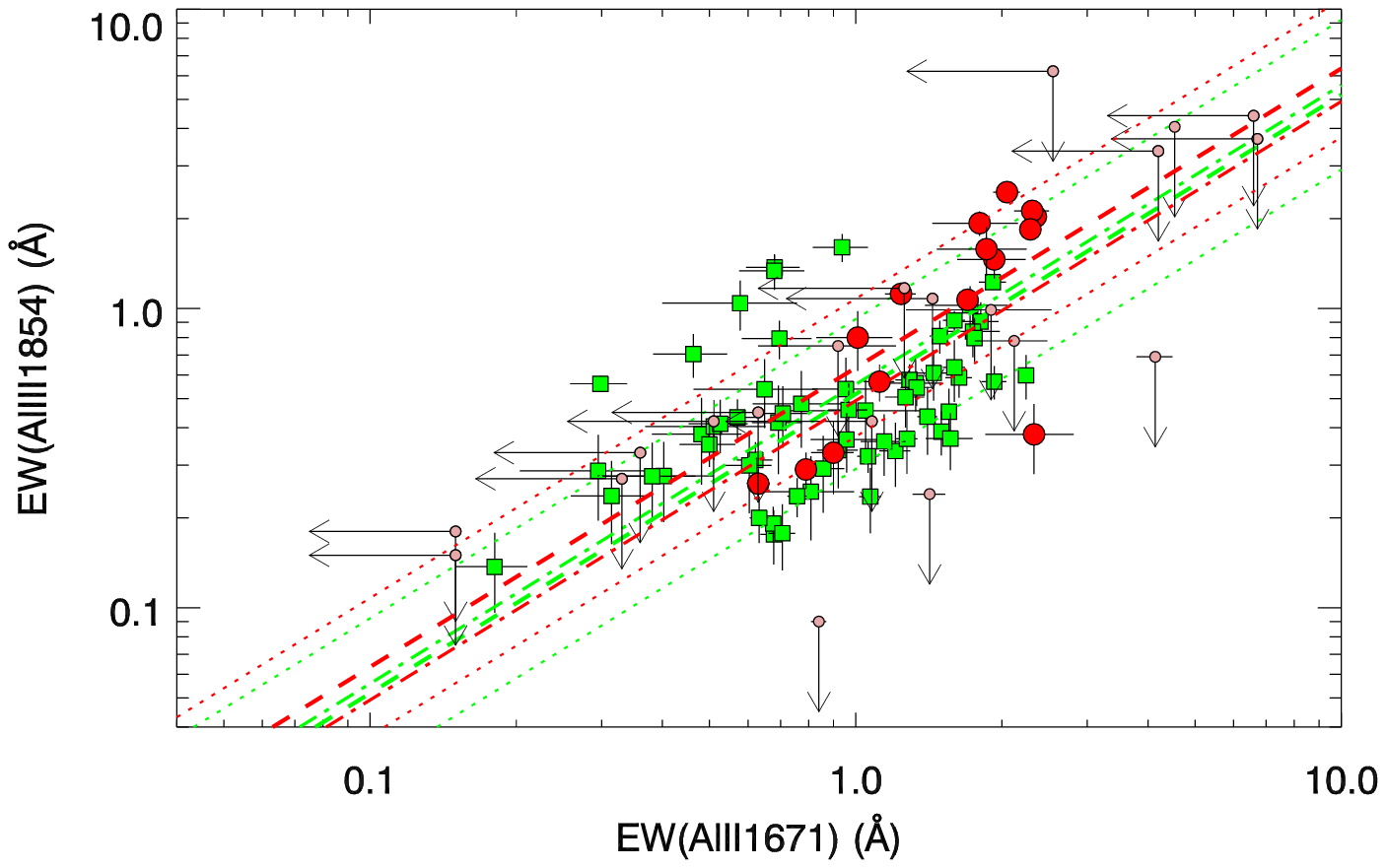}
      \caption{\ion{Al}{iii}$\lambda$1855 vs. \ion{Al}{ii}$\lambda$1671. The colour coding is the same as in Fig.~\ref{Fig:SiIV1402vsSiII1526}.
              }
         \label{Fig:AlIII1854vsAlII1670}
   \end{figure}

In the case of the \ion{Al}{iii}$\lambda$1855 and \ion{Al}{ii}$\lambda$1671 ratio, the ionisation potentials are not as extreme as in the previous cases, meaning that moderate star formation can account for the excitation of Al to these levels (see Fig.~\ref{Fig:AlIII1854vsAlII1670}).

   \begin{figure}[h!]
   \centering
   \includegraphics[width=8cm]{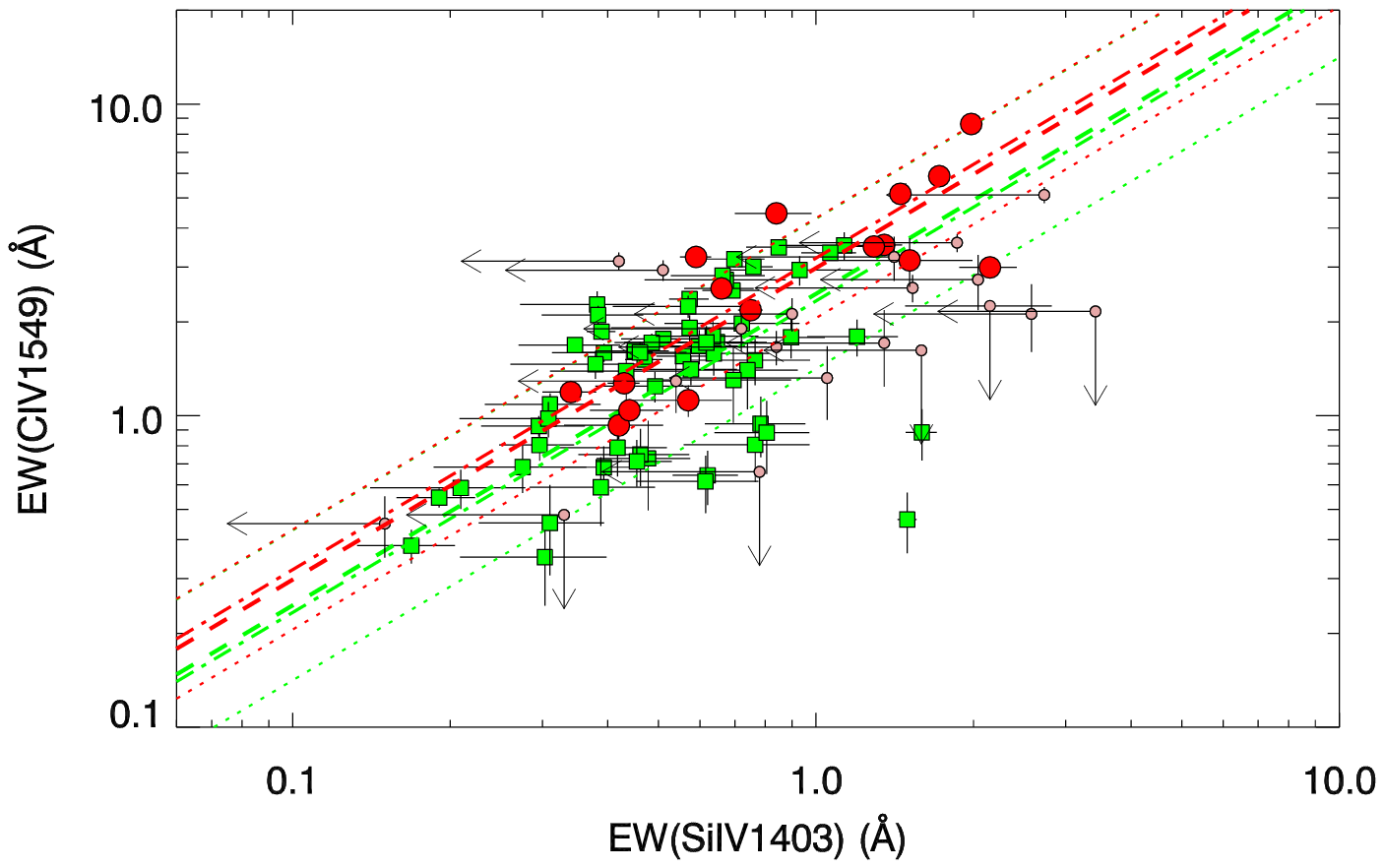}
      \caption{\ion{C}{iv}$\lambda$1549 vs. \ion{Si}{iv}$\lambda$1403. The colour coding is the same as in Fig.~\ref{Fig:SiIV1402vsSiII1526}.
              }
         \label{Fig:CIV1548vsSiIV1402}
   \end{figure}
   
In Fig.~\ref{Fig:CIV1548vsSiIV1402} we plot the ratio of the two more extreme ionised species in our sample, \ion{C}{iv}$\lambda$1549 and \ion{Si}{iv}$\lambda$14032. \citet{ber06} suggest that a Wolf-Rayet outflow model would produce absorbers rich in carbon and low in silicon, making the \ion{C}{iv}/\ion{Si}{iv} a good tracer these kind of massive stars. In fact, this ratio shows a more significant difference with the DLA systems. This indicates a possible excess of Wolf-Rayet stars in the environments of GRBs, consistent with the models that explain long GRBs as the collapse of massive stars.

The fits to all these line ratios are given in Table~\ref{table:ratio}.

\begin{table}[h!]
\caption{EW ratios of selected lines for GRBs and DLAs. Both the average of the detections (when both lines are detected) and the distributions (ratio of the average of the distributions from Tables \ref{table:fits} and \ref{table:dla}) are shown. In the case of the average ratios of detections we give also the dispersion of the distribution.}             
\label{table:ratio}      
\centering                          
\begin{tabular}{|c|c|c|c|c|}        
\hline\hline                 
\multirow{2}{*}{Ratio}	& \multicolumn{2}{c|}{GRB}			& \multicolumn{2}{c|}{DLA}			\\  
\cline{2-5}
					& Det.		& Dist.			& Det.		& Dist.	\\
					& Centre, $\sigma$	& Centre	& Centre, $\sigma$	& Centre	\\
\hline                    
log$\left (\frac{\ion{Si}{iv}\lambda1403}{\ion{Si}{ii}\lambda1527}\right )$	& $-$0.06, 0.23	& $-$0.15			& $-$0.19, 0.24	& $-$0.18		\\
log$\left (\frac{\ion{C}{iv}\lambda1549}{\ion{C}{ii}\lambda1335}\right )$	& 0.13, 0.43	& 0.16			& 0.07, 0.23	& 0.05 			\\
log$\left (\frac{\ion{Al}{iii}\lambda1855}{\ion{Al}{ii}\lambda1671}\right )$	& $-$0.20, 0.23	& $-$0.26			& $-$0.29, 0.25	& $-$0.25			\\
log$\left (\frac{\ion{C}{iv}\lambda1549}{\ion{Si}{iv}\lambda1403}\right )$ 	& 0.47, 0.16	& 0.51			& 0.39, 0.24	& 0.37			\\
\hline                                   
\end{tabular}
\end{table}

\begin{figure}[h]
   \centering
   \includegraphics[width=8cm]{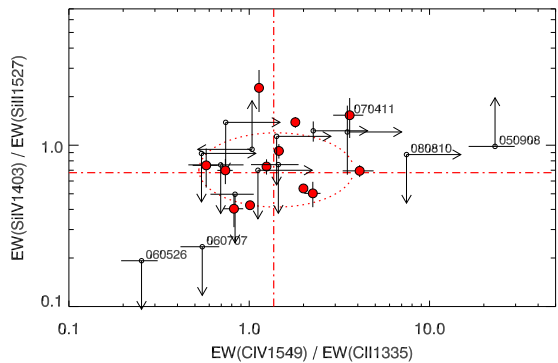}
      \caption{Comparison between the ionisation ratio of carbon and silicon. Dashed dotted lines mark the average values of each ratio and the dotted ellipse the 1-$\sigma$ area.}
         \label{Fig:C_Si}
   \end{figure}

Finally we plot both the ionisation ratios (EW of the highest-ionised species over the EW of the lowest-ionised species) of carbon and silicon in Fig.~\ref{Fig:C_Si}. The ratios of both C and Si and allow us to easily separate highly-ionised (at the top right of the figure) and low-ionised environments (at the lower left) from the average values found in GRBs (centre). It is curious to see that the 4 GRBs with highest ionisation, GRBs 050908, 060124, 080810 and 070411 all have very low neutral hydrogen column densities, with $\log\,(N_{\rm \ion{H}{i}}/cm^{-2}) < 20$ (see also Sect.~\ref{sec:nh}). This relation between high ionisation environments and low column density of neutral hydrogen was already noted by \citet{jak06b} and later modelled by \citet{tho11}. In these very ionised spectra, the photons emitted by the GRB itself or of an environment rich in Wolf-Rayet stars would ionise the hydrogen. In the case of GRB 090426, the amount of absorbing \ion{H}{i} was seen to decrease during the hours that followed the GRB \citep{tho11}, supporting the hypothesis of the ionising GRB.

On the other side, bursts like GRBs 060526 and 060707 show very low ionisation. Given that GRBs are supposed to happen in strong star forming regions, we would expect ionised environments. As a consequence, these bursts with very low ionisation might be mostly probing material that is in the line of sight but unrelated to the star forming regions where GRBs are produced.

   \begin{figure}[h]
   \centering
   \includegraphics[width=8cm]{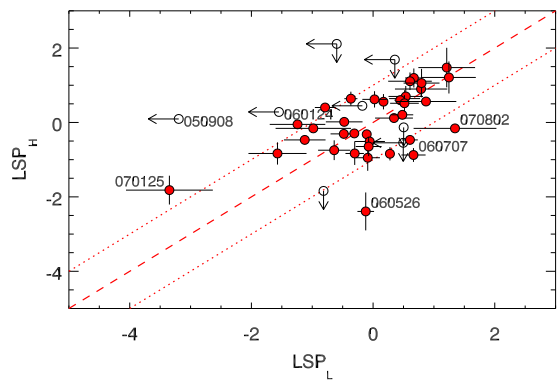}
      \caption{Comparison of the high- and low-ionisation $LSPs$. The dashed line indicates equal low- and high-ionisation $LSP$. The dotted lines mark where they differ by 1.0.}
         \label{Fig:LSPLH}
   \end{figure}

Many GRBs cannot be included in plots like the one in Fig.~\ref{Fig:C_Si} because of the lack of several of the species in the spectrum. In these cases we can study the ionisation of the GRB environment by using the $LSP_{L}$ and $LSP_{H}$. Fig.~\ref{Fig:LSPLH} compares them and shows how some extreme environments, with very low or very high ionisation levels stand out.

\subsection{$LSP$ correlations}

In the following paragraphs we study correlations (or the lack of them) between the strength of the absorption features (through the $LSP$) and several properties of the GRB and its host galaxy environment. For each case we use parametric (Pearson's $r$ correlation coefficient) and non-parametric tests (Spearsman's $\rho$) to determine the strength of the correlations.

\subsubsection{$LSP$ vs. Observation time}

\begin{figure}[h]
   \centering
   \includegraphics[width=8cm]{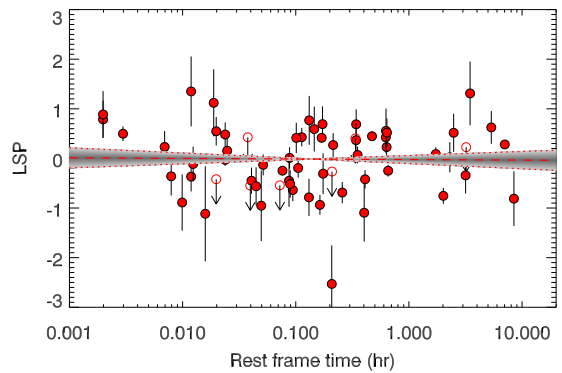}
      \caption{Comparison of the $LSP$ with the time of the observation in rest frame, showing no correlation. The shaded region indicates the uncertainty in the slope of the linear fit.}
         \label{Fig:LSP_t}
   \end{figure}

As a first check, we study how the LSP correlates with the time after the GRB (rest frame) in which the observation was performed. \citet{chr10} find a decrease in the strength of several absorption features of our sample when they compare two composite spectra with average times of 0.4 and 5 hr after the burst. Our data shows no significant correlation of the LSP with time, as shown in Fig.~\ref{Fig:LSP_t}. We find correlation coefficients of $r=-0.015$ (for a sample of 65 data points) and $\rho=-0.03$, implying a probability of correlation of 0.21 ($0.3\sigma$). 

\subsubsection{$LSP$ vs. Extinction}

We here compare the $LSP$ with the extinction, parametrized by A$_V$, the equivalent extinction in the rest-frame $V$-band. $A_V$ is usually derived from SED fitting of power-laws with a superposed extinction law. When available, we use the extinction determined by \citet{zaf11}, using \citet{kan10} and \citet{sch10} for other cases. 

\begin{figure}[h]
   \centering
   \includegraphics[width=8cm]{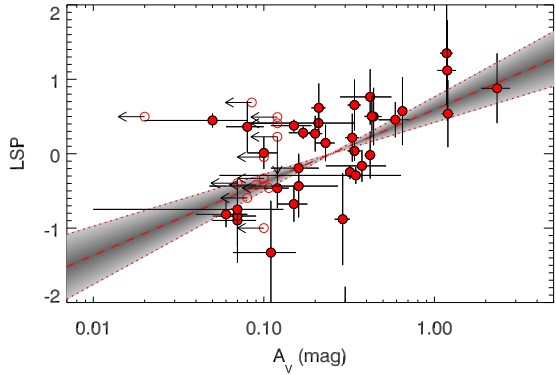}
      \caption{Correlation between the $LSP$ and the intrinsic extinction of the GRB. The shaded region indicates the uncertainty in the slope of the linear fit.}
         \label{Fig:LSP_Av}
   \end{figure}

When a GRB is seen through more material in its host galaxy, the absorption features in the spectrum show higher EWs. Furthermore, if the presence of this material is correlated with the presence of dust, the spectrum would also be more extinguished. This is precisely what we see in Fig.~\ref{Fig:LSP_Av} where the $LSP$ increases with increasing extinction. We obtain correlation coefficients of $r=0.52$ (for a sample of 36 data points) and $\rho=0.63$, indicating a probability of correlation of 0.99997 ($3.7\sigma$). A linear fit to the data gives $LSP = (0.98 \pm 0.28) A_V + (0.59 \pm 0.20)$.

\subsubsection{$LSP$ vs. $N_{\rm H,O}$}
\label{sec:nh}

As the amount of material producing the absorption features increases in the line of sight, together with the increase of the $LSP$, we would expect to see an increase in the column density of neutral hydrogen, measured through the fitting of the absorption Ly$\alpha$ feature. 

\begin{figure}[h]
   \centering
   \includegraphics[width=8cm]{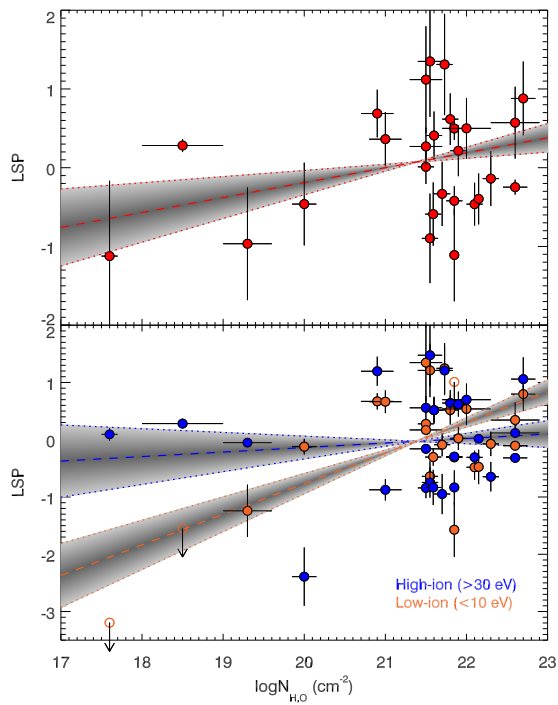}
      \caption{Relation between the $LSP$ and the Ly$\alpha$-derived $N_{\rm H}$ of the GRB. The shaded regions indicate the uncertainty in the slopes of the linear fits.}
         \label{Fig:LSP_nH}
   \end{figure}

Figure~\ref{Fig:LSP_nH} (top) plots the $LSP$ as a function of  the Ly$\alpha$-derived column density of neutral hydrogen, N$_{\rm H,O}$ \citep[from][]{fyn09}. Unfortunately, this is only possible to measure from the ground for GRBs with redshift larger than $z\sim$2, when the 1215 {\AA} feature is redshifted into the visible range. Although there is some tendency to find higher N$_{\rm H}$ in large $LSP$ spectra, there is a great dispersion, pointing to a large dispersion in the metallicities of GRB environments or independent regions generating the absorptions. Correlation coefficients of $r=0.32$ (for a sample of 27 data points) and $\rho=0.15$, with a probability of correlation of 0.56 ($0.7\sigma$) confirms the weakness of the correlation. A linear fit to the data gives $LSP = (0.19 \pm 0.11)\log\,(N_{\rm \ion{H}{i},O}/cm^{-2}) - (3.98 \pm 2.39)$. However, we note that the fit is constrained by those GRBs with $\log\,(N_{\rm \ion{H}{i}} /cm ^{-2}) < 20$. If we exclude those bursts the fit is unconstrained.

It is interesting to see how the plot changes when we consider separately the $LSP_{L}$ and $LSP_{H}$. Figure~\ref{Fig:LSP_nH} (bottom) shows that GRBs with column densities $\log\,(N_{\rm \ion{H}{i},O}/cm^{-2}) < 20$ present extreme ionisation states, showing predominantly very highly ionised environments, although there is also the case of a very low ionised burst. The relation of low neutral hydrogen column densities and high-ionisation has been already seen before \citep{jak06b,cam10,tho11} and interpreted as due to the ionisation of hydrogen in the near GRB environment \citep{laz01,wat07}. 

\subsubsection{$LSP$ vs. $N_{\rm H,X}$}

\begin{figure}[h]
   \centering
   \includegraphics[width=8cm]{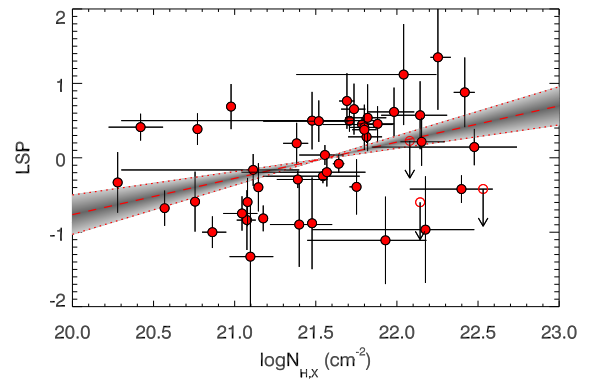}
      \caption{Correlation between the $LSP$ and the X-derived $N_{\rm H}$ of the GRB. The shaded region indicates the uncertainty in the slope of the linear fit.}
         \label{FigLSP_XnH}
   \end{figure}

Using X-ray measurements we can also obtain an estimate of the column density of neutral hydrogen ($N_{\rm H,X}$) by fitting the spectrum of the X-ray afterglow with a power law and an extinction curve \citep{cam10}. However, this extinction is not directly caused by hydrogen but by metals, so it is very dependent on the composition of the host galaxy. The relation between the extinction and $N_{\rm H}$ is well known for the Milky Way, but the uncertainty is large in GRB host galaxies \citep{wat12}. Comparing the $LSP$ with the X-ray derived $N_{\rm H}$ (Fig.~\ref{FigLSP_XnH}) shows a correlation but with large dispersion. We obtain correlation coefficients of $r=0.39$ (for a sample of 44 data points) and $\rho=0.40$, indicating a probability of correlation of 0.993 ($2.6\sigma$). A linear fit gives $LSP = (0.48 \pm 0.17) \log\,(N_{\rm H,X}/cm^{-2}) - (10.51 \pm 3.74)$. In this case we find no significant difference when considering the low- and high-ionised $LSPs$.

\subsubsection{$LSP$ vs. Redshift}

\begin{figure}[h]
   \centering
   \includegraphics[width=8cm]{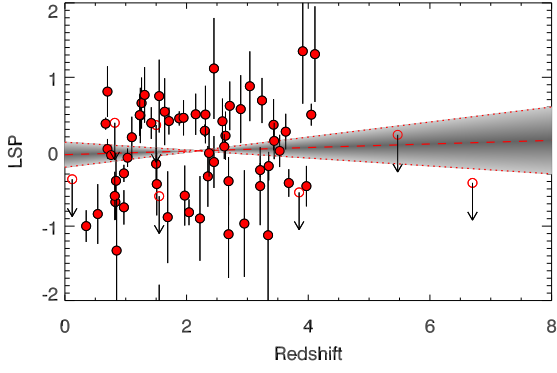}
      \caption{Relation of the $LSP$ with redshift. The shaded region indicates the uncertainty in the slope of the linear fit.}
         \label{Fig_LSP_z}
   \end{figure}

Thanks to the very broad redshift distribution of GRBs, they can help to answer questions related to the evolution of galaxies and their star formation history. In this section we look for a possible evolution of the $LSP$ with the redshift. Figure~\ref{Fig_LSP_z} shows a large dispersion in the $LSPs$ but no significant evolution. A linear fit to the dataset returns a slope of 0.02 $\pm$ 0.08, consistent with no redshift evolution. We obtain correlation coefficients of $r=0.016$ (for a sample of 69 data points) and $\rho=0.06$, indicating a very low correlation probability of 0.41 ($0.5\sigma$).

It has been shown that DLAs show a significant evolution of their metallicity with redshift \citep{wol05}, with lower metallicities for high redshifts. For lower metallicities, one would expect to see, on average, weaker absorption features due to metal lines. For GRBs, this evolution is shallower \citep{fyn06, tho12}, so that it only becomes significant at very high-redshift. In our sample there are only 2 GRBs at $z>5$ and, in those cases, we only have limits for their $LSP$. Although this could be indicative of a lower $LSP$, the sample is not big enough for a high-$z$ study.

\subsubsection{Ionisation vs. $E_{iso}$}

\begin{figure}[h]
   \centering
   \includegraphics[width=8cm]{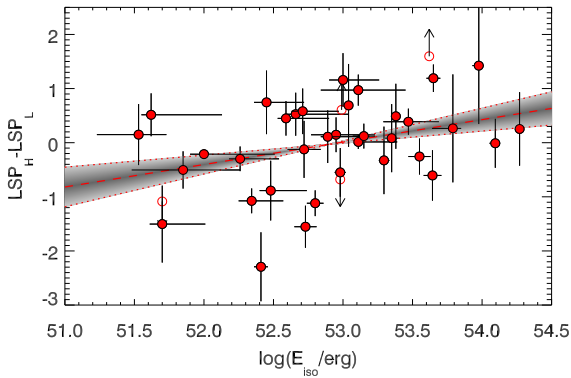}
      \caption{Relation of the ionisation, characterised by $LSP_H-LSP_L$ with $E_{iso}$. The shaded region indicates the uncertainty in the slope of the linear fit.}
         \label{Fig_Eiso}
   \end{figure}

Here we compare the ionisation of the GRB environment, characterised by the difference between the high-ionisation $LSP$ ($LSP_H$) with the low-ionisation $LSP$ ($LSP_L$), with the isotropic equivalent energy release in $\gamma$-rays of the GRB \citep[][]{but07,sak11}. Fig.~\ref{Fig_Eiso} shows that the ionisation grows with $E_{iso}$, with correlation coefficients of $r=0.35$ (for a sample of 35 data points) and $\rho=0.30$, indicating a correlation probability of 0.92 ($1.7\sigma$). A linear fit gives $\left(LSP_H-LSP_L\right) = (0.41 \pm 0.19) \log\,(E_{iso}/erg) - (21.98 \pm 10.17)$.

Although this correlation is weak, it gives a hint of an interesting effect: It could mean that part of the ionisation that we see in afterglow spectra are due to the GRB, or that highly ionised hosts (probably due to intense star formation) produce more energetic GRBs. Part of the dispersion of this correlation can be caused by using the isotropic equivalent energy (where we assume that the energy is emitted isotropically) instead of the beaming-corrected energy (which is the real energy that is released through jets). Unfortunately, we do not have enough data to perform a study with the beaming-corrected energy.

\subsubsection{$LSP$ vs. host galaxy absolute magnitude}

\begin{figure}[h]
   \centering
   \includegraphics[width=8cm]{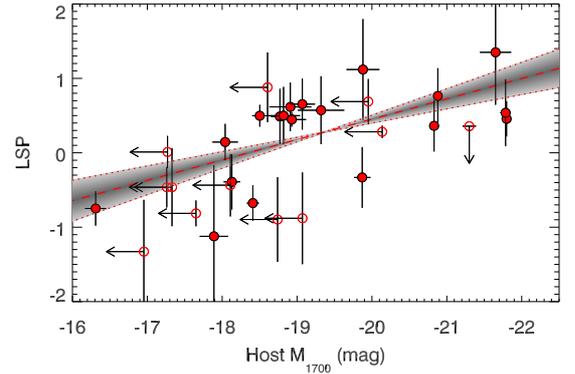}
      \caption{Relation of the $LSP$ with host galaxy magnitude. The shaded region indicates the uncertainty in the slope of the linear fit.}
         \label{Fig_LSP_hosts}
   \end{figure}

In this section we compare the $LSP$ with the absolute host magnitude at a rest-frame wavelength of 1700 {\AA}. These data are obtained from Schulze et al. (in prep.), using the observations from the \textit{TOUGH sample}\footnote{http://www.dark-cosmology.dk/TOUGH} \citep[][Malesani in prep.]{hjo12}, the \textit{GHostS sample}\footnote{http://www.grbhosts.org/} and the \textit{Keck GRB host project imaging catalog}\footnote{http://www.astro.caltech.edu/$\sim$dperley/hosts}. Figure~\ref{Fig_LSP_hosts} shows the fit of the $LSP$ versus host absolute magnitude comparison. In spite of the dispersion, there is a clear correlation that follows $LSP=(-0.27\pm0.08) M_{host,1700} -(5.02\pm1.59)$.  We obtain correlation coefficients of $r=-0.62$ (for a sample of 19 data points) and $\rho=-0.64$, indicating a probability of correlation of 0.997 ($2.7\sigma$).

Larger, more luminous host galaxies, contain more material, and should produce, on average, stronger lines in the spectra of GRBs that they host. Due to the random location of the GRB within the galaxy, the relation between the strength of the lines and the host magnitude will have an intrinsic dispersion, which is also seen in the plot. We have done a similar exercise for $LSP_{L}$ and $LSP_{H}$ looking for differences in the correlation, which could be indicative of different locations for lower and higher ionised species within the host, such as a higher ionisation near the GRB, but found no significant difference.

\subsubsection{$LSP$ vs. galaxy offset}

\begin{figure}[h]
   \centering
   \includegraphics[width=8cm]{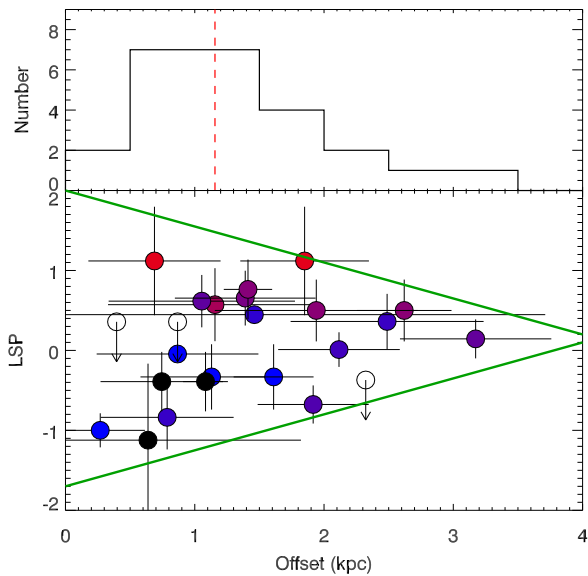}
      \caption{Bottom: Relation of the $LSP$ with the projected offset between the afterglow and the host galaxy. The colour coding indicates the extinction of the afterglow, being red the most extinguished and blue the least ones, those bursts without a reliable extinction measurement are indicated in black. Green lines are qualitative indicators of the distribution limits. Top: Histogram of offsets. The red-dashed line indicates the median offset.}
         \label{Fig_LSP_offset}
   \end{figure}

In this final section we compare the $LSP$ with the projected offset of the GRB afterglow with respect to the centroid of the host galaxy. The offset values have been obtained from Malesani et al. (2012 in prep.). Figure~\ref{Fig_LSP_offset} shows, in its lower panel, the $LSP$ versus the offset, using a colour coding for each object, where redder dots imply higher extinction and bluer ones lower extinction. Those objects for which there is no reliable extinction measurement are coloured black. The top panel shows the histogram of offsets, where we can see that the median offset is 1.16 kpc, similar to the value of 1.3 kpc obtained by \citet{blo02} for an HST sample.

\begin{figure}[h]
   \centering
   \includegraphics[width=8cm]{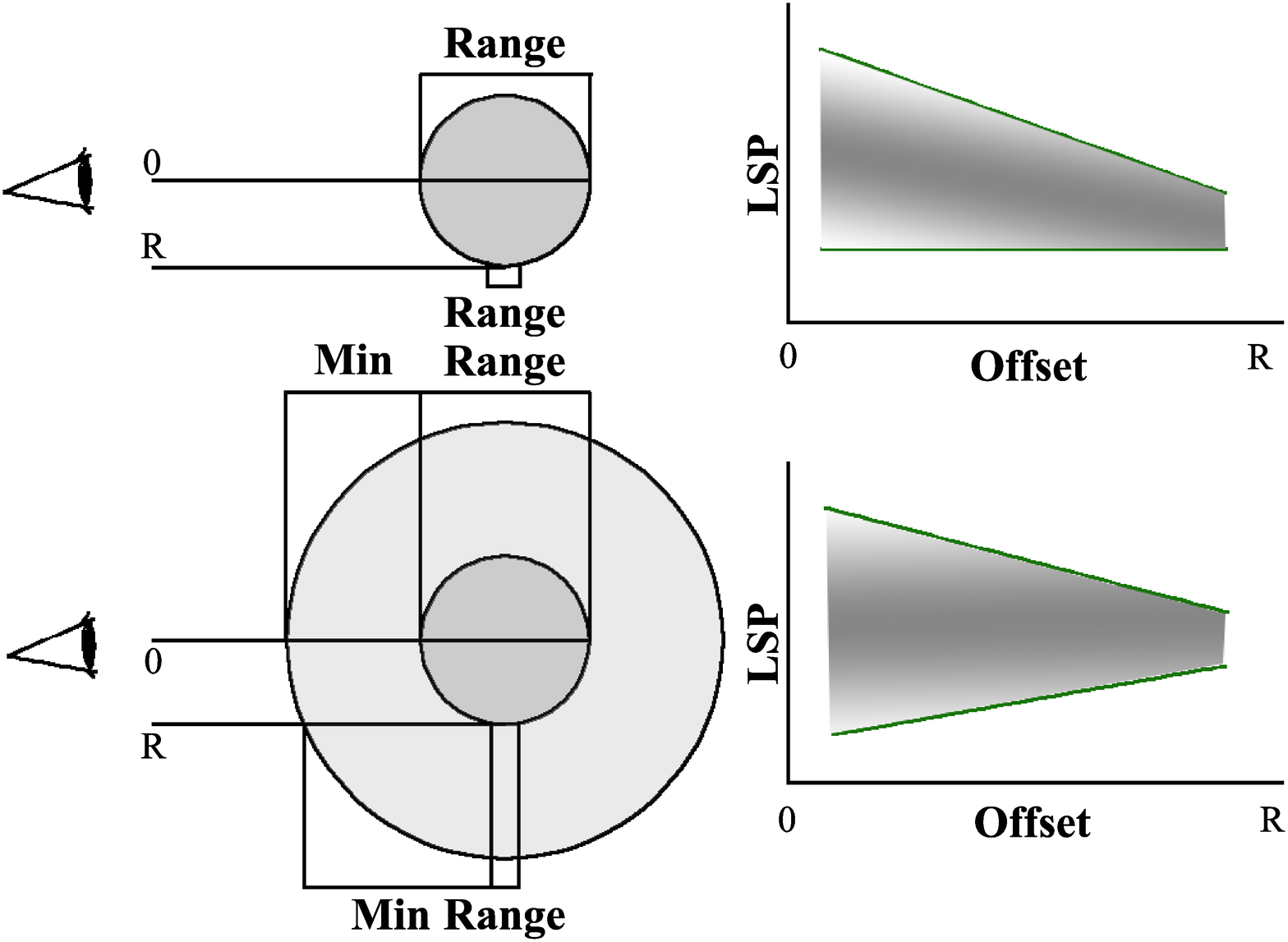}
      \caption{Cartoon showing a limited region of the host in which the GRBs are generated (dark circle) and the effect that an extended region surrounding the previous (or the lack of it) would have in the distribution of $LSPs$ as a function of the galaxy offsets. We indicate how the 'range' of intervening material from the inner region is reduced as the offset increases, while the the 'minimum' material, given by the extended region increases with the offset.}
         \label{Fig_off}
   \end{figure}

If the GRBs could happen at any position within the host galaxy, we would expect that the maximum $LSP$ would decline as the offset increases, whereas the minimum $LSP$ would be limited by the detectability of lines and would not depend on the offset. In Fig.~\ref{Fig_LSP_offset} we see that the first statement is matched, but the second is not. We see how the minimum value of the $LSP$ increases as we increase the offset. A possible explanation to this, is that GRBs only occur within limited areas of the host galaxies (e.g. the most prolific star forming regions). In this case there would be additional material, producing the minimum $LSP$ in an extended region of the galaxy where no GRBs would form. Due to a projection effect, the minimum $LSP$ value produced by this extended region of the host is higher for larger offsets, as depicted in Fig.~\ref{Fig_off} for a very simplified example. Figure~\ref{Fig_LSP_offset} also shows how the bursts at the top of the distribution, which are the ones located behind most of the host galaxy material are also the ones that show the highest extinction, as one would expect. We have also considered if other biases would produce the increase in the minimum $LSP$ with the offset, like having fainter afterglows at higher offsets, differences in the host galaxies, etc. Although we have not found evidence of any significant bias, we cannot discard other effects.

\section{Summary and conclusions}

We have used 69 low-resolution optical spectra of GRB afterglows, covering the redshift range $0.12 <z< 6.7$, to perform a statistical study of the rest-frame equivalent widths of strong absorption features. 61 of those spectra come from the sample presented by \citet{fyn09} while the measurements of the other 8 are presented in this paper. The study includes 22 spectral features produced by 12 different atomic species. To obtain a consistent statistical analysis we include in our sample both detections and detection limits of these spectral features. A summary of the observations is presented in Tables~\ref{table:data1} and \ref{table:data2}. This collection of observations is used for the following:

\begin{itemize}
\item For each spectral feature we calculate the average EW and its standard deviation, from which we obtain a picture of the typical GRB afterglow spectrum and the range of variations that can be expected.
\item Using these values we propose the use of the EW diagrams as graphical tools to compare an individual GRB to the sample and give a catalogue with all our 69 spectra.
\item As compared with a sample of quasar intervening DLA spectra, we find the GRB absorbers to be, on average, 2.5 times stronger and slightly more ionised. The difference of line strength is more significant in the case of the \ion{C}{iv} that is 3.4 times stronger in GRBs. There is also an excess in the \ion{C}{iv}$\lambda$1549/\ion{Si}{iv}$\lambda$1403 ratio. This excess of \ion{C}{iv} can be related to the existence of an excess of Wolf-Rayet stars in the environment of GRBs, consistent with the massive star origin of the explosions.
\item We introduce the $LSP$ as an overall measurement of the strength of the absorption features in a GRB spectrum relative to the sample as a single number. We give the distribution of $LSPs$ so that one can derive, for a specific spectrum the percentile of line strength in which it lies. Similar $LSPs$ are obtained for each GRB afterglow spectrum considering only low- and highly-ionised species respectively.
\item We find correlations between the $LSP$ and the extinction of the GRB, the hydrogen column density and the brightness of the host. However, the $LSP$ is independent of the redshift and the time at which the observation was performed.
\item There is a correlation between the ionisation of the absorbers in the GRB afterglow spectrum and the energy of the $\gamma$-ray emission, indicating that, either the GRB event is responsible for part of the ionisation, or galaxies with higher ionisation (i.e. stronger star formation) produce more energetic GRBs.
\item When comparing the $LSP$ with the projected offset between the host and the afterglow, we detect a decay in the maximum value as the afterglow lies further away from the galaxy, as it would be expected. However, we also detect an increase in the minimum $LSP$, which could be due to an observational bias that we did not consider, but also due to a geometric effect that could provide information on the structure of the host galaxy. Larger samples will be needed to confirm this.
\item Using the distributions of EWs of single species features we use a CoG fitting method to derive the distribution of their column densities and the number of components that make those features. For this we assume that the distribution of velocities of the absorbing atoms is described by $\left\langle\log\,(b/\rm{km\,s}^{-1})\right\rangle=1.1$ and $\sigma_{\log\,(b/\rm{km\,s}^{-1})}=0.3$, as derived from line profile fitting of a sample of high resolution GRB spectra. From the fit we obtain that the average number of components that produce features in GRB afterglow spectra is $6.00_{-1.25}^{+1.00}$. The column densities of each of the species are given in Table~\ref{table:fitN}.
\item The most extreme ionisation ratios in our sample are found for GRBs with low neutral hydrogen column density. Whether this is due to ionisation by the GRB emission will have to be determined in further detailed studies. 

\end{itemize}

\begin{acknowledgements}

We thank the anonymous referee for providing constructive comments. We thank Alexander Kann for fruitful discussions.
The Dark Cosmology Centre is funded by the DNRF.
This research has made use of the GHostS database (www.grbhosts.org), which is partly funded by Spitzer/NASA grant RSA Agreement No. 1287913.
The research activity of AdUP, CT, RSR and JG is supported by Spanish research projects AYA2011-24780/ESP, AYA2009-14000-C03-01/ESP and AYA2010-21887-C04-01.
JPUF and BMJ acknowledge support from the ERC-StG grant EGGS-278202.
SS acknowledges support by a Grant of Excellence from the Icelandic Research Fund.
GL is supported by the Swedish Research Council through grant No. 623-2011-7117.
TZ is funded within the BINGO! (Ôhistory of Baryons: INtergalactic medium/Galaxies cO-evolutionÕ) project by the Agence Nationale de la Recherche (ANR) under the allocation ANR-08-BLAN-0316-01.

\end{acknowledgements}

\bibliographystyle{aa}
\bibliography{ewdist}

\begin{thebibliography}{53}
\expandafter\ifx\csname natexlab\endcsname\relax\def\natexlab#1{#1}\fi

\bibitem[{{Berger} {et~al.}(2008){Berger}, {Fox}, {Cucchiara}, \&
  {Cenko}}]{ber08}
{Berger}, E., {Fox}, D.~B., {Cucchiara}, A., \& {Cenko}, S.~B. 2008, GRB
  Coordinates Network, 8335

\bibitem[{{Berger} {et~al.}(2006){Berger}, {Penprase}, {Cenko}, {Kulkarni},
  {Fox}, {Steidel}, \& {Reddy}}]{ber06}
{Berger}, E., {Penprase}, B.~E., {Cenko}, S.~B., {et~al.} 2006, \apj, 642, 979

\bibitem[{{Bloom} {et~al.}(2002){Bloom}, {Kulkarni}, \& {Djorgovski}}]{blo02}
{Bloom}, J.~S., {Kulkarni}, S.~R., \& {Djorgovski}, S.~G. 2002, \aj, 123, 1111

\bibitem[{{Butler} {et~al.}(2007){Butler}, {Kocevski}, {Bloom}, \&
  {Curtis}}]{but07}
{Butler}, N.~R., {Kocevski}, D., {Bloom}, J.~S., \& {Curtis}, J.~L. 2007, \apj,
  671, 656

\bibitem[{{Campana} {et~al.}(2010){Campana}, {Th{\"o}ne}, {de Ugarte Postigo},
  {Tagliaferri}, {Moretti}, \& {Covino}}]{cam10}
{Campana}, S., {Th{\"o}ne}, C.~C., {de Ugarte Postigo}, A., {et~al.} 2010,
  \mnras, 402, 2429

\bibitem[{{Christensen} {et~al.}(2011){Christensen}, {Fynbo}, {Prochaska},
  {Th{\"o}ne}, {de Ugarte Postigo}, \& {Jakobsson}}]{chr10}
{Christensen}, L., {Fynbo}, J.~P.~U., {Prochaska}, J.~X., {et~al.} 2011, \apj,
  727, 73

\bibitem[{{Christensen} {et~al.}(2004){Christensen}, {Hjorth}, \&
  {Gorosabel}}]{chr04}
{Christensen}, L., {Hjorth}, J., \& {Gorosabel}, J. 2004, \aap, 425, 913

\bibitem[{{Cucchiara} {et~al.}(2011){Cucchiara}, {Levan}, {Fox}, {Tanvir},
  {Ukwatta}, {Berger}, {Kr{\"u}hler}, {K{\"u}pc{\"u} Yolda{\c s}}, {Wu},
  {Toma}, {Greiner}, {Olivares}, {Rowlinson}, {Amati}, {Sakamoto}, {Roth},
  {Stephens}, {Fritz}, {Fynbo}, {Hjorth}, {Malesani}, {Jakobsson}, {Wiersema},
  {O'Brien}, {Soderberg}, {Foley}, {Fruchter}, {Rhoads}, {Rutledge}, {Schmidt},
  {Dopita}, {Podsiadlowski}, {Willingale}, {Wolf}, {Kulkarni}, \&
  {D'Avanzo}}]{cuc11}
{Cucchiara}, A., {Levan}, A.~J., {Fox}, D.~B., {et~al.} 2011, \apj, 736, 7

\bibitem[{{D'Avanzo} {et~al.}(2008){D'Avanzo}, {D'Elia}, \& {Covino}}]{dav08}
{D'Avanzo}, P., {D'Elia}, V., \& {Covino}, S. 2008, GRB Coordinates Network,
  8345

\bibitem[{{de Ugarte Postigo} {et~al.}(2009{\natexlab{a}}){de Ugarte Postigo},
  {Gorosabel}, {Fynbo}, {Wiersema}, \& {Tanvir}}]{deu09d}
{de Ugarte Postigo}, A., {Gorosabel}, J., {Fynbo}, J.~P.~U., {Wiersema}, K., \&
  {Tanvir}, N. 2009{\natexlab{a}}, GRB Coordinates Network, 9771

\bibitem[{{de Ugarte Postigo} {et~al.}(2009{\natexlab{b}}){de Ugarte Postigo},
  {Gorosabel}, {Malesani}, {Fynbo}, \& {Levan}}]{deu09b}
{de Ugarte Postigo}, A., {Gorosabel}, J., {Malesani}, D., {Fynbo}, J.~P.~U., \&
  {Levan}, A.~J. 2009{\natexlab{b}}, GRB Coordinates Network, 9381

\bibitem[{{de Ugarte Postigo} {et~al.}(2009{\natexlab{c}}){de Ugarte Postigo},
  {Gorosabel}, {Malesani}, {Fynbo}, \& {Levan}}]{deu09c}
{de Ugarte Postigo}, A., {Gorosabel}, J., {Malesani}, D., {Fynbo}, J.~P.~U., \&
  {Levan}, A.~J. 2009{\natexlab{c}}, GRB Coordinates Network, 9383

\bibitem[{{de Ugarte Postigo} {et~al.}(2009{\natexlab{d}}){de Ugarte Postigo},
  {Jakobsson}, {Malesani}, {Fynbo}, {Simpson}, \& {Barros}}]{deu09}
{de Ugarte Postigo}, A., {Jakobsson}, P., {Malesani}, D., {et~al.}
  2009{\natexlab{d}}, GRB Coordinates Network, 8766

\bibitem[{{D'Elia} {et~al.}(2011){D'Elia}, {Campana}, {Covino}, {D'Avanzo},
  {Piranomonte}, \& {Tagliaferri}}]{del11}
{D'Elia}, V., {Campana}, S., {Covino}, S., {et~al.} 2011, \mnras, 418, 680

\bibitem[{{D'Elia} {et~al.}(2007){D'Elia}, {Fiore}, {Meurs}, {Chincarini},
  {Melandri}, {Norci}, {Pellizza}, {Perna}, {Piranomonte}, {Sbordone},
  {Stella}, {Tagliaferri}, {Vergani}, {Ward}, {Angelini}, {Antonelli},
  {Burrows}, {Campana}, {Capalbi}, {Cimatti}, {Costa}, {Cusumano}, {Della
  Valle}, {Filliatre}, {Fontana}, {Frontera}, {Fugazza}, {Gehrels}, {Giannini},
  {Giommi}, {Goldoni}, {Guetta}, {Israel}, {Lazzati}, {Malesani}, {Marconi},
  {Mason}, {Mereghetti}, {Mirabel}, {Molinari}, {Moretti}, {Nousek}, {Perri},
  {Piro}, {Stratta}, {Testa}, \& {Vietri}}]{del07}
{D'Elia}, V., {Fiore}, F., {Meurs}, E.~J.~A., {et~al.} 2007, \aap, 467, 629

\bibitem[{{D'Elia} {et~al.}(2009){D'Elia}, {Fiore}, {Perna}, {Krongold},
  {Covino}, {Fugazza}, {Lazzati}, {Nicastro}, {Antonelli}, {Campana},
  {Chincarini}, {D'Avanzo}, {Della Valle}, {Goldoni}, {Guetta}, {Guidorzi},
  {Meurs}, {Mirabel}, {Molinari}, {Norci}, {Piranomonte}, {Stella}, {Stratta},
  {Tagliaferri}, \& {Ward}}]{del09}
{D'Elia}, V., {Fiore}, F., {Perna}, R., {et~al.} 2009, \apj, 694, 332

\bibitem[{{Fox} {et~al.}(2008){Fox}, {Ledoux}, {Vreeswijk}, {Smette}, \&
  {Jaunsen}}]{fox08}
{Fox}, A.~J., {Ledoux}, C., {Vreeswijk}, P.~M., {Smette}, A., \& {Jaunsen},
  A.~O. 2008, \aap, 491, 189

\bibitem[{{Fruchter} {et~al.}(2006){Fruchter}, {Levan}, {Strolger},
  {Vreeswijk}, {Thorsett}, {Bersier}, {Burud}, {Castro Cer{\'o}n},
  {Castro-Tirado}, {Conselice}, {Dahlen}, {Ferguson}, {Fynbo}, {Garnavich},
  {Gibbons}, {Gorosabel}, {Gull}, {Hjorth}, {Holland}, {Kouveliotou}, {Levay},
  {Livio}, {Metzger}, {Nugent}, {Petro}, {Pian}, {Rhoads}, {Riess}, {Sahu},
  {Smette}, {Tanvir}, {Wijers}, \& {Woosley}}]{fru06}
{Fruchter}, A.~S., {Levan}, A.~J., {Strolger}, L., {et~al.} 2006, \nat, 441,
  463

\bibitem[{{Fynbo} {et~al.}(2009){Fynbo}, {Jakobsson}, {Prochaska}, {Malesani},
  {Ledoux}, {de Ugarte Postigo}, {Nardini}, {Vreeswijk}, {Wiersema}, {Hjorth},
  {Sollerman}, {Chen}, {Th{\"o}ne}, {Bj{\"o}rnsson}, {Bloom}, {Castro-Tirado},
  {Christensen}, {De Cia}, {Fruchter}, {Gorosabel}, {Graham}, {Jaunsen},
  {Jensen}, {Kann}, {Kouveliotou}, {Levan}, {Maund}, {Masetti},
  {Milvang-Jensen}, {Palazzi}, {Perley}, {Pian}, {Rol}, {Schady}, {Starling},
  {Tanvir}, {Watson}, {Xu}, {Augusteijn}, {Grundahl}, {Telting}, \&
  {Quirion}}]{fyn09}
{Fynbo}, J.~P.~U., {Jakobsson}, P., {Prochaska}, J.~X., {et~al.} 2009, \apjs,
  185, 526

\bibitem[{{Fynbo} {et~al.}(2008){Fynbo}, {Prochaska}, {Sommer-Larsen},
  {Dessauges-Zavadsky}, \& {M{\o}ller}}]{fyn08}
{Fynbo}, J.~P.~U., {Prochaska}, J.~X., {Sommer-Larsen}, J.,
  {Dessauges-Zavadsky}, M., \& {M{\o}ller}, P. 2008, \apj, 683, 321

\bibitem[{{Fynbo} {et~al.}(2006){Fynbo}, {Starling}, {Ledoux}, {Wiersema},
  {Th{\"o}ne}, {Sollerman}, {Jakobsson}, {Hjorth}, {Watson}, {Vreeswijk},
  {M{\o}ller}, {Rol}, {Gorosabel}, {N{\"a}r{\"a}nen}, {Wijers},
  {Bj{\"o}rnsson}, {Castro Cer{\'o}n}, {Curran}, {Hartmann}, {Holland},
  {Jensen}, {Levan}, {Limousin}, {Kouveliotou}, {Nelemans}, {Pedersen},
  {Priddey}, \& {Tanvir}}]{fyn06}
{Fynbo}, J.~P.~U., {Starling}, R.~L.~C., {Ledoux}, C., {et~al.} 2006, \aap,
  451, L47

\bibitem[{{Galama} {et~al.}(1998){Galama}, {Vreeswijk}, {van Paradijs},
  {Kouveliotou}, {Augusteijn}, {B{\"o}hnhardt}, {Brewer}, {Doublier},
  {Gonzalez}, {Leibundgut}, {Lidman}, {Hainaut}, {Patat}, {Heise}, {in't Zand},
  {Hurley}, {Groot}, {Strom}, {Mazzali}, {Iwamoto}, {Nomoto}, {Umeda},
  {Nakamura}, {Young}, {Suzuki}, {Shigeyama}, {Koshut}, {Kippen}, {Robinson},
  {de Wildt}, {Wijers}, {Tanvir}, {Greiner}, {Pian}, {Palazzi}, {Frontera},
  {Masetti}, {Nicastro}, {Feroci}, {Costa}, {Piro}, {Peterson}, {Tinney},
  {Boyle}, {Cannon}, {Stathakis}, {Sadler}, {Begam}, \& {Ianna}}]{gal98}
{Galama}, T.~J., {Vreeswijk}, P.~M., {van Paradijs}, J., {et~al.} 1998, \nat,
  395, 670

\bibitem[{{Hjorth} {et~al.}(2012){Hjorth}, {Malesani}, {Jakobsson}, {Jaunsen},
  {Fynbo}, {Gorosabel}, {Kr{\"u}hler}, {Levan}, {Micha{\l}owski},
  {Milvang-Jensen}, {M{\o}ller}, {Schulze}, {Tanvir}, \& {Watson}}]{hjo12}
{Hjorth}, J., {Malesani}, D., {Jakobsson}, P., {et~al.} 2012, \apj, 756, 187

\bibitem[{{Jakobsson} {et~al.}(2009){Jakobsson}, {de Ugarte Postigo},
  {Gorosabel}, {Tanvir}, {Christensen}, \& {Fynbo}}]{jak09}
{Jakobsson}, P., {de Ugarte Postigo}, A., {Gorosabel}, J., {et~al.} 2009, GRB
  Coordinates Network, 9797

\bibitem[{{Jakobsson} {et~al.}(2006{\natexlab{a}}){Jakobsson}, {Fynbo},
  {Ledoux}, {Vreeswijk}, {Kann}, {Hjorth}, {Priddey}, {Tanvir}, {Reichart},
  {Gorosabel}, {Klose}, {Watson}, {Sollerman}, {Fruchter}, {de Ugarte Postigo},
  {Wiersema}, {Bj{\"o}rnsson}, {Chapman}, {Th{\"o}ne}, {Pedersen}, \&
  {Jensen}}]{jak06b}
{Jakobsson}, P., {Fynbo}, J.~P.~U., {Ledoux}, C., {et~al.} 2006{\natexlab{a}},
  \aap, 460, L13

\bibitem[{{Jakobsson} {et~al.}(2006{\natexlab{b}}){Jakobsson}, {Levan},
  {Fynbo}, {Priddey}, {Hjorth}, {Tanvir}, {Watson}, {Jensen}, {Sollerman},
  {Natarajan}, {Gorosabel}, {Castro Cer{\'o}n}, {Pedersen}, {Pursimo},
  {{\'A}rnad{\'o}ttir}, {Castro-Tirado}, {Davis}, {Deeg}, {Fiuza},
  {Mikolaitis}, \& {Sousa}}]{jak06}
{Jakobsson}, P., {Levan}, A., {Fynbo}, J.~P.~U., {et~al.} 2006{\natexlab{b}},
  \aap, 447, 897

\bibitem[{{Kann} {et~al.}(2010){Kann}, {Klose}, {Zhang}, {Malesani}, {Nakar},
  {Pozanenko}, {Wilson}, {Butler}, {Jakobsson}, {Schulze}, {Andreev},
  {Antonelli}, {Bikmaev}, {Biryukov}, {B{\"o}ttcher}, {Burenin}, {Castro
  Cer{\'o}n}, {Castro-Tirado}, {Chincarini}, {Cobb}, {Covino}, {D'Avanzo},
  {D'Elia}, {Della Valle}, {de Ugarte Postigo}, {Efimov}, {Ferrero}, {Fugazza},
  {Fynbo}, {G{\aa}lfalk}, {Grundahl}, {Gorosabel}, {Gupta}, {Guziy}, {Hafizov},
  {Hjorth}, {Holhjem}, {Ibrahimov}, {Im}, {Israel}, {Je{\'l}inek}, {Jensen},
  {Karimov}, {Khamitov}, {Kizilo{\v g}lu}, {Klunko}, {Kub{\'a}nek}, {Kutyrev},
  {Laursen}, {Levan}, {Mannucci}, {Martin}, {Mescheryakov}, {Mirabal},
  {Norris}, {Ovaldsen}, {Paraficz}, {Pavlenko}, {Piranomonte}, {Rossi},
  {Rumyantsev}, {Salinas}, {Sergeev}, {Sharapov}, {Sollerman}, {Stecklum},
  {Stella}, {Tagliaferri}, {Tanvir}, {Telting}, {Testa}, {Updike}, {Volnova},
  {Watson}, {Wiersema}, \& {Xu}}]{kan10}
{Kann}, D.~A., {Klose}, S., {Zhang}, B., {et~al.} 2010, \apj, 720, 1513

\bibitem[{{Kann} {et~al.}(2007){Kann}, {Masetti}, \& {Klose}}]{kan07}
{Kann}, D.~A., {Masetti}, N., \& {Klose}, S. 2007, \aj, 133, 1187

\bibitem[{{Lazzati} {et~al.}(2001){Lazzati}, {Perna}, \& {Ghisellini}}]{laz01}
{Lazzati}, D., {Perna}, R., \& {Ghisellini}, G. 2001, \mnras, 325, L19

\bibitem[{{Malesani} {et~al.}(2009){Malesani}, {Fynbo}, {D'Elia}, {de Ugarte
  Postigo}, {Jakobsson}, \& {Thoene}}]{mal09}
{Malesani}, D., {Fynbo}, J.~P.~U., {D'Elia}, V., {et~al.} 2009, GRB Coordinates
  Network, 9457

\bibitem[{{Morton}(2003)}]{mor03}
{Morton}, D.~C. 2003, \apjs, 149, 205

\bibitem[{{Noterdaeme} {et~al.}(2009){Noterdaeme}, {Petitjean}, {Ledoux}, \&
  {Srianand}}]{not09}
{Noterdaeme}, P., {Petitjean}, P., {Ledoux}, C., \& {Srianand}, R. 2009, \aap,
  505, 1087

\bibitem[{{Piranomonte} {et~al.}(2008){Piranomonte}, {Ward}, {Fiore},
  {Vergani}, {D'Elia}, {Krongold}, {Nicastro}, {Meurs}, {Chincarini}, {Covino},
  {Della Valle}, {Fugazza}, {Norci}, {Sbordone}, {Stella}, {Tagliaferri},
  {Burrows}, {Gehrels}, {Goldoni}, {Malesani}, {Mirabel}, {Pellizza}, \&
  {Perna}}]{pir08}
{Piranomonte}, S., {Ward}, P.~A., {Fiore}, F., {et~al.} 2008, \aap, 492, 775

\bibitem[{{Pontzen} {et~al.}(2010){Pontzen}, {Deason}, {Governato}, {Pettini},
  {Wadsley}, {Quinn}, {Brooks}, {Bellovary}, \& {Fynbo}}]{pon10}
{Pontzen}, A., {Deason}, A., {Governato}, F., {et~al.} 2010, \mnras, 402, 1523

\bibitem[{{Prochaska} {et~al.}(2007){Prochaska}, {Chen}, {Dessauges-Zavadsky},
  \& {Bloom}}]{pro07}
{Prochaska}, J.~X., {Chen}, H.-W., {Dessauges-Zavadsky}, M., \& {Bloom}, J.~S.
  2007, \apj, 666, 267

\bibitem[{{Prochter} {et~al.}(2006){Prochter}, {Prochaska}, {Chen}, {Bloom},
  {Dessauges-Zavadsky}, {Foley}, {Lopez}, {Pettini}, {Dupree}, \&
  {Guhathakurta}}]{pro06}
{Prochter}, G.~E., {Prochaska}, J.~X., {Chen}, H.-W., {et~al.} 2006, \apjl,
  648, L93

\bibitem[{{Racusin} {et~al.}(2008){Racusin}, {Karpov}, {Sokolowski}, {Granot},
  {Wu}, {Pal'Shin}, {Covino}, {van der Horst}, {Oates}, {Schady}, {Smith},
  {Cummings}, {Starling}, {Piotrowski}, {Zhang}, {Evans}, {Holland}, {Malek},
  {Page}, {Vetere}, {Margutti}, {Guidorzi}, {Kamble}, {Curran}, {Beardmore},
  {Kouveliotou}, {Mankiewicz}, {Melandri}, {O'Brien}, {Page}, {Piran},
  {Tanvir}, {Wrochna}, {Aptekar}, {Barthelmy}, {Bartolini}, {Beskin}, {Bondar},
  {Bremer}, {Campana}, {Castro-Tirado}, {Cucchiara}, {Cwiok}, {D'Avanzo},
  {D'Elia}, {Della Valle}, {de Ugarte Postigo}, {Dominik}, {Falcone}, {Fiore},
  {Fox}, {Frederiks}, {Fruchter}, {Fugazza}, {Garrett}, {Gehrels},
  {Golenetskii}, {Gomboc}, {Gorosabel}, {Greco}, {Guarnieri}, {Immler},
  {Jelinek}, {Kasprowicz}, {La Parola}, {Levan}, {Mangano}, {Mazets},
  {Molinari}, {Moretti}, {Nawrocki}, {Oleynik}, {Osborne}, {Pagani}, {Pandey},
  {Paragi}, {Perri}, {Piccioni}, {Ramirez-Ruiz}, {Roming}, {Steele}, {Strom},
  {Testa}, {Tosti}, {Ulanov}, {Wiersema}, {Wijers}, {Winters}, {Zarnecki},
  {Zerbi}, {M{\'e}sz{\'a}ros}, {Chincarini}, \& {Burrows}}]{rac08}
{Racusin}, J.~L., {Karpov}, S.~V., {Sokolowski}, M., {et~al.} 2008, \nat, 455,
  183

\bibitem[{{Sakamoto} {et~al.}(2011){Sakamoto}, {Barthelmy}, {Baumgartner},
  {Cummings}, {Fenimore}, {Gehrels}, {Krimm}, {Markwardt}, {Palmer}, {Parsons},
  {Sato}, {Stamatikos}, {Tueller}, {Ukwatta}, \& {Zhang}}]{sak11}
{Sakamoto}, T., {Barthelmy}, S.~D., {Baumgartner}, W.~H., {et~al.} 2011, \apjs,
  195, 2

\bibitem[{{Salvaterra} {et~al.}(2009){Salvaterra}, {Della Valle}, {Campana},
  {Chincarini}, {Covino}, {D'Avanzo}, {Fern{\'a}ndez-Soto}, {Guidorzi},
  {Mannucci}, {Margutti}, {Th{\"o}ne}, {Antonelli}, {Barthelmy}, {de Pasquale},
  {D'Elia}, {Fiore}, {Fugazza}, {Hunt}, {Maiorano}, {Marinoni}, {Marshall},
  {Molinari}, {Nousek}, {Pian}, {Racusin}, {Stella}, {Amati}, {Andreuzzi},
  {Cusumano}, {Fenimore}, {Ferrero}, {Giommi}, {Guetta}, {Holland}, {Hurley},
  {Israel}, {Mao}, {Markwardt}, {Masetti}, {Pagani}, {Palazzi}, {Palmer},
  {Piranomonte}, {Tagliaferri}, \& {Testa}}]{sal09}
{Salvaterra}, R., {Della Valle}, M., {Campana}, S., {et~al.} 2009, \nat, 461,
  1258

\bibitem[{{Savaglio} {et~al.}(2003){Savaglio}, {Fall}, \& {Fiore}}]{sav03}
{Savaglio}, S., {Fall}, S.~M., \& {Fiore}, F. 2003, \apj, 585, 638

\bibitem[{{Schady} {et~al.}(2010){Schady}, {Page}, {Oates}, {Still}, {de
  Pasquale}, {Dwelly}, {Kuin}, {Holland}, {Marshall}, \& {Roming}}]{sch10}
{Schady}, P., {Page}, M.~J., {Oates}, S.~R., {et~al.} 2010, \mnras, 401, 2773

\bibitem[{{Svensson} {et~al.}(2010){Svensson}, {Levan}, {Tanvir}, {Fruchter},
  \& {Strolger}}]{sve10}
{Svensson}, K.~M., {Levan}, A.~J., {Tanvir}, N.~R., {Fruchter}, A.~S., \&
  {Strolger}, L.-G. 2010, \mnras, 405, 57

\bibitem[{{Tanvir} {et~al.}(2009){Tanvir}, {Fox}, {Levan}, {Berger},
  {Wiersema}, {Fynbo}, {Cucchiara}, {Kr{\"u}hler}, {Gehrels}, {Bloom},
  {Greiner}, {Evans}, {Rol}, {Olivares}, {Hjorth}, {Jakobsson}, {Farihi},
  {Willingale}, {Starling}, {Cenko}, {Perley}, {Maund}, {Duke}, {Wijers},
  {Adamson}, {Allan}, {Bremer}, {Burrows}, {Castro-Tirado}, {Cavanagh}, {de
  Ugarte Postigo}, {Dopita}, {Fatkhullin}, {Fruchter}, {Foley}, {Gorosabel},
  {Kennea}, {Kerr}, {Klose}, {Krimm}, {Komarova}, {Kulkarni}, {Moskvitin},
  {Mundell}, {Naylor}, {Page}, {Penprase}, {Perri}, {Podsiadlowski}, {Roth},
  {Rutledge}, {Sakamoto}, {Schady}, {Schmidt}, {Soderberg}, {Sollerman},
  {Stephens}, {Stratta}, {Ukwatta}, {Watson}, {Westra}, {Wold}, \&
  {Wolf}}]{tan09}
{Tanvir}, N.~R., {Fox}, D.~B., {Levan}, A.~J., {et~al.} 2009, \nat, 461, 1254

\bibitem[{{Tejos} {et~al.}(2007){Tejos}, {Lopez}, {Prochaska}, {Chen}, \&
  {Dessauges-Zavadsky}}]{tej07}
{Tejos}, N., {Lopez}, S., {Prochaska}, J.~X., {Chen}, H.-W., \&
  {Dessauges-Zavadsky}, M. 2007, \apj, 671, 622

\bibitem[{{Th{\"o}ne} {et~al.}(2011){Th{\"o}ne}, {Campana}, {Lazzati}, {de
  Ugarte Postigo}, {Fynbo}, {Christensen}, {Levan}, {Aloy}, {Hjorth},
  {Jakobsson}, {Levesque}, {Malesani}, {Milvang-Jensen}, {Roming}, {Tanvir},
  {Wiersema}, {Gladders}, {Wuyts}, \& {Dahle}}]{tho11}
{Th{\"o}ne}, C.~C., {Campana}, S., {Lazzati}, D., {et~al.} 2011, \mnras, 414,
  479

\bibitem[{{Th{\"o}ne} {et~al.}(2012){Th{\"o}ne}, {Fynbo}, {Goldoni}, {de Ugarte
  Postigo}, {Campana}, {Vergani}, {Covino}, {Kruehler}, {Kaper}, {Tanvir},
  {Zafar}, {D'Elia}, {Gorosabel}, {Greiner}, {Groot}, {Hammer}, {Jakobsson},
  {Klose}, {Levan}, {Milvang-Jensen}, {Nicuesa Guelbenzu}, {Palazzi},
  {Piranomonte}, {Tagliaferri}, {Watson}, {Wiersema}, \& {Wijers}}]{tho12}
{Th{\"o}ne}, C.~C., {Fynbo}, J.~P.~U., {Goldoni}, P., {et~al.} 2012,
  arXiv:1206.2337

\bibitem[{{Th{\"o}ne} {et~al.}(2009){Th{\"o}ne}, {Jakobsson}, {De Cia},
  {Levan}, {Fynbo}, {Hjorth}, {Malesani}, {Tanvir}, {Fugazza}, \&
  {D'Avanzo}}]{tho09}
{Th{\"o}ne}, C.~C., {Jakobsson}, P., {De Cia}, A., {et~al.} 2009, GRB
  Coordinates Network, 9409

\bibitem[{{Vergani} {et~al.}(2009){Vergani}, {Petitjean}, {Ledoux},
  {Vreeswijk}, {Smette}, \& {Meurs}}]{ver09}
{Vergani}, S.~D., {Petitjean}, P., {Ledoux}, C., {et~al.} 2009, \aap, 503, 771

\bibitem[{{Vreeswijk} {et~al.}(2007){Vreeswijk}, {Ledoux}, {Smette}, {Ellison},
  {Jaunsen}, {Andersen}, {Fruchter}, {Fynbo}, {Hjorth}, {Kaufer}, {M{\o}ller},
  {Petitjean}, {Savaglio}, \& {Wijers}}]{vre07}
{Vreeswijk}, P.~M., {Ledoux}, C., {Smette}, A., {et~al.} 2007, \aap, 468, 83

\bibitem[{{Watson} {et~al.}(2007){Watson}, {Hjorth}, {Fynbo}, {Jakobsson},
  {Foley}, {Sollerman}, \& {Wijers}}]{wat07}
{Watson}, D., {Hjorth}, J., {Fynbo}, J.~P.~U., {et~al.} 2007, \apjl, 660, L101

\bibitem[{{Watson} \& {Jakobsson}(2012)}]{wat12}
{Watson}, D. \& {Jakobsson}, P. 2012, \apj, 754, 89

\bibitem[{{Wolfe} {et~al.}(2005){Wolfe}, {Gawiser}, \& {Prochaska}}]{wol05}
{Wolfe}, A.~M., {Gawiser}, E., \& {Prochaska}, J.~X. 2005, \araa, 43, 861

\bibitem[{{Zafar} {et~al.}(2011){Zafar}, {Watson}, {Fynbo}, {Malesani},
  {Jakobsson}, \& {de Ugarte Postigo}}]{zaf11}
{Zafar}, T., {Watson}, D., {Fynbo}, J.~P.~U., {et~al.} 2011, \aap, 532, A143

\end{thebibliography}

\longtabL{1}{
\begin{landscape}
\begin{footnotesize}
\begin{longtable}{l c c c c c c c c c c c c}        
\caption{\label{table:data1}   Rest-frame EWs (in \AA) of lines with wavelength below 2000\AA. The table displays detections or upper limits (3-$\sigma$) whenever the spectrum covers the range of the feature and the feature is not affected by strong telluric absorptions.}\\          
\hline\hline
GRB    & \textit{z}      & 1260         & 1303         & 1335         & 1394         & 1403         & 1527         & 1549         & 1608         & 1671         & 1855         & 1863         \\
       &        & \ion{Si}{ii}            & \ion{O}{i}/\ion{Si}{ii}/\ion{O}{i}*     & \ion{C}{ii}/\ion{C}{ii}*        & \ion{Si}{iv}            & \ion{Si}{iv}            & \ion{Si}{ii}            & \ion{C}{iv}/\ion{C}{iv}         & \ion{Fe}{ii}            & \ion{Al}{ii}            & \ion{Al}{iii}           & \ion{Al}{iii}           \\
\hline                        
\endfirsthead
\caption{continued.}\\
\hline\hline 
GRB    & \textit{z}      & 1260         & 1303         & 1335         & 1394         & 1403         & 1527         & 1549         & 1608         & 1671         & 1855         & 1863         \\
       &        & \ion{Si}{ii}            & \ion{O}{i}/\ion{Si}{ii}/\ion{O}{i}*     & \ion{C}{ii}/\ion{C}{ii}*        & \ion{Si}{iv}            & \ion{Si}{iv}            & \ion{Si}{ii}            & \ion{C}{iv}/\ion{C}{iv}         & \ion{Fe}{ii}            & \ion{Al}{ii}            & \ion{Al}{iii}           & \ion{Al}{iii}           \\
\hline                        
\endhead
\hline
\endfoot
050319  &3.24 &   1.82$\pm$   0.55  &   2.70$\pm$   0.75  &   2.17$\pm$   0.55  &$<$   2.05 &   2.15$\pm$   0.67  &$<$   2.27 &$<$   2.26 &$<$   2.57 &--- &--- & --- \\
050401  &2.89 &$<$   3.39 &   3.60$\pm$   1.03  &   3.08$\pm$   0.77  &$<$   3.60  &$<$   1.53 &   3.08$\pm$   0.51  &   2.57$\pm$   0.26  &   2.19$\pm$   0.39  &   1.80$\pm$   0.36  &   1.93$\pm$   0.18  &   1.93 $\pm$   0.21  \\
050408  &1.24 &--- &--- &--- &--- &--- &--- &--- &--- &--- &--- & --- \\
050730  &3.97 &   1.11$\pm$   0.06  &$<$   1.43  &   0.99$\pm$   0.04  &   0.82$\pm$   0.03  &   0.46$\pm$   0.05  &--- &--- &   0.26$\pm$   0.04  &   0.62$\pm$   0.06  &--- & --- \\
050802  &1.71 &--- &$<$  14.18 &$<$   4.66 &$<$   2.48 &$<$   2.04 &$<$   1.58 &   2.73$\pm$   0.55  &$<$   1.63 &$<$   1.44 &$<$   1.08 &$<$   0.97 \\
050824  &0.83 &--- &--- &--- &--- &--- &--- &--- &--- &--- &--- & --- \\
050908  &3.34 &$<$   0.08 &   0.30$\pm$   0.05  &   0.14$\pm$   0.04  &$<$   1.50  &   0.59$\pm$   0.04  &$<$   0.61  &   3.23$\pm$   0.05  &--- &$<$   0.15 &$<$   0.18 &$<$   0.18 \\
050922C &2.22 &   0.65$\pm$   0.06  &$<$   2.17  &   0.31$\pm$   0.06  &   0.47$\pm$   0.03  &   0.43$\pm$   0.03  &   0.62$\pm$   0.03  &   1.27$\pm$   0.03  &   0.19$\pm$   0.03  &   0.84$\pm$   0.03  &$<$   0.10 &$<$   0.11 \\
060115  &3.53 &$<$   1.19 &   2.54$\pm$   0.40  &   1.90$\pm$   0.40  &$<$   1.02 &$<$   1.04 &   1.39$\pm$   0.35  &   1.32$\pm$   0.35  &$<$   1.29 &--- &$<$   2.63 &$<$   2.29 \\
060124  &2.30 &$<$   0.85 &$<$   0.64 &$<$   0.70 &   1.09$\pm$   0.28  &$<$   0.50 &$<$   0.39 &   2.93$\pm$   0.22  &$<$   0.32 &$<$   0.35 &$<$   0.33 &$<$   0.33 \\
060206  &4.05 &$<$   2.37  &   2.78$\pm$   0.02  &   2.27$\pm$   0.03  &--- &--- &--- &--- &--- &--- &--- & --- \\
060210  &3.91 &   4.42$\pm$   0.09  &   4.85$\pm$   0.08  &   4.32$\pm$   0.08  &   3.05$\pm$   0.06  &   1.98$\pm$   0.03  &   3.67$\pm$   0.04  &   8.63$\pm$   0.05  &   1.61$\pm$   0.04  &--- &--- & --- \\
060502A &1.50 &--- &--- &--- &--- &--- &--- &--- &--- &--- &$<$   2.95 &$<$   2.78 \\
060526  &3.21 &   1.43$\pm$   0.07  &   1.78$\pm$   0.10  &   1.78$\pm$   0.10  &$<$   0.18 &$<$   0.16 &   0.78$\pm$   0.10  &   0.45$\pm$   0.10  &$<$   0.15 &   0.86$\pm$   0.10  &--- & $<$   0.33 \\
060614  &0.12 &--- &--- &--- &--- &--- &--- &--- &--- &--- &--- & --- \\
060707  &3.43 &$<$   1.00 &   3.43$\pm$   0.70  &   2.35$\pm$   0.29  &$<$   0.53 &$<$   0.53 &   2.30$\pm$   0.32  &   1.29$\pm$   0.27  &$<$   0.58 &   2.12$\pm$   0.36  &$<$   0.78 &$<$   0.79 \\
060714  &2.71 &   1.94$\pm$   0.16  &   3.42$\pm$   0.16  &   2.83$\pm$   0.13  &   1.56$\pm$   0.11  &   1.35$\pm$   0.11  &   1.83$\pm$   0.13  &   3.53$\pm$   0.11  &   1.35$\pm$   0.13  &   2.05$\pm$   0.13  &   2.45$\pm$   0.13  &   1.16 $\pm$   0.13  \\
060729  &0.54 &--- &--- &--- &--- &--- &--- &--- &--- &--- &--- & --- \\
060904B &0.70 &--- &--- &--- &--- &--- &--- &--- &--- &--- &--- & --- \\
060906  &3.68 &   0.85$\pm$   0.21  &   1.18$\pm$   0.24  &   1.30$\pm$   0.19  &   0.68$\pm$   0.17  &$<$   0.34 &$<$   0.42 &$<$   0.47 &$<$   0.47 &--- &$<$   0.76 &$<$   0.67 \\
060908  &1.88 &--- &--- &--- &$<$   1.33 &$<$   1.40 &$<$   0.89 &   3.23$\pm$   0.52  &$<$   0.71 &$<$   1.08 &$<$   0.43 &$<$   0.43 \\
060926  &3.21 &$<$   0.74 &$<$   0.72 &   1.31$\pm$   0.33  &$<$   0.73 &$<$   0.73 &   0.95$\pm$   0.26  &   1.90$\pm$   0.29  &$<$   0.41 &--- &--- & --- \\
060927  &5.47 &$<$   1.29 &$<$   2.96 &$<$   2.67 &--- &--- &--- &--- &--- &--- &--- & --- \\
061007  &1.26 &--- &--- &--- &--- &--- &--- &--- &$<$   8.18 &$<$   6.73 &$<$   3.70 &$<$   3.62 \\
061021  &0.35 &--- &--- &--- &--- &--- &--- &--- &--- &--- &--- & --- \\
061110A &0.76 &--- &--- &--- &--- &--- &--- &--- &--- &--- &$<$  12.17 &$<$  11.69 \\
061110B &3.44 &$<$   2.28 &   2.61$\pm$   0.47  &   1.51$\pm$   0.32  &   1.49$\pm$   0.27  &   1.33$\pm$   0.29  &   0.72$\pm$   0.18  &--- &$<$   0.44 &   0.92$\pm$   0.29  &$<$   0.76 &$<$   0.80 \\
061121  &1.31 &--- &--- &--- &--- &--- &   2.05$\pm$   0.06  &   4.42$\pm$   0.06  &   1.93$\pm$   0.03  &   2.35$\pm$   0.03  &   2.03$\pm$   0.03  &   1.62 $\pm$   0.03  \\
070110  &2.35 &   1.34$\pm$   0.12  &   2.12$\pm$   0.12  &   1.13$\pm$   0.09  &   0.42$\pm$   0.09  &   0.42$\pm$   0.09  &   1.04$\pm$   0.09  &   0.93$\pm$   0.09  &   0.75$\pm$   0.09  &   0.90$\pm$   0.09  &   0.33$\pm$   0.09  &   0.30 $\pm$   0.09  \\
070125  &1.55 &--- &--- &--- &--- &--- &$<$   0.25 &   0.67$\pm$   0.12  &$<$   0.19 &$<$   0.16 &$<$   0.14 &$<$   0.14 \\
070306  &1.50 &--- &--- &--- &--- &--- &$<$   9.25 &$<$   7.64 &$<$   5.84 &$<$   4.21 &$<$   3.37 &$<$   3.31 \\
070318  &0.84 &--- &--- &--- &--- &--- &--- &--- &--- &--- &--- & --- \\
070411  &2.95 &   0.28$\pm$   0.08  &   0.33$\pm$   0.08  &   0.71$\pm$   0.13  &   0.78$\pm$   0.10  &   0.66$\pm$   0.10  &   0.43$\pm$   0.10  &   2.56$\pm$   0.13  &$<$   0.13 &   0.46$\pm$   0.08  &--- & --- \\
070419A &0.97 &--- &--- &--- &--- &--- &--- &--- &--- &--- &--- & --- \\
070506  &2.31 &   2.72$\pm$   0.30  &   3.23$\pm$   0.33  &$<$   1.31 &   1.42$\pm$   0.21  &   2.15$\pm$   0.27  &   1.75$\pm$   0.12  &   2.99$\pm$   0.15  &   1.42$\pm$   0.09  &   1.24$\pm$   0.09  &   1.12$\pm$   0.06  &   0.51 $\pm$   0.06  \\
070508  &0.82 &--- &--- &--- &--- &--- &--- &--- &--- &--- &--- & --- \\
070611  &2.04 &$<$   0.92 &$<$   0.92 &$<$   0.94 &$<$   0.78 &$<$   0.77 &$<$   0.67 &$<$   0.65 &$<$   0.59 &$<$   0.52 &$<$   0.41 &$<$   0.40 \\
070721B &3.63 &$<$   1.30 &   1.40$\pm$   0.41  &$<$   0.89 &   1.30$\pm$   0.41  &   1.51$\pm$   0.48  &   1.25$\pm$   0.41  &   3.15$\pm$   0.56  &$<$   1.02 &--- &$<$   1.33 &$<$   1.91 \\
070802  &2.45 &$<$   5.70 &$<$   4.56 &$<$   3.91 &$<$   2.70 &$<$   2.57 &   2.90$\pm$   0.58  &   2.12$\pm$   0.52  &   3.04$\pm$   0.46  &   4.14$\pm$   0.35  &$<$   0.70 &$<$   0.82 \\
071020  &2.15 &$<$   5.33 &$<$   4.87 &$<$   4.08 &$<$   3.19 &$<$   3.42 &$<$   2.36 &$<$   2.16 &$<$   1.84 &   1.90$\pm$   0.63  &$<$   0.99 &$<$   1.01 \\
071031  &2.69 &   0.72$\pm$   0.04  &   1.33$\pm$   0.03  &   1.21$\pm$   0.04  &   1.00$\pm$   0.04  &   0.75$\pm$   0.04  &   0.54$\pm$   0.03  &   2.18$\pm$   0.04  &   0.88$\pm$   0.04  &   0.63$\pm$   0.03  &   0.26$\pm$   0.03  &--- \\
071112C &0.82 &--- &--- &--- &--- &--- &--- &--- &--- &--- &--- & --- \\
080210  &2.64 &   1.15$\pm$   0.19  &   2.66$\pm$   0.16  &   2.39$\pm$   0.14  &   1.57$\pm$   0.11  &   1.29$\pm$   0.11  &   1.40$\pm$   0.08  &   3.49$\pm$   0.11  &   1.02$\pm$   0.08  &   1.70$\pm$   0.08  &   1.07$\pm$   0.08  &   0.66 $\pm$   0.08  \\
080319B &0.97 &--- &--- &--- &--- &--- &--- &--- &--- &--- &$<$   3.44 &$<$   3.23 \\
080319C &1.95 &--- &--- &$<$   2.88 &$<$   1.98 &$<$   1.87 &$<$   0.95 &   3.59$\pm$   0.24  &$<$   0.67 &   1.93$\pm$   0.31  &   1.46$\pm$   0.20  &   1.42 $\pm$   0.20  \\
080330  &1.51 &--- &--- &--- &--- &--- &$<$   0.92 &   1.71$\pm$   0.28  &$<$   0.74 &$<$   0.62 &$<$   0.44 &$<$   0.44 \\
080411  &1.03 &--- &--- &--- &--- &--- &--- &--- &--- &--- &--- &    0.70 $\pm$   0.05  \\
080413B &1.10 &--- &--- &--- &--- &--- &--- &--- &$<$   0.09 &   2.33$\pm$   0.48  &   0.38$\pm$   0.10  &   0.38 $\pm$   0.10  \\
080520  &1.55 &--- &--- &--- &--- &--- &$<$   8.69 &$<$  10.25 &$<$   4.74 &$<$   6.59 &$<$   4.40 &$<$   4.35 \\
080603B &2.69 &$<$   0.11 &   0.72$\pm$   0.04  &   1.05$\pm$   0.04  &   0.51$\pm$   0.04  &   0.34$\pm$   0.04  &   0.15$\pm$   0.04  &   1.19$\pm$   0.04  &$<$   0.33  &   0.42$\pm$   0.04  &--- & --- \\
080604  &1.42 &--- &--- &--- &--- &--- &--- &--- &--- &--- &--- & --- \\
080605  &1.64 &--- &--- &$<$   6.89 &$<$   3.89 &$<$   2.74 &   1.97$\pm$   0.23  &   5.11$\pm$   0.30  &   0.61$\pm$   0.19  &   2.31$\pm$   0.19  &   2.12$\pm$   0.11  &   1.55 $\pm$   0.11  \\
080607  &3.04 &--- &--- &--- &   2.67$\pm$   0.10  &   1.45$\pm$   0.07  &   1.94$\pm$   0.04  &   5.15$\pm$   0.06  &   3.67$\pm$   0.05  &   2.29$\pm$   0.02  &   1.84$\pm$   0.03  &   1.52 $\pm$   0.04  \\
080707  &1.23 &--- &--- &--- &--- &--- &--- &--- &--- &--- &$<$   1.28 &$<$   1.26 \\
080710  &0.85 &--- &--- &--- &--- &--- &--- &--- &--- &--- &--- & --- \\
080721  &2.59 &   1.75$\pm$   0.19  &   1.70$\pm$   0.19  &   1.98$\pm$   0.19  &   1.34$\pm$   0.14  &   0.84$\pm$   0.14  &   1.67$\pm$   0.11  &   4.46$\pm$   0.17  &   2.45$\pm$   0.14  &   1.42$\pm$   0.11  &$<$   0.23 &$<$   0.24 \\
080810  &3.35 &$<$   0.42 &$<$   0.43 &$<$   0.43 &$<$   0.39 &$<$   0.41 &   0.48$\pm$   0.14  &   3.13$\pm$   0.21  &--- &--- &--- & --- \\
080905B &2.37 &$<$   1.86 &   2.67$\pm$   0.59  &$<$   1.18 &$<$   0.87 &$<$   0.84 &   0.74$\pm$   0.24  &   1.66$\pm$   0.21  &$<$   0.43 &   1.01$\pm$   0.18  &   0.80$\pm$   0.18  &$<$   0.41 \\
080913  &6.70 &$<$   0.78 &$<$   1.51 &$<$   7.87 &--- &--- &--- &--- &--- &--- &--- & --- \\
080916A &0.67 &--- &--- &--- &--- &--- &--- &--- &--- &--- &--- & --- \\
080928  &1.69 &--- &$<$   2.14 &$<$   1.44 &$<$   1.01 &$<$   0.90 &$<$   0.49 &   2.12$\pm$   0.26  &   0.52$\pm$   0.15  &$<$   0.32 &$<$   0.26 &$<$   0.26 \\
081007  &0.53 &--- &--- &--- &--- &--- &--- &--- &--- &--- &--- & --- \\
081008  &1.97 &   1.04$\pm$   0.24  &   1.58$\pm$   0.17  &   1.41$\pm$   0.10  &   0.47$\pm$   0.07  &   0.44$\pm$   0.07  &   0.63$\pm$   0.05  &   1.04$\pm$   0.07  &   0.59$\pm$   0.05  &   0.79$\pm$   0.04  &   0.29$\pm$   0.04  &   0.17 $\pm$   0.04  \\
090102  &1.55 &--- &--- &--- &--- &--- &$<$   2.19 &   3.50$\pm$   0.65  &   3.05$\pm$   0.48  &   1.86$\pm$   0.39  &   1.58$\pm$   0.29  &   0.67 $\pm$   0.20  \\
090516  &4.11 &$<$   6.83  &$<$   6.83  &   5.81$\pm$   0.04  &   2.84$\pm$   0.04  &   1.72$\pm$   0.04  &   4.05$\pm$   0.06  &   5.87$\pm$   0.10  &   3.09$\pm$   0.06  &   4.54$\pm$   0.06  &$<$   4.04  &$<$   2.39  \\
090519  &3.85 &$<$   1.59 &$<$   2.24 &$<$   1.95 &$<$   1.45 &$<$   1.59 &$<$   2.78 &$<$   1.63 &--- &$<$   2.55 &$<$   6.22 &$<$   3.72 \\
090529  &2.62 &$<$   2.04 &$<$   1.64 &$<$   1.52 &$<$   1.50 &$<$   1.34 &   1.93$\pm$   0.44  &   1.71$\pm$   0.47  &$<$   1.43 &$<$   1.25 &$<$   1.18 &$<$   1.17 \\
090812  &2.45 &   1.11$\pm$   0.24  &   2.20$\pm$   0.30  &   1.94$\pm$   0.21  &   0.54$\pm$   0.13  &   0.57$\pm$   0.12  &   0.76$\pm$   0.13  &   1.12$\pm$   0.13  &   1.87$\pm$   0.17  &   1.12$\pm$   0.08  &   0.57$\pm$   0.08  &   0.75 $\pm$   0.14  \\
090814A &0.70 &--- &--- &--- &--- &--- &--- &--- &--- &--- &--- & --- \\
\end{longtable}
\end{footnotesize}
\end{landscape}
}

\longtabL{2}{
\begin{landscape}
\begin{footnotesize}
\begin{longtable}{l c c c c c c c c c c c c}        
\caption{\label{table:data2}  Rest-frame EWs of lines with wavelength above 2000\AA. The table displays detections or upper limits (3-$\sigma$) whenever the spectrum covers the range of the feature and the feature is not affected by strong telluric absorptions.}  \\          
\hline\hline
GRB    & \textit{z}      & 2026         & 2062         & 2344         & 2374         & 2383         & 2587         & 2600         & 2800 	& 2852	& 3935          & 3970 \\
       &        & \ion{Zn}{ii}/\ion{Cr}{ii}       & \ion{Cr}{ii}/\ion{Zn}{ii}       & \ion{Fe}{ii}/\ion{Fe}{ii}*      & \ion{Fe}{ii}            & \ion{Fe}{ii}/\ion{Fe}{ii}*      & \ion{Fe}{ii}            & \ion{Fe}{ii}*/\ion{Fe}{ii}      & \ion{Mg}{ii}            & \ion{Mg}{i}		& \ion{Ca}{ii}             & \ion{Ca}{ii} \\
\hline                        
\endfirsthead
\caption{continued.}\\
\hline\hline 
GRB    & \textit{z}      & 2026         & 2062         & 2344         & 2374         & 2383         & 2587         & 2600         & 2800 	& 2852	& 3935          & 3970 \\
       &        & \ion{Zn}{ii}/\ion{Cr}{ii}       & \ion{Cr}{ii}/\ion{Zn}{ii}       & \ion{Fe}{ii}/\ion{Fe}{ii}*      & \ion{Fe}{ii}            & \ion{Fe}{ii}/\ion{Fe}{ii}*      & \ion{Fe}{ii}            & \ion{Fe}{ii}*/\ion{Fe}{ii}      & \ion{Mg}{ii}            & \ion{Mg}{i}		& \ion{Ca}{ii}             & \ion{Ca}{ii} \\
\hline                        
\endhead
\hline
\endfoot
050319  &3.24 &--- &--- &--- &--- &--- &--- &--- &--- &--- &--- &--- \\
050401  &2.89 &   1.47 $\pm$   0.21  &   1.44 $\pm$   0.23  &   1.41 $\pm$   0.18  &   1.59 $\pm$   0.26  &   1.59 $\pm$   0.26  &--- &--- &--- &--- &--- &--- \\
050408  &1.24 &$<$   1.58 &   2.75 $\pm$   0.67  &   2.45 $\pm$   0.34  &   1.34 $\pm$   0.30  &   1.88 $\pm$   0.29  &   1.58 $\pm$   0.19  &   3.50 $\pm$   0.25  &   4.87 $\pm$   0.29  &   1.55 $\pm$   0.17  &$<$   0.35 &--- \\
050730  &3.97 &--- &--- &--- &--- &--- &--- &--- &--- &--- &--- &--- \\
050802  &1.71 &$<$   0.90 &$<$   1.59 &   2.62 $\pm$   0.33  &$<$   0.87 &   2.69 $\pm$   0.33  &   1.85 $\pm$   0.41  &--- &--- &--- &--- &--- \\
050824  &0.83 &$<$   4.84 &$<$   2.36 &   0.98 $\pm$   0.27  &$<$   2.46  &$<$   2.46  &   0.82 $\pm$   0.22  &$<$   0.57 &   2.19 $\pm$   0.33  &$<$   0.49 &$<$   0.49 &$<$   0.41 \\
050908  &3.34 &   1.15 $\pm$   0.23  &--- &--- &--- &--- &--- &--- &--- &--- &--- &--- \\
050922C &2.22 &$<$   0.14 &$<$   0.13 &--- &--- &--- &--- &--- &--- &--- &--- &--- \\
060115  &3.53 &--- &--- &--- &--- &--- &--- &--- &--- &--- &--- &--- \\
060124  &2.30 &$<$   0.23 &$<$   0.34 &--- &--- &$<$   0.80 &--- &--- &--- &--- &--- &--- \\
060206  &4.05 &--- &--- &--- &--- &--- &--- &--- &--- &--- &--- &--- \\
060210  &3.91 &--- &--- &--- &--- &--- &--- &--- &--- &--- &--- &--- \\
060502A &1.50 &$<$   0.71 &$<$   0.59 &   0.92 $\pm$   0.22  &   1.09 $\pm$   0.12  &   2.40 $\pm$   0.11  &   0.83 $\pm$   0.10  &   1.56 $\pm$   0.09  &   3.60 $\pm$   0.11  &   0.90 $\pm$   0.08  &$<$   0.17 &--- \\
060526  &3.21 &$<$   0.36 &$<$   0.36 &--- &--- &--- &--- &--- &--- &--- &--- &--- \\
060614  &0.12 &--- &--- &--- &--- &--- &--- &--- &--- &--- &--- &$<$   0.99 \\
060707  &3.43 &--- &--- &--- &--- &--- &--- &--- &--- &--- &--- &--- \\
060714  &2.71 &$<$   0.16 &--- &   2.16 $\pm$   0.30  &   1.83 $\pm$   0.27  &   2.59 $\pm$   0.27  &--- &--- &--- &--- &--- &--- \\
060729  &0.54 &--- &--- &$<$   0.37 &$<$   0.29 &$<$   0.25 &$<$   0.05  &   0.45 $\pm$   0.13  &   1.82 $\pm$   0.04  &   0.59 $\pm$   0.04  &   0.75 $\pm$   0.03  &   0.55 $\pm$   0.03  \\
060904B &0.70 &--- &--- &--- &--- &--- &   0.94 $\pm$   0.18  &   1.82 $\pm$   0.18  &   4.65 $\pm$   0.12  &   1.00 $\pm$   0.12  &   1.35 $\pm$   0.12  &   1.06 $\pm$   0.12  \\
060906  &3.68 &--- &--- &--- &--- &--- &--- &--- &--- &--- &--- &--- \\
060908  &1.88 &$<$   0.36 &$<$   0.34 &--- &--- &--- &--- &--- &--- &--- &--- &--- \\
060926  &3.21 &$<$   0.76 &$<$   1.05 &--- &--- &--- &--- &--- &--- &--- &--- &--- \\
060927  &5.47 &--- &--- &--- &--- &--- &--- &--- &--- &--- &--- &--- \\
061007  &1.26 &$<$   2.84 &$<$   2.64 &   2.83 $\pm$   0.75  &   3.54 $\pm$   0.88  &   2.26 $\pm$   0.66  &   3.23 $\pm$   0.58  &   2.61 $\pm$   0.62  &   5.40 $\pm$   0.97  &$<$   1.10 &$<$   1.04 &--- \\
061021  &0.35 &--- &--- &--- &--- &--- &--- &--- &   1.85 $\pm$   0.52  &$<$   0.96 &$<$   0.82 &$<$   0.49 \\
061110A &0.76 &$<$   6.60 &$<$   6.27 &$<$   3.41 &$<$   3.36 &$<$   3.46 &$<$   2.85 &$<$   2.81 &   3.64 $\pm$   0.80  &$<$   2.23 &$<$   2.32 &--- \\
061110B &3.44 &--- &--- &--- &--- &--- &--- &--- &--- &--- &--- &--- \\
061121  &1.31 &   1.05 $\pm$   0.03  &   1.00 $\pm$   0.03  &   2.83 $\pm$   0.03  &   2.60 $\pm$   0.03  &   3.69 $\pm$   0.03  &   3.18 $\pm$   0.04  &   3.81 $\pm$   0.03  &   8.64 $\pm$   0.05  &   2.47 $\pm$   0.04  &   1.76 $\pm$   0.08  &   1.43 $\pm$   0.07  \\
070110  &2.35 &   0.60 $\pm$   0.12  &--- &   0.87 $\pm$   0.24  &   0.99 $\pm$   0.18  &   2.12 $\pm$   0.24  &$<$   0.82 &   2.03 $\pm$   0.18  &$<$   1.76 &$<$   1.24 &--- &--- \\
070125  &1.55 &$<$   0.13 &$<$   0.13 &$<$   0.13 &$<$   0.13 &$<$   0.13 &$<$   0.13 &$<$   0.13 &   0.35 $\pm$   0.08  &$<$   0.14 &$<$   0.19 &--- \\
070306  &1.50 &$<$   2.71 &$<$   2.65 &$<$   2.70 &$<$   2.68 &$<$   2.60 &$<$   2.50 &$<$   2.52 &--- &--- &$<$   2.35 &--- \\
070318  &0.84 &$<$   0.91 &$<$   0.70 &   0.76 $\pm$   0.16  &   0.43 $\pm$   0.11  &   1.20 $\pm$   0.16  &$<$   3.64  &$<$   3.64  &   2.61 $\pm$   0.16  &   0.65 $\pm$   0.11  &   2.07 $\pm$   0.22  &   0.98 $\pm$   0.22  \\
070411  &2.95 &--- &--- &--- &--- &--- &--- &--- &--- &--- &--- &--- \\
070419A &0.97 &--- &--- &--- &--- &--- &--- &--- &--- &--- &   1.25 $\pm$   0.16  &   0.61 $\pm$   0.15  \\
070506  &2.31 &   0.94 $\pm$   0.21  &   0.88 $\pm$   0.24  &--- &--- &--- &--- &--- &--- &--- &--- &--- \\
070508  &0.82 &--- &--- &$<$   8.79 &$<$   8.21 &$<$   7.80 &$<$   5.72 &$<$   5.72 &$<$   4.96 &$<$   4.95 &$<$   4.83 &$<$   3.56 \\
070611  &2.04 &$<$   0.40 &$<$   0.42 &$<$   0.53 &$<$   0.55 &$<$   0.61 &$<$   0.82 &$<$   0.68 &   2.11 $\pm$   0.69  &$<$   0.65 &$<$   0.87 &--- \\
070721B &3.63 &$<$   8.24 &--- &--- &--- &--- &--- &--- &--- &--- &--- &--- \\
070802  &2.45 &   1.22 $\pm$   0.29  &$<$   0.86 &   4.78 $\pm$   0.58  &   3.68 $\pm$   0.43  &   5.45 $\pm$   0.43  &   5.65 $\pm$   0.38  &   8.41 $\pm$   0.38  &--- &--- &--- &--- \\
071020  &2.15 &$<$   1.06 &$<$   1.10 &$<$   1.38 &$<$   1.44 &   1.90 $\pm$   0.63  &   2.22 $\pm$   0.63  &   3.49 $\pm$   0.63  &   7.62 $\pm$   0.13  &$<$   2.36 &$<$   1.57 &--- \\
071031  &2.69 &$<$   0.10 &--- &   1.17 $\pm$   0.07  &   0.66 $\pm$   0.09  &   1.21 $\pm$   0.05  &   0.93 $\pm$   0.16  &   1.40 $\pm$   0.09  &--- &--- &--- &--- \\
071112C &0.82 &$<$   2.77 &$<$   1.99 &   1.10 $\pm$   0.27  &$<$   0.71 &   0.88 $\pm$   0.27  &   1.10 $\pm$   0.27  &   1.32 $\pm$   0.27  &   2.69 $\pm$   0.33  &$<$   0.58 &$<$   0.51 &$<$   0.73 \\
080210  &2.64 &   0.63 $\pm$   0.11  &$<$   0.16 &   1.73 $\pm$   0.11  &   0.49 $\pm$   0.14  &   1.87 $\pm$   0.14  &$<$   0.54 &$<$   0.77 &--- &--- &--- &--- \\
080319B &0.97 &$<$   0.90 &$<$   0.94 &$<$   0.67 &$<$   0.64 &$<$   0.62 &   0.91 $\pm$   0.25  &$<$   0.53 &   1.88 $\pm$   0.51  &$<$   0.50 &$<$   0.49 &--- \\
080319C &1.95 &   0.78 $\pm$   0.24  &   0.78 $\pm$   0.20  &--- &--- &--- &--- &--- &--- &--- &--- &--- \\
080330  &1.51 &   1.71 $\pm$   0.16  &$<$   0.37 &$<$   0.53 &$<$   0.36 &   0.88 $\pm$   0.20  &$<$   0.36 &$<$   0.36 &   2.07 $\pm$   0.12  &--- &$<$   0.39 &--- \\
080411  &1.03 &   0.64 $\pm$   0.03  &   0.55 $\pm$   0.04  &$<$   1.41  &$<$   4.59  &$<$   4.59  &$<$   4.12  &$<$   4.12  &   3.17 $\pm$   0.07  &   0.58 $\pm$   0.05  &   1.08 $\pm$   0.15  &   1.11 $\pm$   0.15  \\
080413B &1.10 &   0.67 $\pm$   0.10  &   0.67 $\pm$   0.05  &   2.17 $\pm$   0.04  &   1.45 $\pm$   0.07  &   2.42 $\pm$   0.07  &$<$   7.54  &$<$   7.54  &   6.38 $\pm$   0.10  &   1.04 $\pm$   0.05  &--- &--- \\
080520  &1.55 &$<$   3.36 &$<$   3.46 &$<$   3.21 &$<$   3.15 &$<$   3.09 &$<$   2.65 &$<$   2.62 &$<$   2.45 &$<$   2.46 &$<$   2.61 &--- \\
080603B &2.69 &--- &--- &--- &--- &--- &--- &--- &--- &--- &--- &--- \\
080604  &1.42 &--- &--- &   1.98 $\pm$   0.29  &   1.78 $\pm$   0.25  &   1.94 $\pm$   0.25  &   2.64 $\pm$   0.21  &   2.64 $\pm$   0.21  &   5.37 $\pm$   0.37  &$<$   0.37 &--- &--- \\
080605  &1.64 &   0.61 $\pm$   0.11  &   0.38 $\pm$   0.11  &   2.50 $\pm$   0.11  &$<$   4.39  &$<$   4.39  &   2.12 $\pm$   0.11  &--- &   8.56 $\pm$   0.23  &   2.31 $\pm$   0.11  &$<$   0.25 &--- \\
080607  &3.04 &   1.57 $\pm$   0.04  &--- &--- &--- &--- &--- &--- &--- &--- &--- &--- \\
080707  &1.23 &$<$   1.25 &$<$   1.16 &   2.29 $\pm$   0.40  &$<$   0.79 &   2.38 $\pm$   0.40  &   1.75 $\pm$   0.36  &   2.06 $\pm$   0.36  &   7.35 $\pm$   0.49  &   2.20 $\pm$   0.36  &$<$   0.85 &$<$   3.28 \\
080710  &0.85 &--- &--- &   0.42 $\pm$   0.03  &   0.21 $\pm$   0.02  &   0.57 $\pm$   0.02  &   0.41 $\pm$   0.02  &   0.71 $\pm$   0.02  &   2.26 $\pm$   0.02  &   0.36 $\pm$   0.02  &$<$   0.03 &--- \\
080721  &2.59 &$<$   0.33 &$<$   0.31 &--- &--- &--- &--- &--- &--- &--- &--- &--- \\
080810  &3.35 &--- &--- &--- &--- &--- &--- &--- &--- &--- &--- &--- \\
080905B &2.37 &   0.59 $\pm$   0.18  &   0.83 $\pm$   0.15  &   1.51 $\pm$   0.30  &   1.57 $\pm$   0.18  &   1.28 $\pm$   0.21  &   1.51 $\pm$   0.18  &   1.66 $\pm$   0.42  &   5.93 $\pm$   1.19  &$<$   1.55 &$<$   1.08 &--- \\
080913  &6.70 &--- &--- &--- &--- &--- &--- &--- &--- &--- &--- &--- \\
080916A &0.67 &--- &--- &$<$   3.05 &$<$   3.86 &$<$   3.74 &$<$   2.55 &$<$   2.46 &   4.91 $\pm$   0.84  &$<$   2.12 &$<$   1.80 &$<$   1.51 \\
080928  &1.69 &$<$   0.22 &$<$   0.23 &   0.74 $\pm$   0.15  &   0.63 $\pm$   0.11  &   0.45 $\pm$   0.07  &   0.52 $\pm$   0.07  &   0.82 $\pm$   0.11  &   2.38 $\pm$   0.22  &$<$   0.25 &--- &--- \\
081007  &0.53 &--- &--- &--- &--- &--- &--- &--- &--- &--- &--- &$<$  18.80 \\
081008  &1.97 &   0.29 $\pm$   0.05  &   0.24 $\pm$   0.05  &--- &--- &--- &--- &--- &--- &--- &--- &--- \\
090102  &1.55 &$<$   0.95 &$<$   0.94 &   2.97 $\pm$   0.87  &$<$   1.11 &   2.49 $\pm$   0.78  &   2.46 $\pm$   0.45  &   3.66 $\pm$   0.51  &  10.13 $\pm$   0.63  &   5.00 $\pm$   0.93  &$<$   1.93 &--- \\
090516  &4.11 &--- &--- &--- &--- &--- &--- &--- &--- &--- &--- &--- \\
090519  &3.85 &--- &--- &--- &--- &--- &--- &--- &--- &--- &--- &--- \\
090529  &2.62 &$<$   1.72 &$<$   1.67 &$<$   2.13 &$<$   2.40 &$<$   2.94 &$<$   5.25 &$<$   7.27 &--- &--- &--- &--- \\
090812  &2.45 &   0.55 $\pm$   0.08  &   0.50 $\pm$   0.07  &   1.04 $\pm$   0.09  &   1.19 $\pm$   0.10  &   1.81 $\pm$   0.10  &   1.24 $\pm$   0.21  &   1.61 $\pm$   0.14  &$<$  23.03 &$<$  43.12 &--- &--- \\
090814A &0.70 &--- &--- &$<$  14.05 &$<$  12.83 &$<$  12.83 &$<$  12.83 &$<$  14.05 &   4.16 $\pm$   0.79  &$<$  12.83 &   2.51 $\pm$   0.66  &   2.69 $\pm$   0.72  \\
\end{longtable}
\end{footnotesize}
\end{landscape}
}

\begin{appendix} 

\section{Additional spectra}
In this section we present further spectra complementing the sample of \citet{fyn09}, analysed in an equivalent way.

\subsection{GRB\,081007}

\begin{figure*}[h!]
   \centering
   \includegraphics[width=11cm]{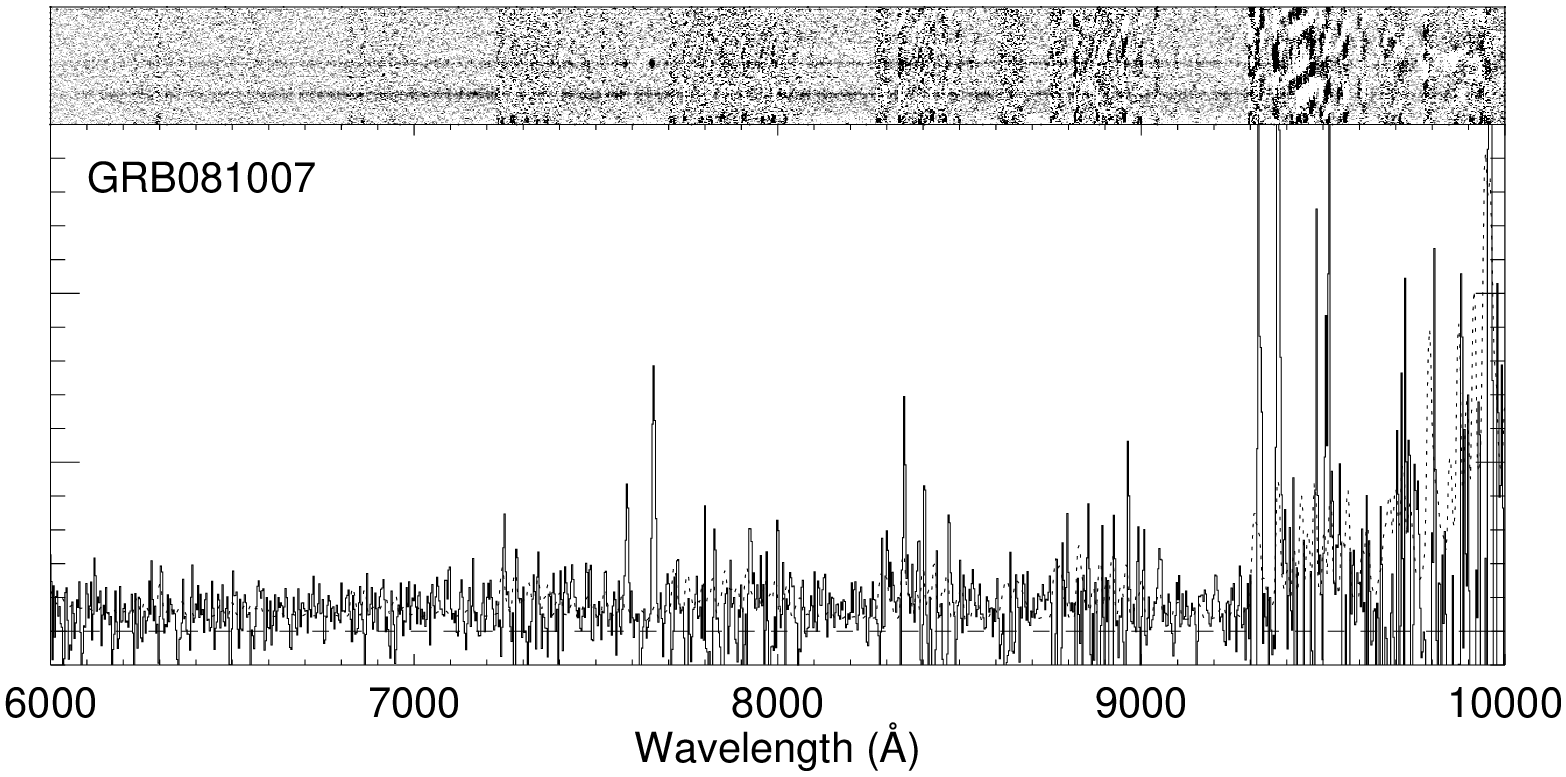}
      \caption{Spectrum of GRB\,081007. The GRB afterglow is the central trace in the image.}
         \label{Fig:spec081007}
   \end{figure*}

This spectrum has not yet been published. It was a 2700s exposure obtained with FORS2 at the VLT using the 300I grating. It shows evidence of 2 emission lines (see Table~\ref{table:081007} and Fig.~\ref{Fig:spec081007}) at a redshift of $z=0.530$, in agreement of the redshift suggested by \citet{ber08}. The lack of [\ion{O}{iii}]$\lambda5008$ is probably due to contamination by a sky line.

\begin{table}[h!]
\caption{GRB\,081007 ($z=0.530$). Observed-frame equivalent widths.}             
\label{table:081007}      
\centering                          
\begin{tabular}{c c c c}        
\hline\hline                 
Wavelength & Feature & $z$ & EW \\    
\hline                        
7438.2	& H-$\beta$ 4862.68	& 0.5297	& emission  \\
7589.5	& [OIII] 4960.29			& 0.5300	& emission  \\
\hline                                   
\end{tabular}
\end{table}

\subsection{GRB\,081008}

\begin{figure*}[h!]
   \centering
   \includegraphics[width=11cm]{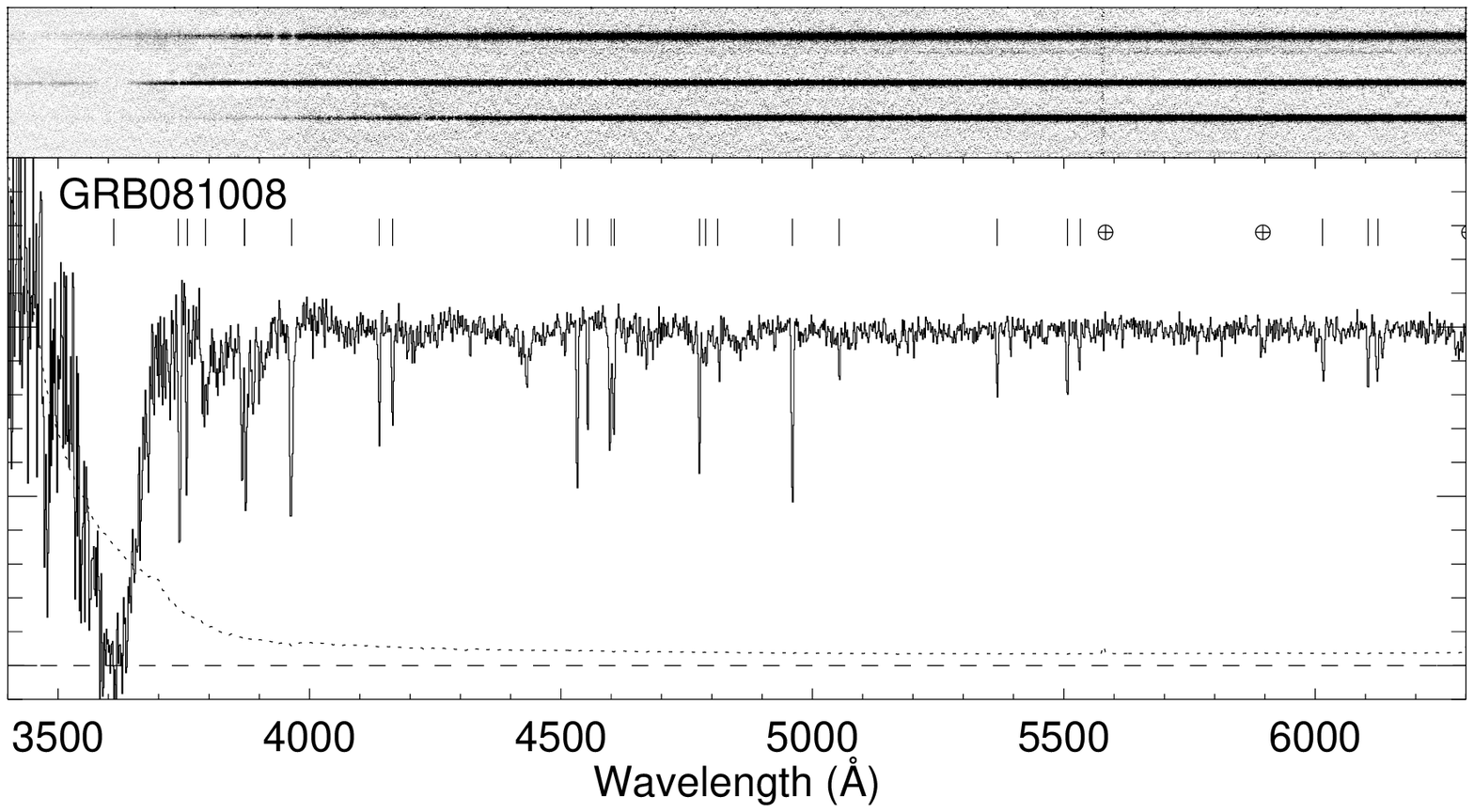}
      \caption{Spectrum of GRB\,081008. The GRB afterglow is the central trace in the image.}
         \label{Fig:spec081008}
   \end{figure*}

The data presented here (see Table~\ref{table:081008} and Fig.~\ref{Fig:spec081008}) were initially published as a GCN circular \citep{dav08} and later in a more detailed study \citep{del11}. The observation was performed using FORS2 at the VLT with the 600B grating and consisted of 3$\times$900 s exposure. We obtain a redshift of 1.969$\pm$0.001, consistent with the value determined by \citet{del11}. From the fit of the Ly$\alpha$ absorption, we determine a neutral hydrogen column density of $\log\,(N_{\rm \ion{H}{i}}/cm^{-2}) = 21.59 \pm 0.10$.

\begin{table}[h!]
\caption{GRB\,081008 ($z=1.969$). Observed-frame equivalent widths.}             
\label{table:081008}      
\centering                          
\begin{tabular}{c c c c}        
\hline\hline                 
Wavelength & Feature & $z$ & EW \\    
\hline                        
3741.5 	& SII+SiII 1259.9 & 1.9696& 3.1$\pm$0.7	\\
3755.7   	& SII* 1264.74  	& 1.9695 	& 1.5$\pm$0.4 \\
3793.1   	& CI 1277.25	& 1.9697	& 3.1$\pm$0.7 \\
3865.9   	& OI 1302.17   	& 1.9689	& 4.7$\pm$0.5	\\
3874.1   	& SiII 1304.37  	& 1.9701	& 			\\
3887.5	& SiII*1309.28	& 1.9692	& 1.2$\pm$0.3 	\\
3964.3   	& CII 1334.53   	& 1.9706	& 4.2$\pm$0.3 \\
4138.5   	& SiIV 1393.76 	& 1.9693	& 1.4$\pm$0.2	\\
4165.4   	& SiIV 1402.77 	& 1.9694	& 1.3$\pm$0.2	\\
4434.0	& ?			& 		& 1.7$\pm$0.3	\\
4532.5   	& SiII 1526.71  	& 1.9688	& 1.88$\pm$0.14 \\
4553.0   	& SiII* 1533.43 	& 1.9692	& 1.13$\pm$0.13 \\
4600.    	& CIV+CIV 1549	& 		& 3.1$\pm$0.2	\\
4775.4   	& FeII 1608.45 	& 1.9690	& 1.74$\pm$0.15 \\
4786.2   	& FeII 1611.20 	& 1.9706	& 0.90$\pm$0.19 \\
4811.6   	&?   		& 		& 0.84$\pm$0.19 \\
4960.5   	& AlII 1670.79  	& 1.9690	& 2.34$\pm$0.13 \\
5053.9    	& ?  		& 		& 0.51$\pm$0.11\\
5367.9  	& SiII 1808.01  	& 1.9690	& 0.76$\pm$0.12\\
5506.8	& AlIII 1854.72 	& 1.9691	& 0.87$\pm$0.12\\
5530.8   	& AlIII 1862.79 	& 1.9691	& 0.51$\pm$0.13\\
6015.7   	& ZnII 2026.14	& 1.9690	& 0.85$\pm$0.15\\
6105.5   	& CrII 2056.26	& 1.9692	& 0.76$\pm$0.14\\
6123.2   	& CrII+ZnII 2062.4 & 	1.9690& 0.72$\pm$0.14\\
6133.9	& CrII 2066.16	& 1.9687	& 0.43$\pm$0.15\\
\hline                                   
\end{tabular}
\end{table}

\subsection{GRB\,090102}

\begin{figure*}[h!]
   \centering
   \includegraphics[width=11cm]{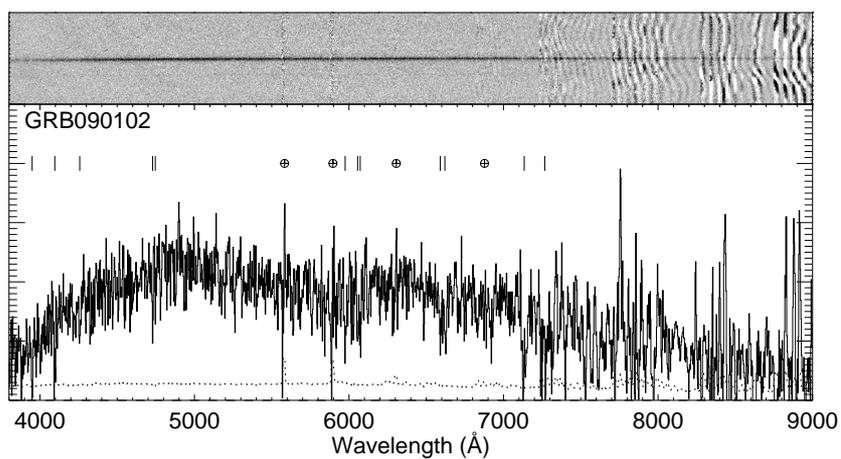}
      \caption{Spectrum of GRB\,090102.}
         \label{Fig:spec090102}
   \end{figure*}

The spectrum presented here (see Table~\ref{table:090102} and Fig.~\ref{Fig:spec090102}) has only been published in a GCN \citep{deu09}. The spectrum an exposure time of 2700s and was obtained with AlFOSC at the 2.5m NOT telescope. From the absorption features we determine a redshift of 1.548$\pm$0.001.

\begin{table}[h!]
\caption{GRB\,090102 ($z=1.548$). Observed-frame equivalent widths.}             
\label{table:090102}      
\centering                          
\begin{tabular}{c c c c}        
\hline\hline                 
Wavelength & Feature & $z$ & EW \\    
\hline                        
3949.6    	& CIV+CIV 1549	& 		&8.9$\pm$01.7	\\
4096.0   	& FeII 1608.45 	& 1.5466	& 7.8$\pm$1.2 \\
4257.7   	& AlII 1670.79  	& 1.5483	& 4.7$\pm$1.0 \\
4728.8	& AlIII 1854.72 	& 1.5496	& 4.0$\pm$0.7\\
4746.4   	& AlIII 1862.79 	& 1.5480	& 1.7$\pm$0.5\\
5975.1   	& FeII 2344.2  	& 1.5489	& 7.6$\pm$2.2 \\
6055.8  	& FeII 2374.5 	& 1.5503	&   5.3$\pm$2.0 \\
6072.4  	& FeII 2382.8 	& 1.5484	& 6.4$\pm$2.0 \\
6590.9	& FeII 2587 	& 1.5480	& 6.3$\pm$1.1  \\
6620.4	& FeII 2600 	& 1.5461	& 9.3$\pm$1.3  \\
7134.0	& MgII+MgII 2800& 		& 25.8$\pm$1.6  \\
7266.9	& MgI 2852	& 1.5471	& 12.74$\pm$2.4  \\
\hline                                   
\end{tabular}
\end{table}

\subsection{GRB\,090516}

\begin{figure*}[h!]
   \centering
   \includegraphics[width=11cm]{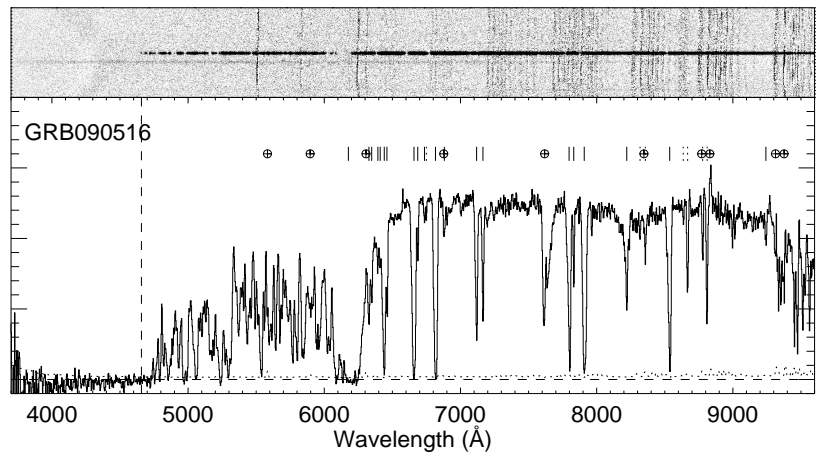}
      \caption{Spectrum of GRB\,090516. The GRB afterglow is the bright trace in the centre of the image.}
         \label{Fig:spec090516}
   \end{figure*}

The spectrum of GRB\,090516 (see Table~\ref{table:090516} and Fig.~\ref{Fig:spec090516}) has, until now, only been reported in GCNs \citep{deu09b,deu09c}. It was obtained with FORS2 at the VLT with the 300V grism and an exposure of 2$\times$1800s. From the absorption features we determine a redshift of 4.111$\pm$0.006. By fitting the Ly$\alpha$ absorption we determine a neutral hydrogen column density of$\log\,(N_{\rm \ion{H}{i}}/cm^{-2}) = 21.70 \pm 0.10$.

\begin{table}[h!]
\caption{GRB\,090516 ($z=4.111\pm0.006$). Observed-frame equivalent widths.}             
\label{table:090516}      
\centering                          
\begin{tabular}{c c c c}        
\hline\hline                 
Wavelength & Feature & $z$ & EW \\    
\hline                        

6327.8  &  NV 1232.8  & 4.1329 & 12.4 +/- 0.3 \\
6347.9   & NV 1242.9    &4.1073 &    \\

6393.6   & SII 1250.6   & 4.1124 &   35.6 +\- 0.3 \\
6410.6  &  SII 1253.8 & 4.1129&  \\
6440.4  &  SiII 1260.4 & 4.1098 &  \\
6460.1   & SiII* 1264.7  & 4.1080 &  \\

6658.5  &  OI 1302.2, SiII 1304.4  &  & 34.9 +/- 0.2 \\
6686.2  &  SiII* 1309.3 & 4.1067 &  \\

6735.7  &  NiII 1317.2   & 4.1136 & 2.3 +/- 0.2 \\
6749.2  &  ? & &  \\

6816.6   & CII+CII* 1335    &  &  29.7 +/- 0.2 \\

6881.7   & NiII 1345.9  & 4.1131 & 7.2 +/- 0.3 \\

7118.3  &  SiIV 1393.8  & 4.1071 & 14.5 +/- 0.2 \\
7165.0  &  SiIV 1402.8  & 4.1076 &  8.8 +/- 0.2 \\
7797.3  &  SiII 1526.7  & 4.1073 & 20.7 +/- 0.3 \\
7831.8  &  SiII* 1533.4  & 4.1075  & 3.8 +/- 0.2 \\
7908.8  &  CIV+CIV 1549 &  &  30.0 +/- 0.5 \\
7963.0  &  ?  & & 1.1 +/- 0.2 \\
8220.9  &  FeII 1608.5 & 4.1109 &   15.8 +/- 0.3 \\
8318.8  &  FeII 2249.9 &2.6974 &   0.8 +/- 0.2 \\
8357.6  &  FeII 2260.8 & 2.6967&   1.9 +/- 0.3 \\
8536.2   & AlII 1670.8 & 4.1090 &  23.2 +/- 0.3 \\
8635.5  &  ?  & &  1.0 +/- 0.3 \\
8667.7   & FeII 2344.2  &  2.6975  &6.3 +/- 0.3 \\
8778.1  &  FeII 2374.5 & 2.6968  &    3.7 +/- 0.3 \\
8811.5  &  FeII 2382.8 & 2.6980   & 10.5 +/- 0.3 \\
9242.7  &  SiII 1808.0 & 4.1121 & 2.4 +/- 0.3 \\
\hline                                   
\end{tabular}
\end{table}

\subsection{GRB\,090519}

\begin{figure*}[h!]
   \centering
   \includegraphics[width=11cm]{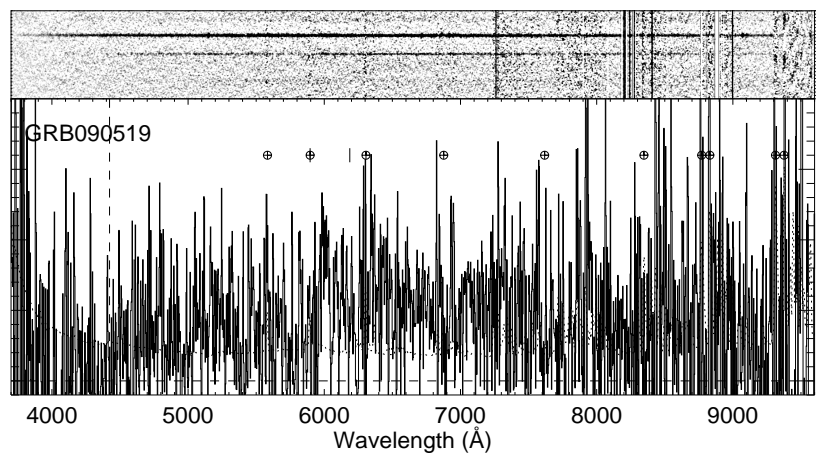}
      \caption{Spectrum of GRB\,090519. The GRB afterglow is the weak trace in the centre of the image.}
         \label{Fig:spec090519}
   \end{figure*}

This spectrum has been only reported in a GCN circular \citep{tho09}. It was obtained with FORS2 at the VLT and grism 300V with an exposure time of 3$\times$1800s. The spectrum has low signal to noise ratio and shows no absorption features (see Fig.~\ref{Fig:spec090519}) and the redshift is determined by the identification of the Lyman break as well as broad absorptions of Ly$\alpha$ and Ly$\beta$.

\subsection{GRB\,090529}

\begin{figure*}[h!]
   \centering
   \includegraphics[width=11cm]{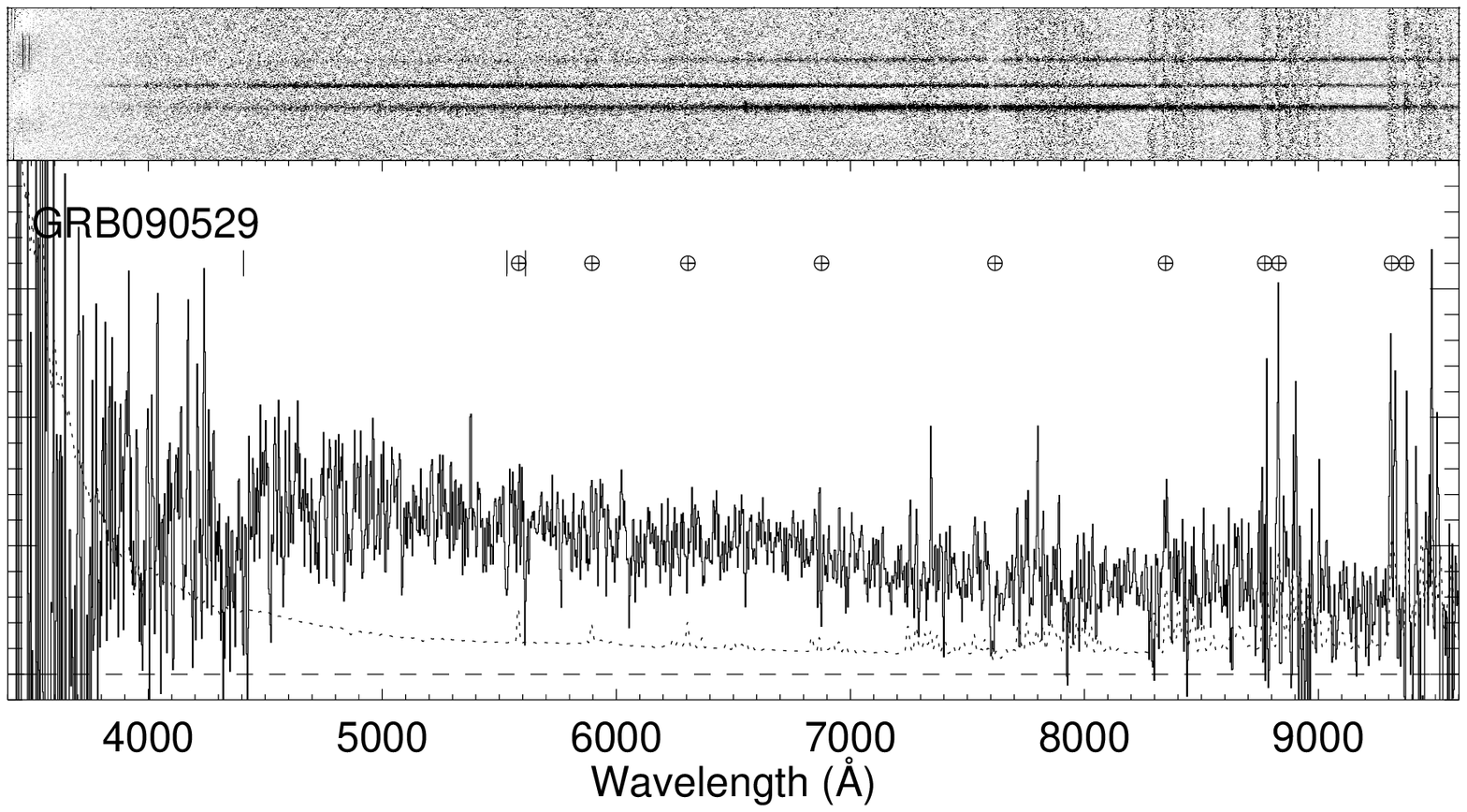}
      \caption{Spectrum of GRB\,090529. The GRB afterglow is the central trace in the image.}
         \label{Fig:spec090529}
   \end{figure*}

These data have been only reported in a GCN \citep{mal09}. The spectrum is a 2$\times$1800 s exposure obtained using FORS2 with grism 300V at the VLT. From a limited amount of absorption features (see Table~\ref{table:090529} and Fig.~\ref{Fig:spec090529}) we determine a redshift of 2.624$\pm$0.003.

\begin{table}[h!]
\caption{GRB\,090529. Observed-frame equivalent widths. $z=2.624$}             
\label{table:090529}      
\centering                          
\begin{tabular}{c c c c}        
\hline\hline                 
Wavelength & Feature & $z$ & EW \\    
\hline                        
5532.3 & SiII1526.7 & 2.6237 & 7.0 $\pm$ 1.6\\
5611.7 & CIV1548+1550 &  &6.2 $\pm$ 1.7\\
\hline                                   
\end{tabular}
\end{table}

\subsection{GRB\,090812}

\begin{figure*}[h!]
   \centering
   \includegraphics[width=11cm]{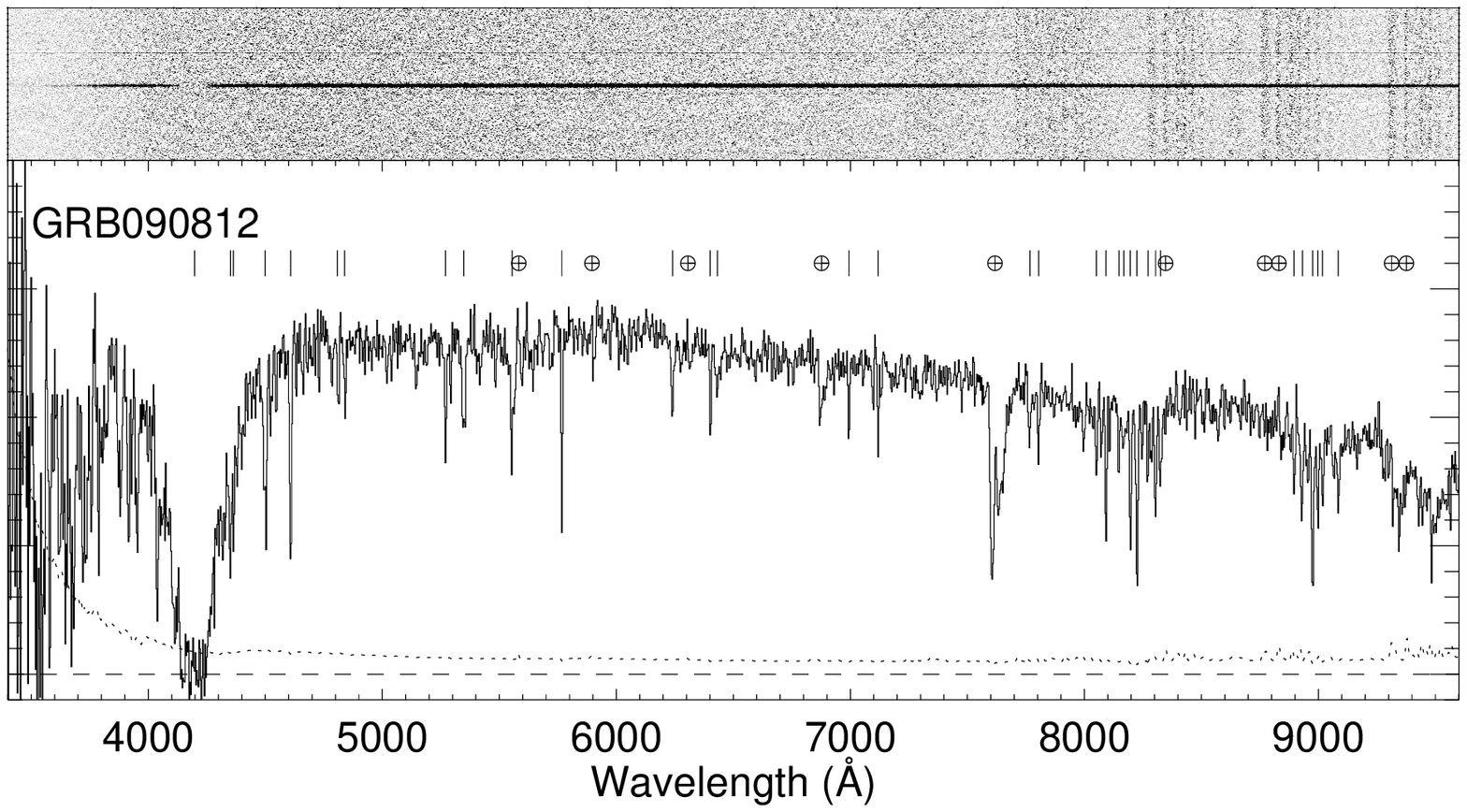}
      \caption{Spectrum of GRB\,090812. The GRB afterglow is the central trace in the image.}
         \label{Fig:spec090812}
   \end{figure*}

The data presented here (see Table~\ref{table:090812} and Fig.~\ref{Fig:spec090812}) have previously been published only in the GCN circulars \citep{deu09d}. It is a FORS2 spectrum from the VLT, which shows multiple absorption features at a common redshift of 2.4521$\pm$0.0010. From the Ly$\alpha$ absorption we determine a neutral hydrogen column density of $\log\,(N_{\rm \ion{H}{i}}/cm^{-2}) = 22.30 \pm 0.10$.

\begin{table}[h!]
\caption{GRB\,090812. Observed-frame equivalent widths. $z=2.452$}             
\label{table:090812}      
\centering                          
\begin{tabular}{c c c c}        
\hline\hline                 
Wavelength & Feature & $z$ & EW \\    
\hline                        
4351.3	& SiII 1260	.42		& 2.4523	& 3.83$\pm$0.83	\\
4365.2 	& SiII* 1264.74		& 2.4515	& 2.03$\pm$0.50	\\
4499.2 	& OI+SiII 1303		& 		& 7.59$\pm$1.02	\\
4608.3 	& CII 1334.53		& 2.4531	& 6.71$\pm$0.72	\\
4811.9 	& SiIV 1393.76		& 2.4525	& 1.87$\pm$0.44	\\
4840.5 	& SiIV 1402.77		& 2.4507	& 1.97$\pm$0.43	\\
5270.2 	& SiII 1526.71		& 2.4520	& 2.62$\pm$0.45	\\
5292.3 	& SiII* 1533.43		& 2.4513	& 0.96$\pm$0.28	\\
5348.3 	& CIV+CIV 1549	& 		& 3.85$\pm$0.44	\\
5414.1 	& ?				& 		& 1.48$\pm$0.46	\\
5483.9 	& ?				& 		& 0.88$\pm$0.26	\\
5554.4 	& FeII 1608.45		& 2.4533	& 6.44$\pm$0.58	\\
5596.7 	& FeII* 1621.69		& 2.4512	& 1.29$\pm$0.33	\\
5638.7 	& FeII*?			& 		& 1.73$\pm$0.45	\\
5722.5 	& CI 1656.93		& 2.4537 	& 1.42$\pm$0.37	\\
5767.6 	& AlII 1670.79		& 2.4520 	& 3.88$\pm$0.28	\\
6241.1 	& SI+Si II1808		& 		& 2.92$\pm$0.42	\\
6276.3 	& SiII* 1817.45		& 2.4534	& 1.10$\pm$0.37	\\
6315.7 	& MgI 1827.94?	& 		& 1.05$\pm$0.35	\\
6402.2 	& AlIII 1854.72		& 2.4518	& 1.95$\pm$0.26	\\
6434.3 	& AlIII 1862.79		& 2.4541	& 2.58$\pm$0.47	\\
6880.7 	& ?				& 		& 4.31$\pm$0.47	\\
9663.6 	& ZnII+CrII+MgI 2026	& 		& 1.89$\pm$0.26	\\
7096.8 	& CrII 2056.26		& 2.4513	& 0.99$\pm$0.27	\\
7118.8 	& CrII+ZnII 2062	& 		& 1.74$\pm$0.23	\\
7131.4 	& CrII 2066.16		& 2.4515	& 0.76$\pm$0.24	\\
7766.2 	& FeII 2249.88		& 2.4518	& 1.88$\pm$0.39	\\
7805.1 	& FeII 2260.78		& 2.4524	& 1.84$\pm$0.32	\\
7998.0 	& NiII 2316.70		& 2.4523	& 1.47$\pm$0.46	\\
8051.7 	& FeII* 2333.52		& 2.4505	& 1.86$\pm$0.33	\\
8072.8 	& FeII* 2338.72		& 2.4518	& 1.16$\pm$0.33	\\
8092.6 	& FeII 2344.21		& 2.4522	& 3.58$\pm$0.30	\\
8108.0 	& FeII* 2349.02		& 2.4517	& 0.85$\pm$0.23	\\
8147.4 	& FeII* 2359.83		& 2.4525	& 1.46$\pm$0.28	\\
8168.7 	& FeII* 2365.55		& 2.4532	& 1.42$\pm$0.45	\\
8196.9 	& FeII 2374.46		& 2.4521	& 4.09$\pm$0.34	\\
8224.8 	& FeII 2382.77		& 2.4518	& 6.24$\pm$0.36	\\
8246.7 	& FeII* 2389.36 	& 2.4514	& 0.85$\pm$0.28	\\
8270.8 	& FeII* 2396.15		& 2.4517	& 1.70$\pm$0.34	\\
8284.7 	& FeII* 2399.98 	& 2.4520	& 1.55$\pm$0.49	\\
8304.7 	& FeII* 2405.16		& 2.4529	& 3.80$\pm$0.46	\\
8325.8 	& FeII* 2411.25		& 2.4529	& 3.14$\pm$0.42	\\
8697.3 	& ?				& 		& 2.18$\pm$0.48	\\
8930.5 	& FeII 2586.49		& 2.4527	& 4.27$\pm$0.74	\\
8952.7 	& MnII 2594.36		& 2.4508	& 2.60$\pm$0.63	\\
8975.3 	& FeII 2600.23		& 2.4517	& 5.57$\pm$0.48	\\
8997.7 	& MnII+FeII* 2607	& 		& 2.76$\pm$0.50	\\
9018.1 	& FeII*+FeII* 2612	& 		& 2.32$\pm$0.48	\\
9078.6 	& FeII* 2629.08		& 2.4531	& 5.02$\pm$0.71	\\
9280.6 	& ?				& 		& 1.82$\pm$0.51	\\
9347.5 	& ?				& 		& 10.10$\pm$1.38	\\
\hline                                   
\end{tabular}
\end{table}

\subsection{GRB\,090814A}

\begin{figure*}[h!]
   \centering
   \includegraphics[width=12cm]{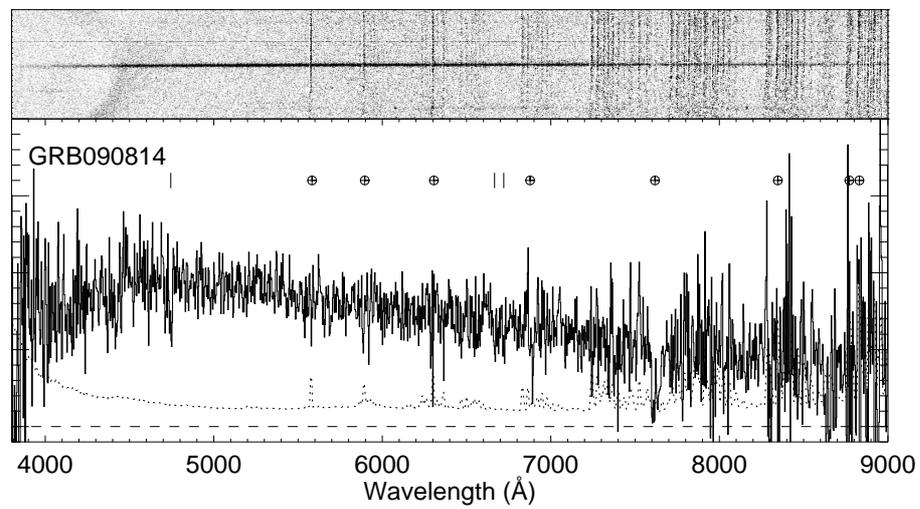}
      \caption{Spectrum of GRB\,090814A. The GRB afterglow is the central trace in the image.}
         \label{Fig:spec090814A}
   \end{figure*}

This spectrum (see Table~\ref{table:090814A} and Fig.~\ref{Fig:spec090814A}) has only been published in a GCN circular \citep{jak09}. It was obtained with FORS2 at VLT. Using a limited amount of absorption features we determine a redshift of 0.694$\pm$0.002.

\begin{table}[h!]
\caption{GRB\,090814A. Observed-frame equivalent widths. $z=0.694$}             
\label{table:090814A}      
\centering                          
\begin{tabular}{c c c c}        
\hline\hline                 
Wavelength & Feature & $z$ & EW \\    
\hline                        
4744.6 	& MgII+MgII 2800 		& 	& 7.08 $\pm$ 1.34\\
6666.1 	& CaII 3934.78		& 0.6941	& 4.27 $\pm$ 1.13\\
6720.7	& CaII 3969.59		& 0.6930	& 4.58 $\pm$ 1.22\\
\hline                                   
\end{tabular}
\end{table}

\clearpage
   \newpage

   \clearpage
   \newpage

\section{Line fitting}

\begin{figure*}[h]
\centering
\begin{tabular}{cc}
\includegraphics[width=8cm,clip]{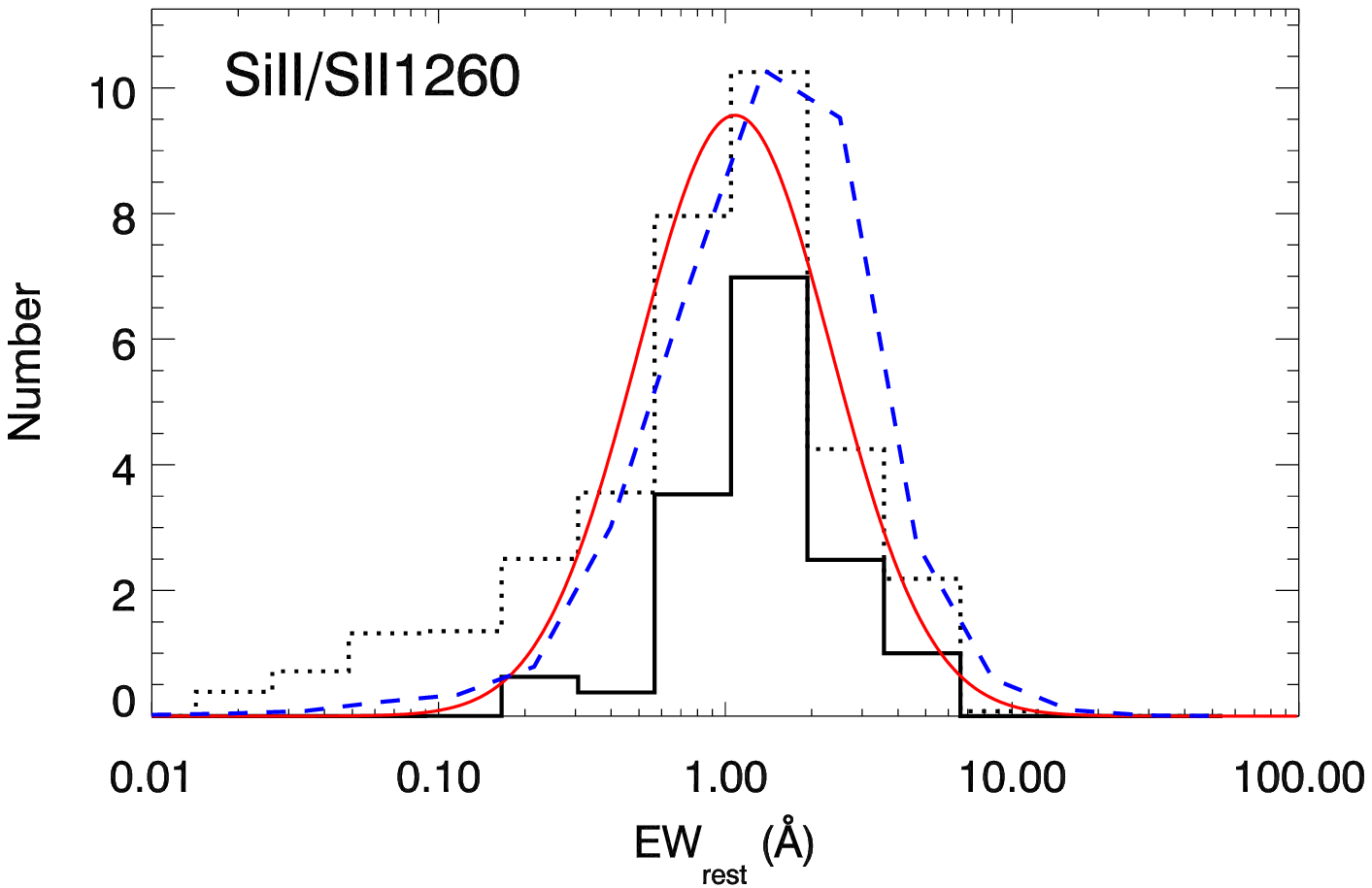} & \includegraphics[width=8cm,clip]{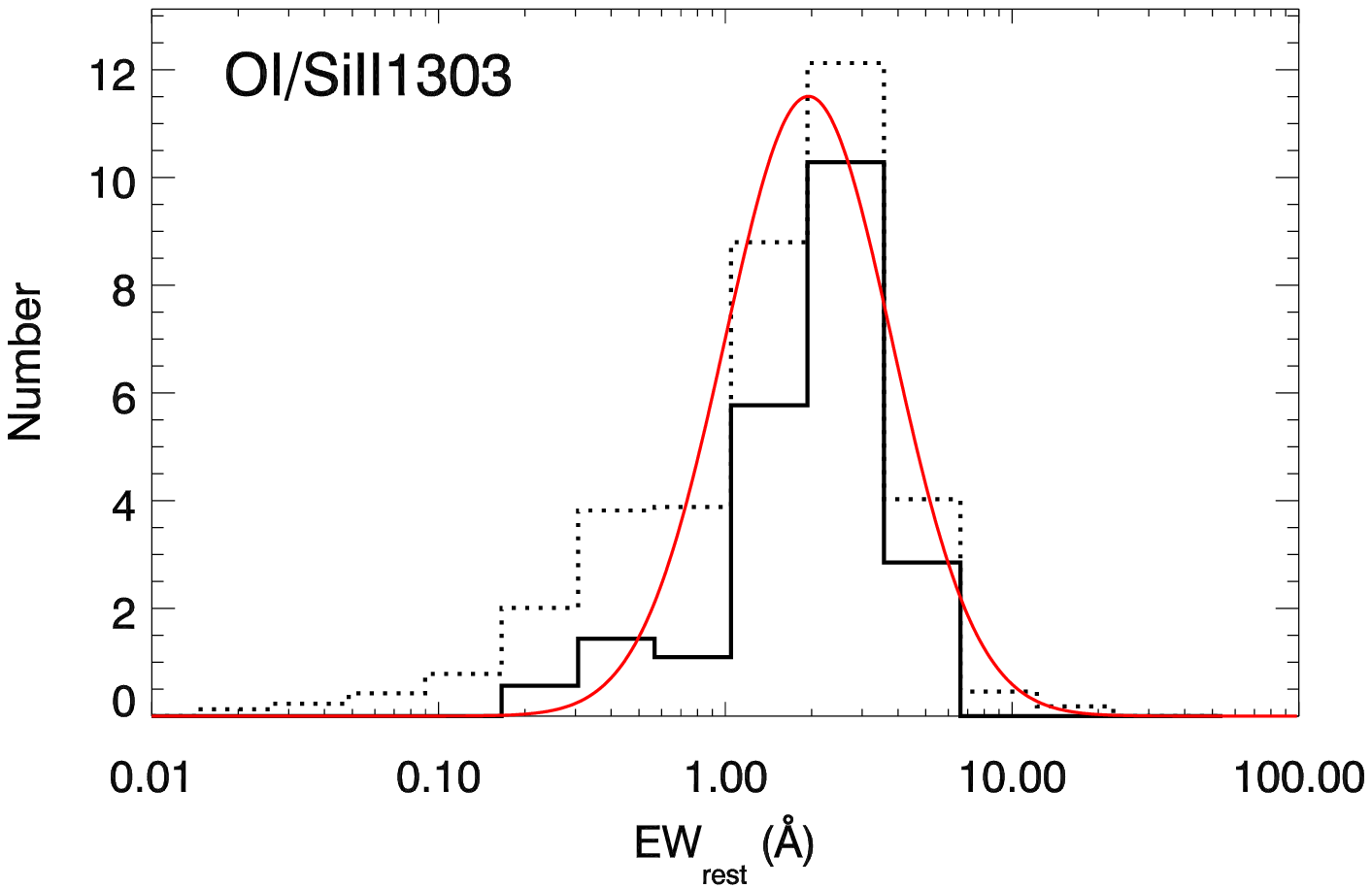} \\
\includegraphics[width=8cm,clip]{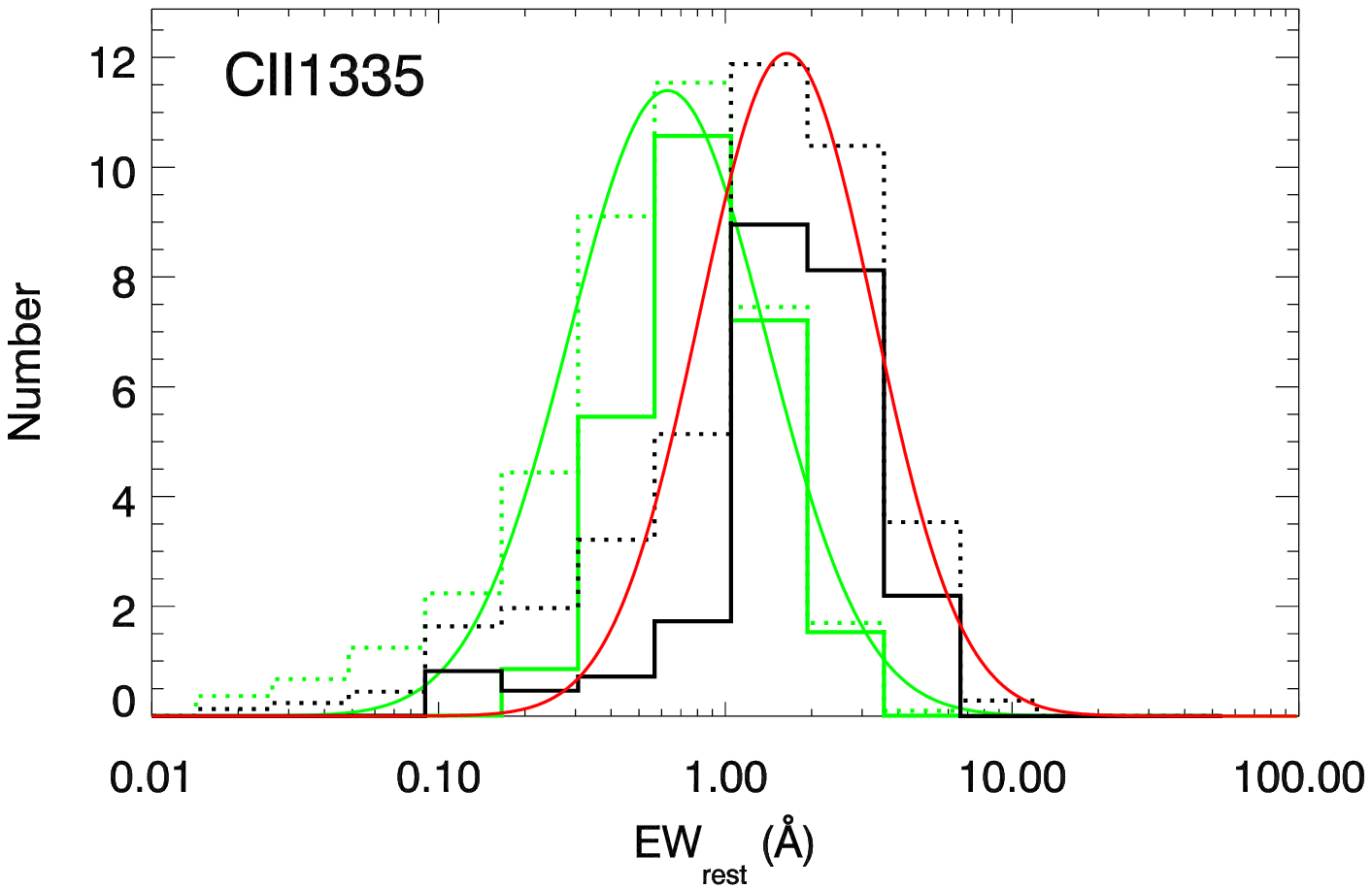} & \includegraphics[width=8cm,clip]{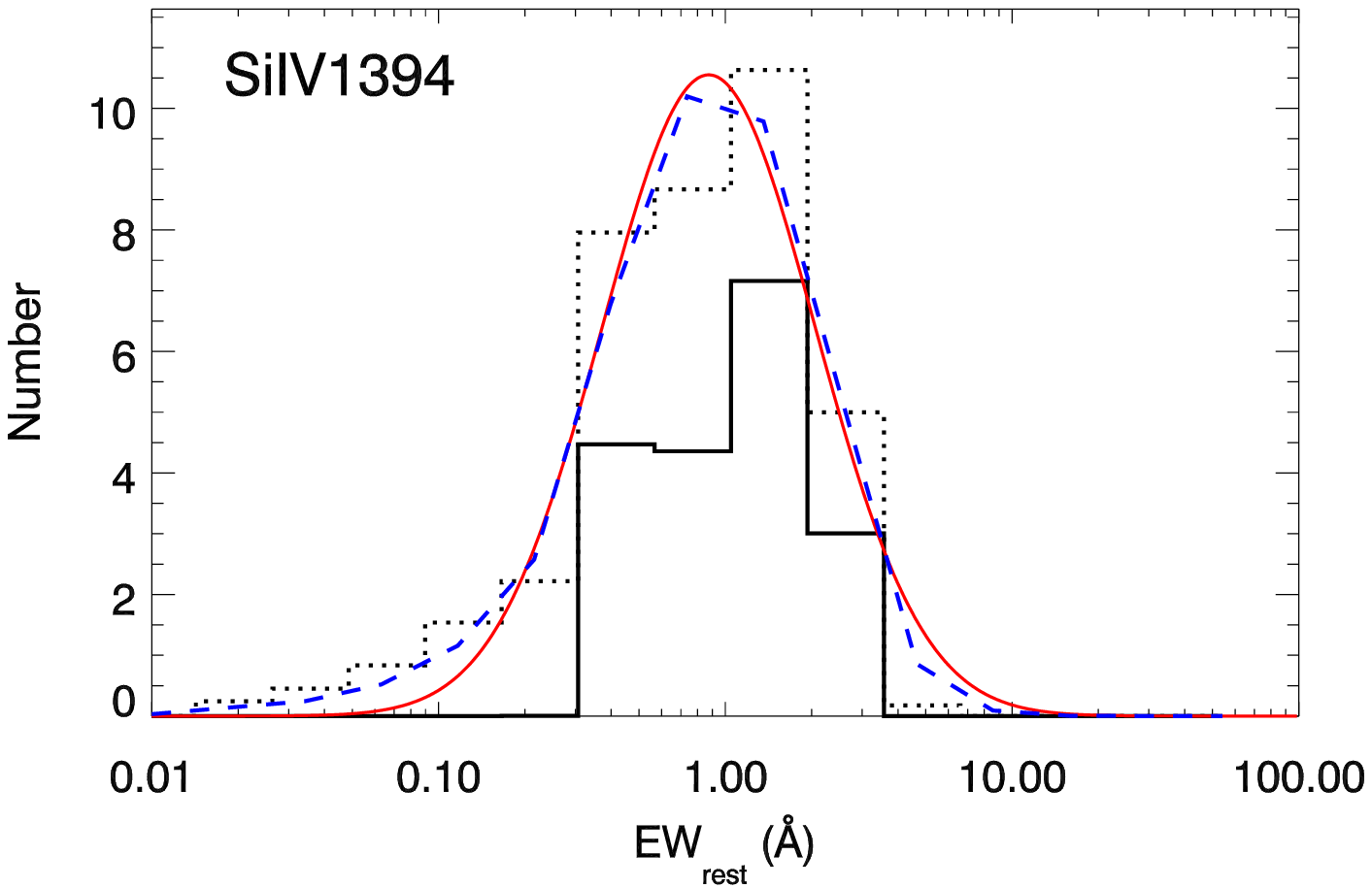} \\
\includegraphics[width=8cm,clip]{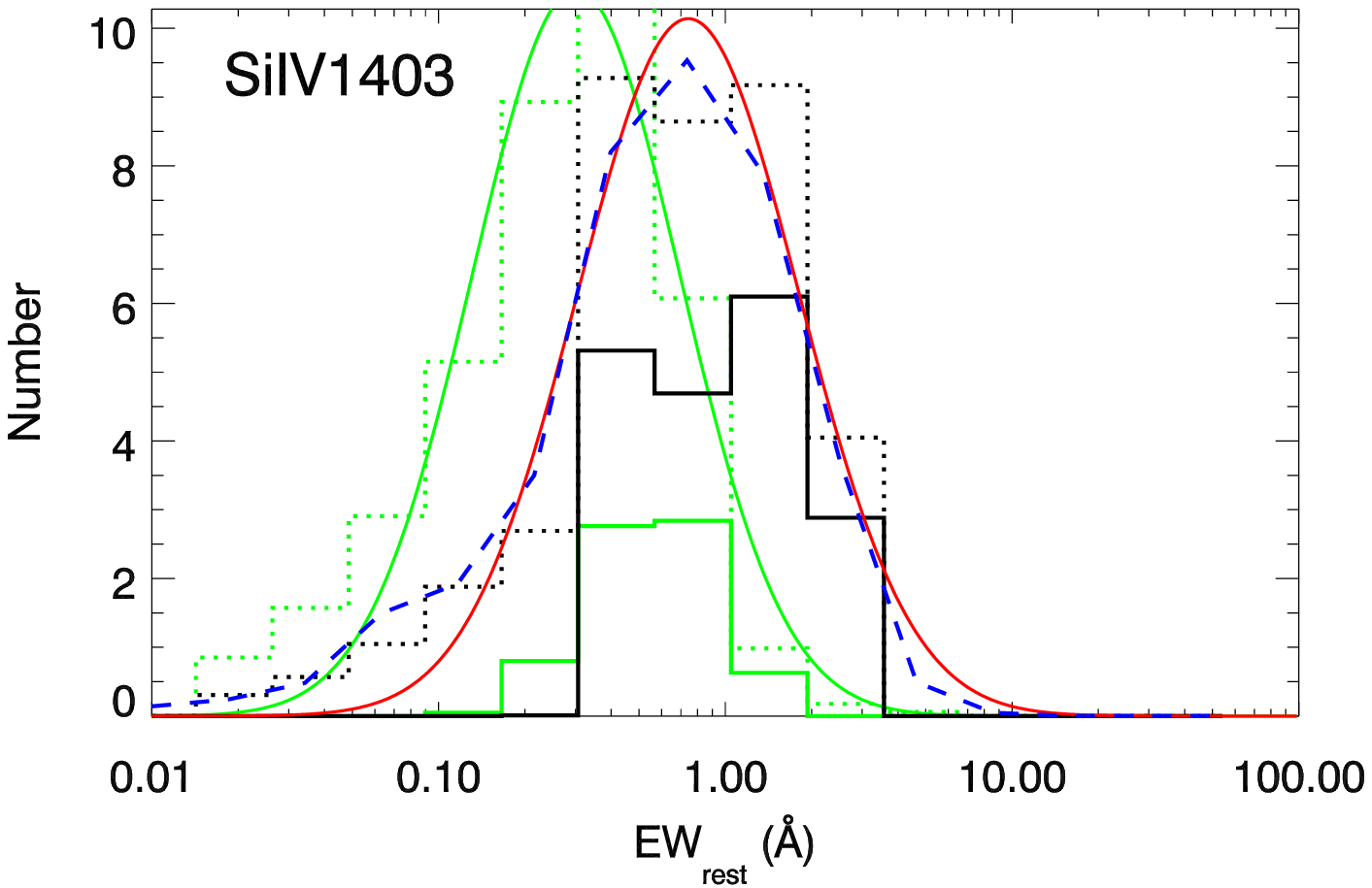} & \includegraphics[width=8cm,clip]{SiII1526_log.eps} \\
\includegraphics[width=8cm,clip]{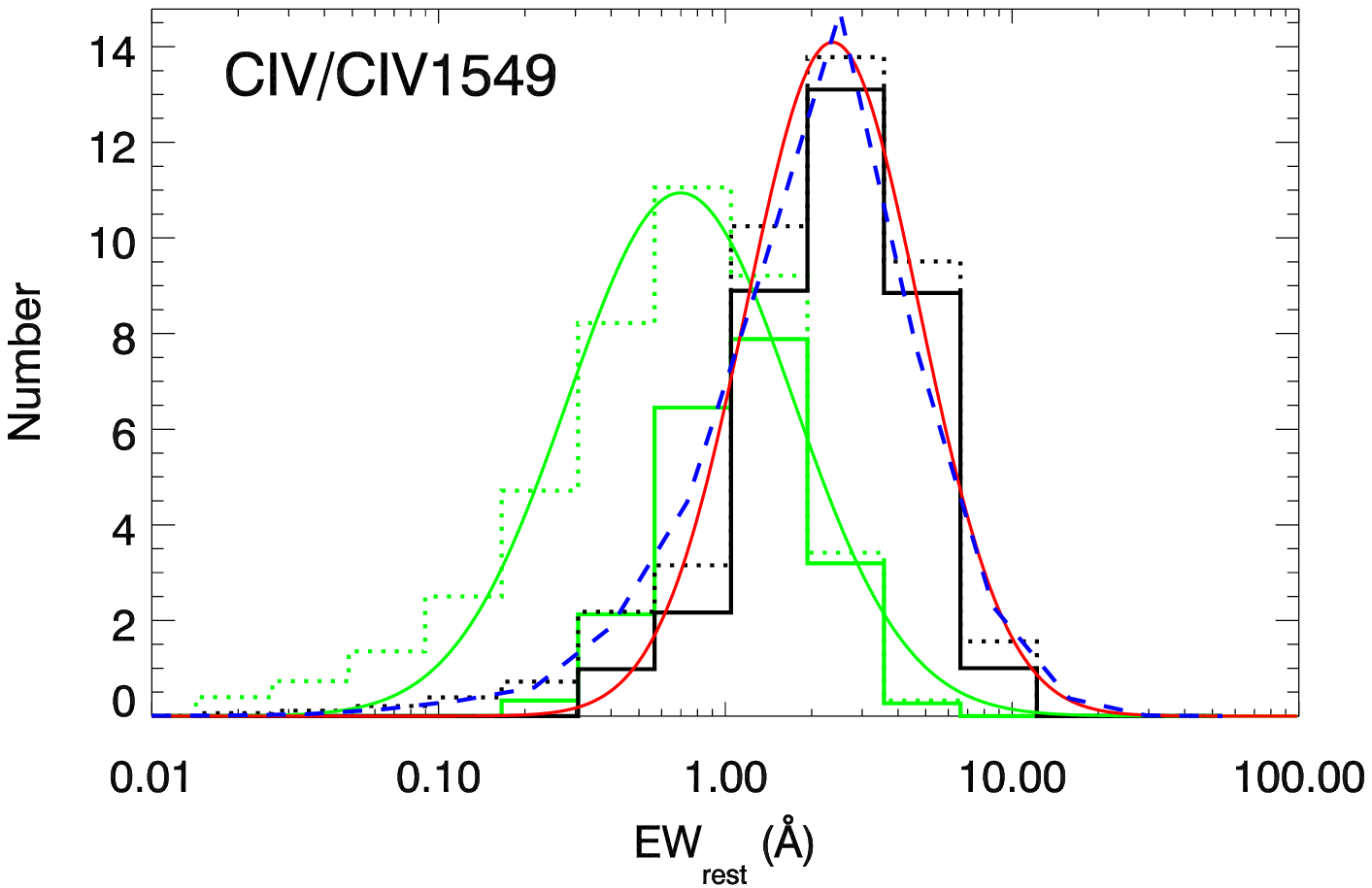} & \includegraphics[width=8cm,clip]{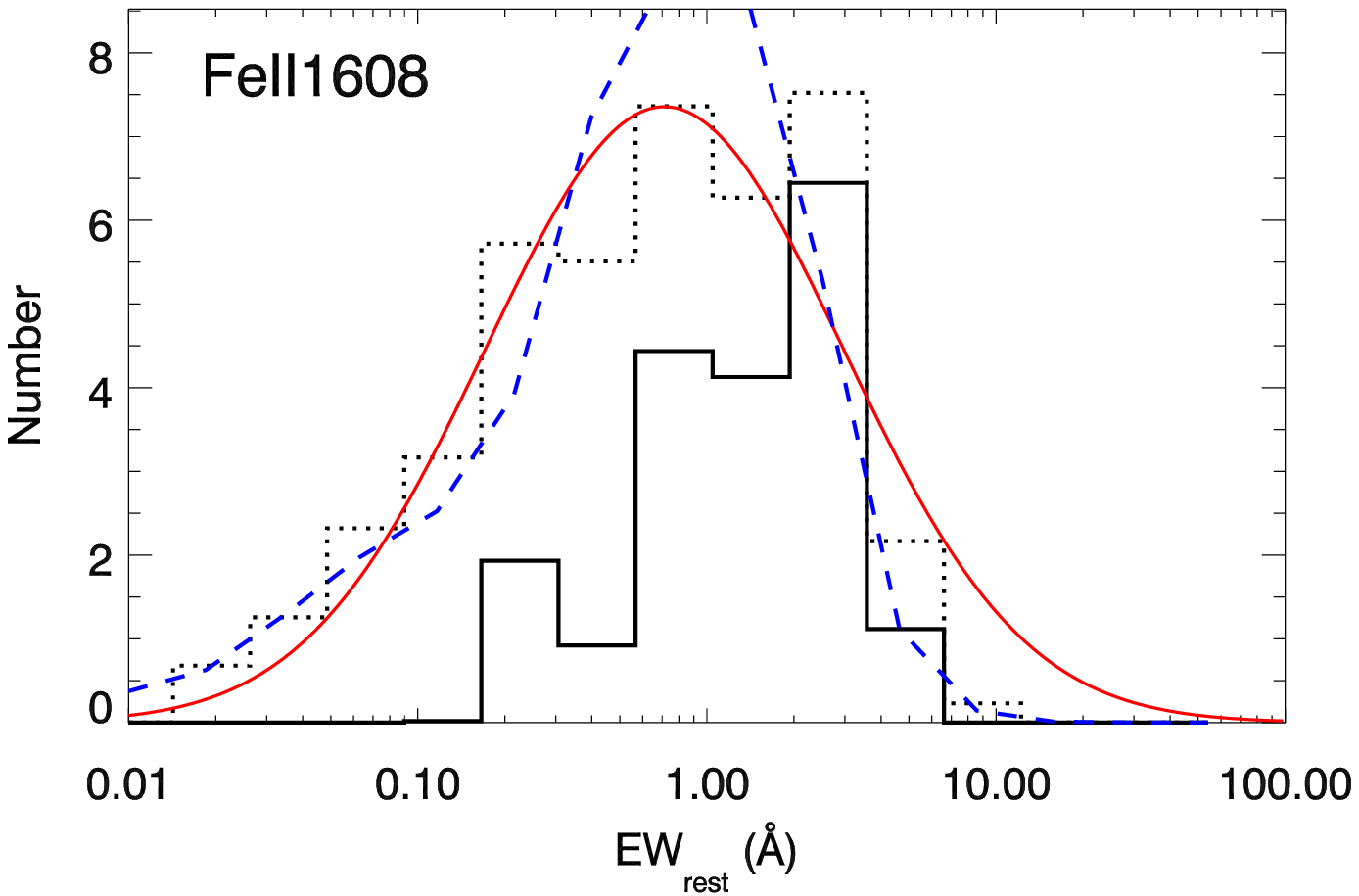} \\
\end{tabular}
\caption{
EW distribution of the different spectral features. The thick black line histogram indicates only detections, whereas the dotted histogram includes also the limits. The red line is the best fit of the complete histogram with a lognormal distribution. In some cases there is a dashed blue line for the physical fit, as described in Section~\ref{sec:fit}. In green we show a histogram of EWs, and its correspondent lognormal fit, of a sample of DLA systems (see Section~\ref{sec:dla}) for a subsample of lines.}
\label{Fig:fits}
\end{figure*}

Figure~\ref{Fig:fits} shows the distribution of the EWs of each the lines. The thick black line histogram indicates only detections, whereas the dotted histogram includes also the limits. The red line is the best fit of the complete histogram with a lognormal distribution and the dashed blue line (when available) is a physical fit as described in Section~\ref{sec:fit}. In a limited amount of features we also show in green a histogram of EWs in a sample of damped Lyman-$\alpha$ systems (see Section\ref{sec:dla}).

   \clearpage
   \newpage

\clearpage
   \newpage

\begin{figure*}[h]
\centering
\begin{tabular}{cc}
\includegraphics[width=8cm,clip]{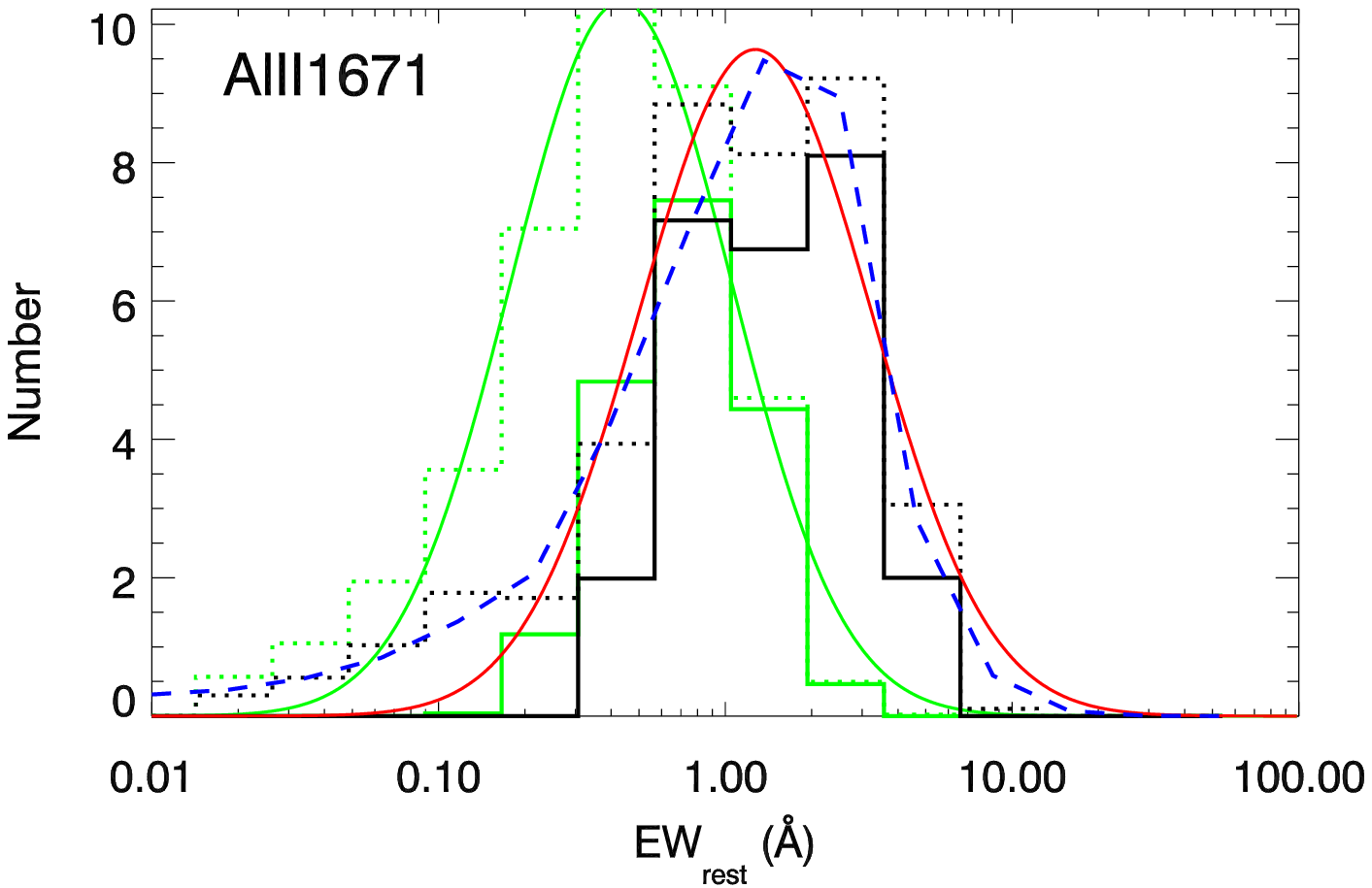} & \includegraphics[width=8cm,clip]{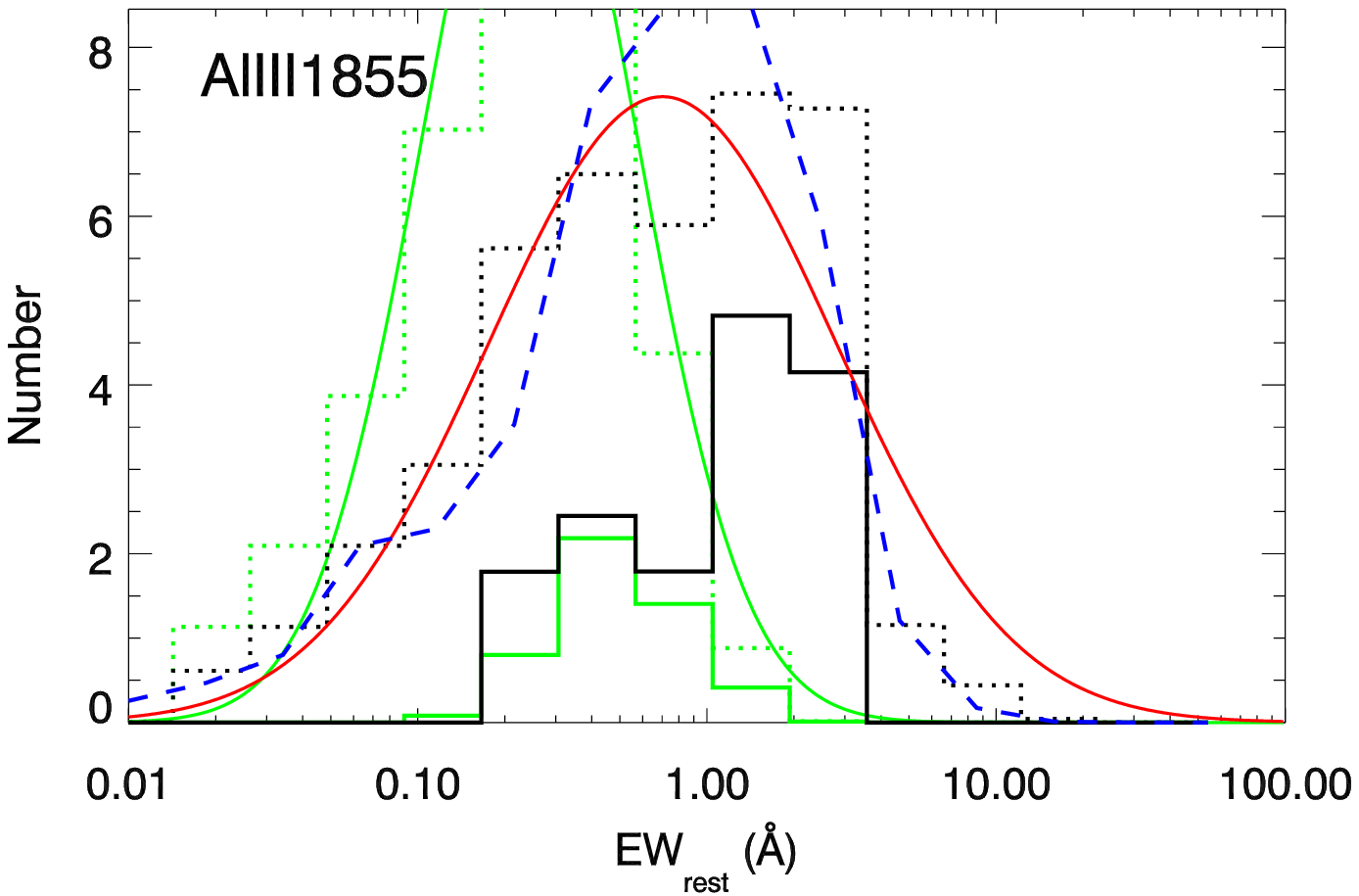} \\
\includegraphics[width=8cm,clip]{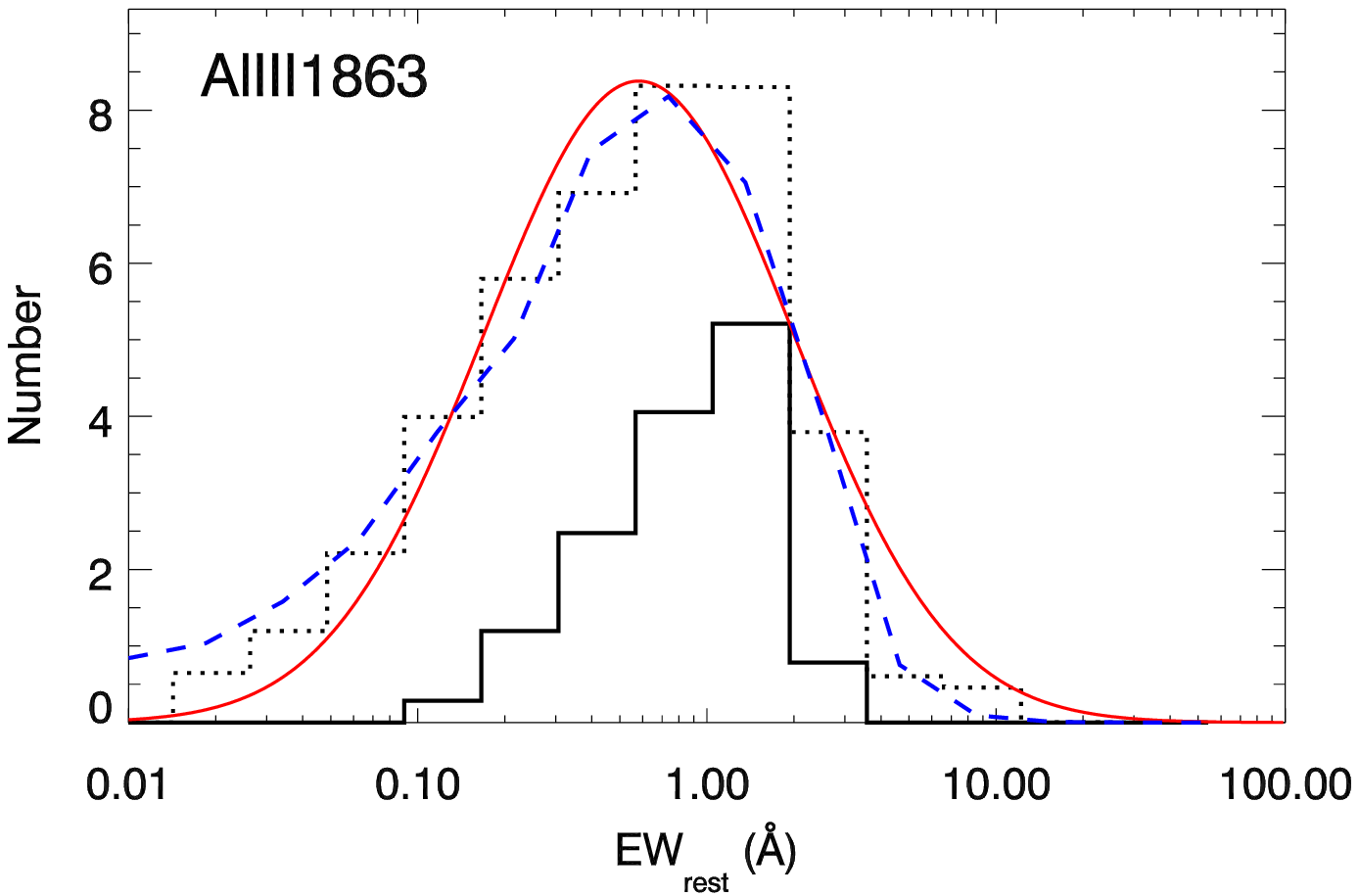} & \includegraphics[width=8cm,clip]{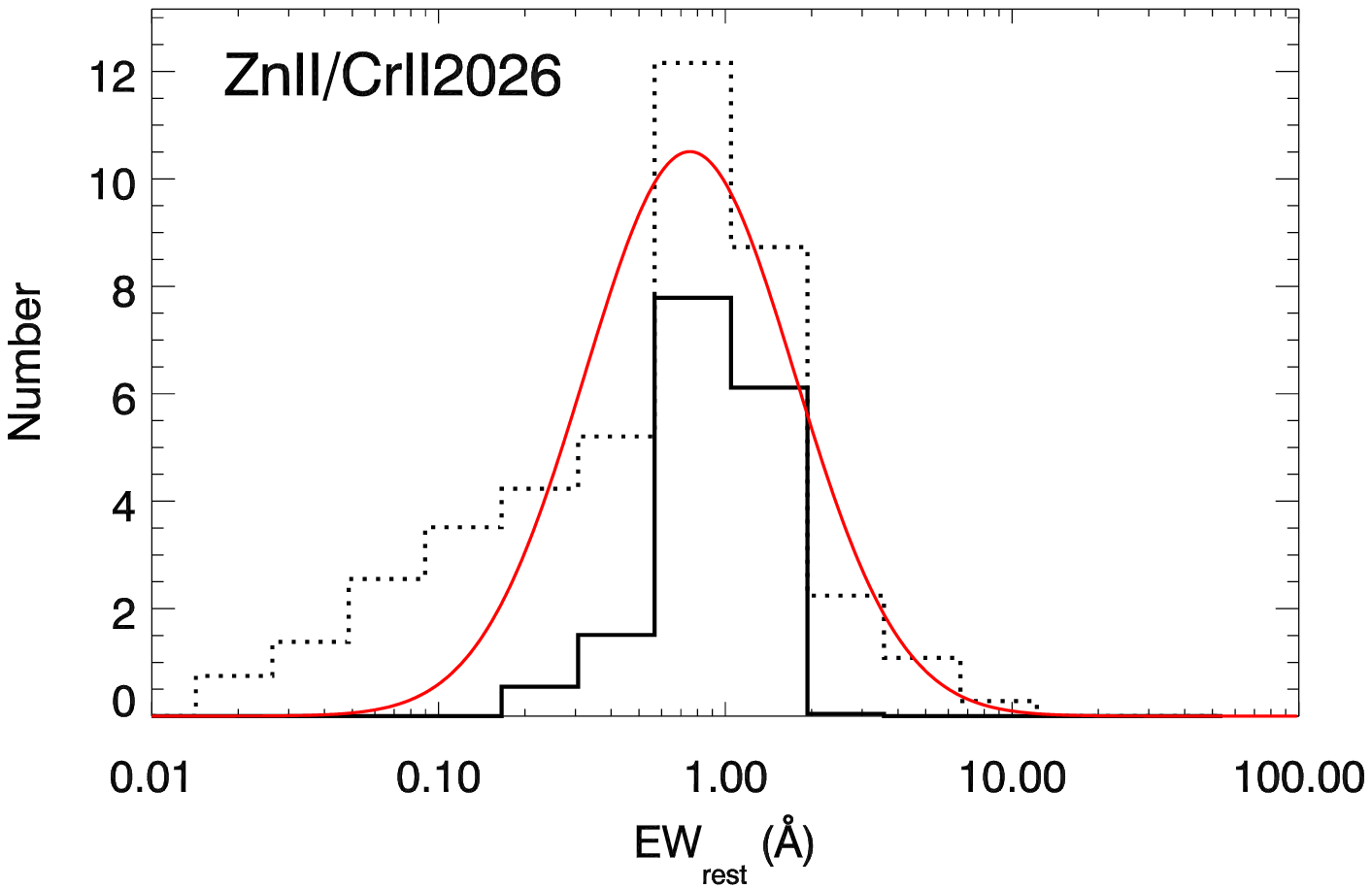} \\
\includegraphics[width=8cm,clip]{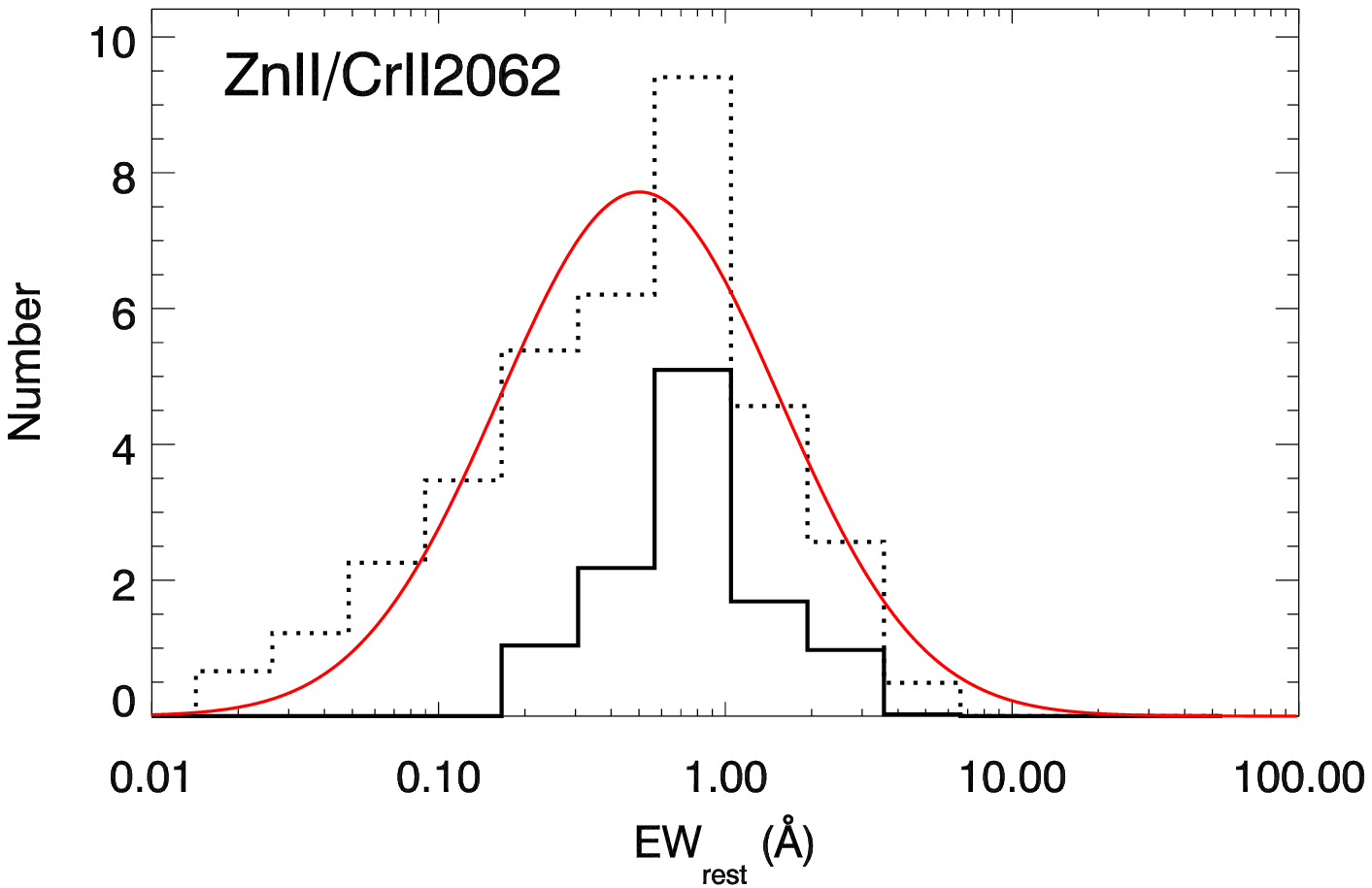} & \includegraphics[width=8cm,clip]{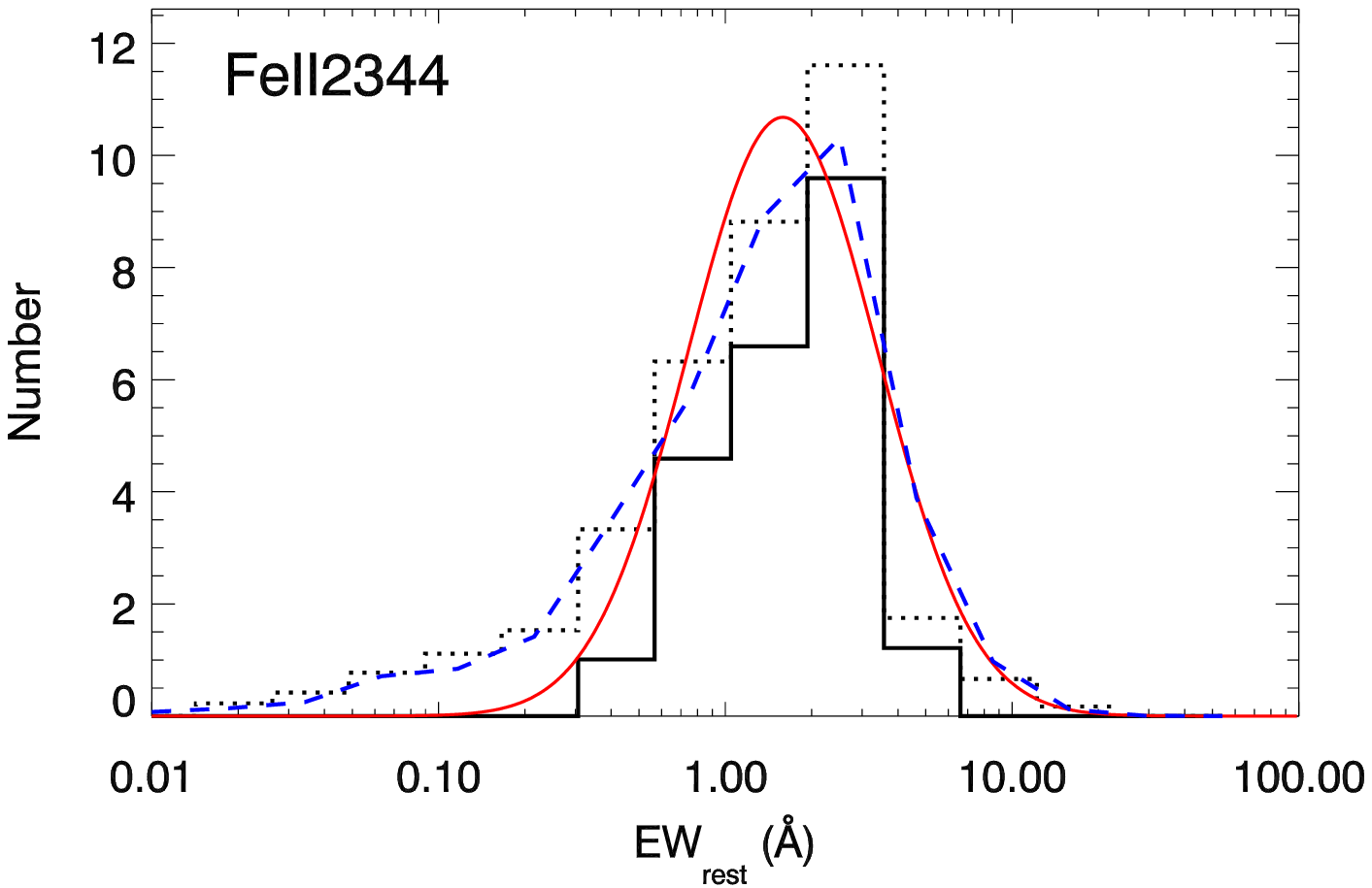} \\
\includegraphics[width=8cm,clip]{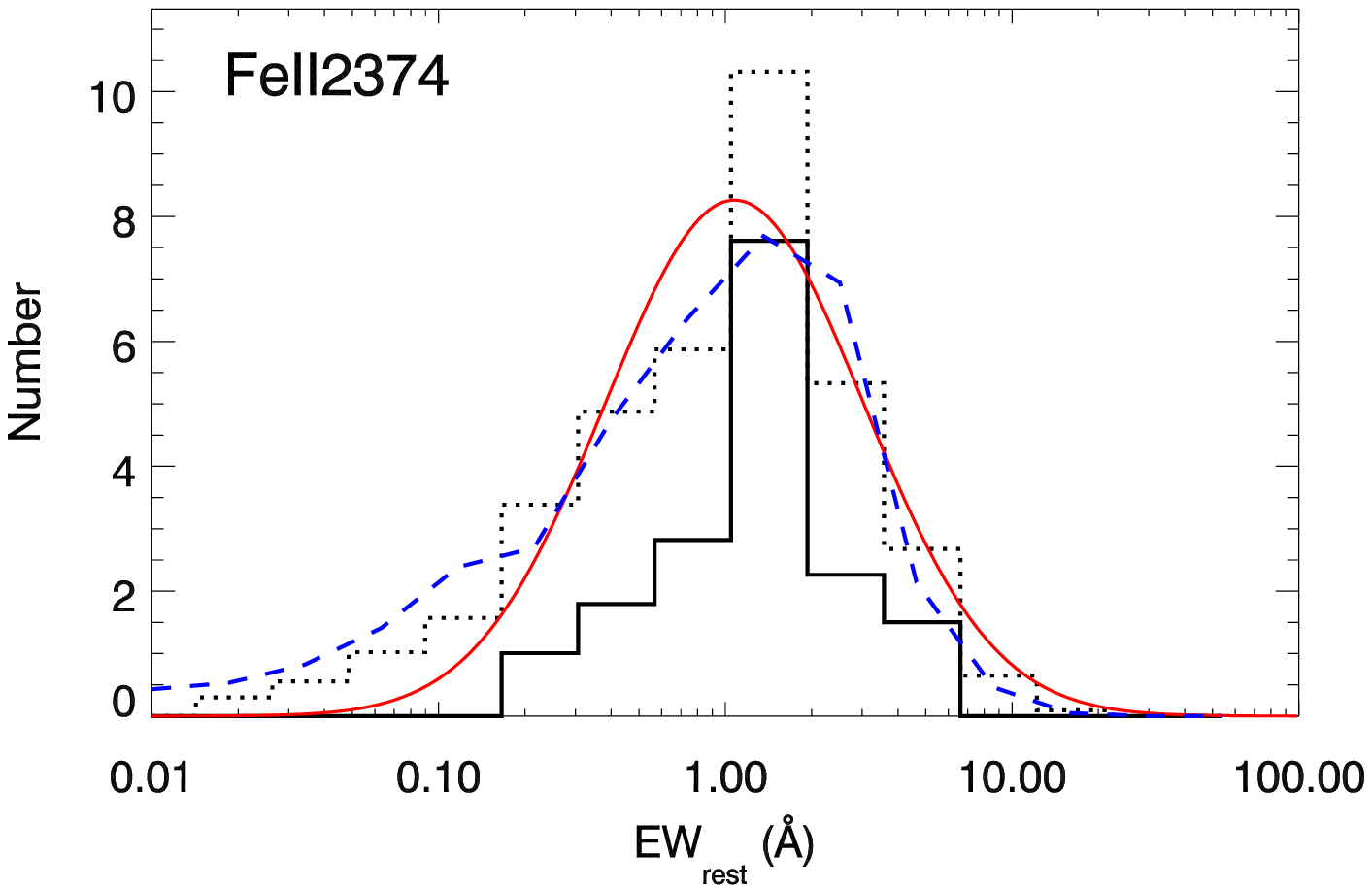} & \includegraphics[width=8cm,clip]{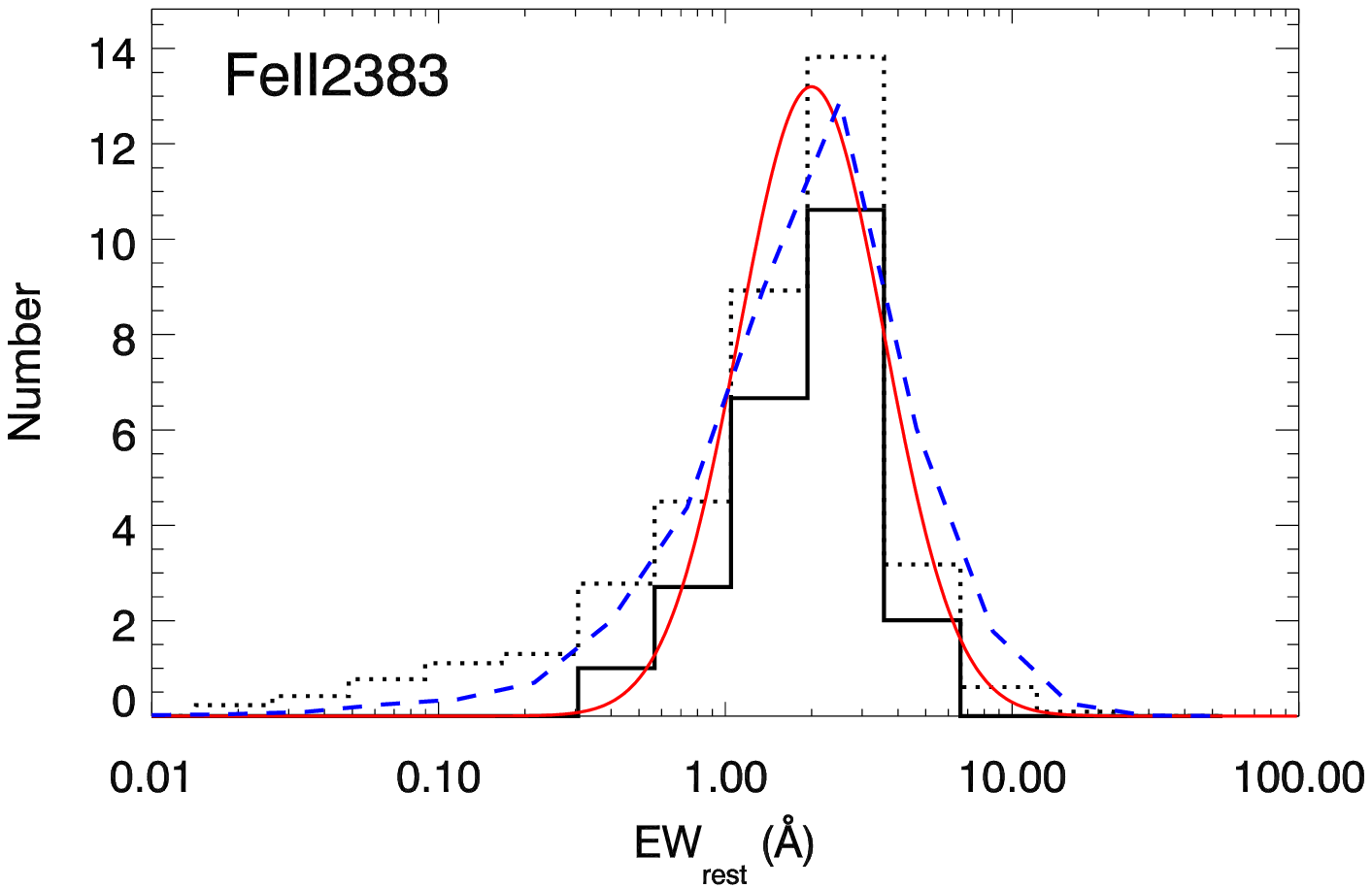} \\
\end{tabular}
\end{figure*}

\clearpage
   \newpage

\begin{figure*}[h]
\centering
\begin{tabular}{cc}
\includegraphics[width=8cm,clip]{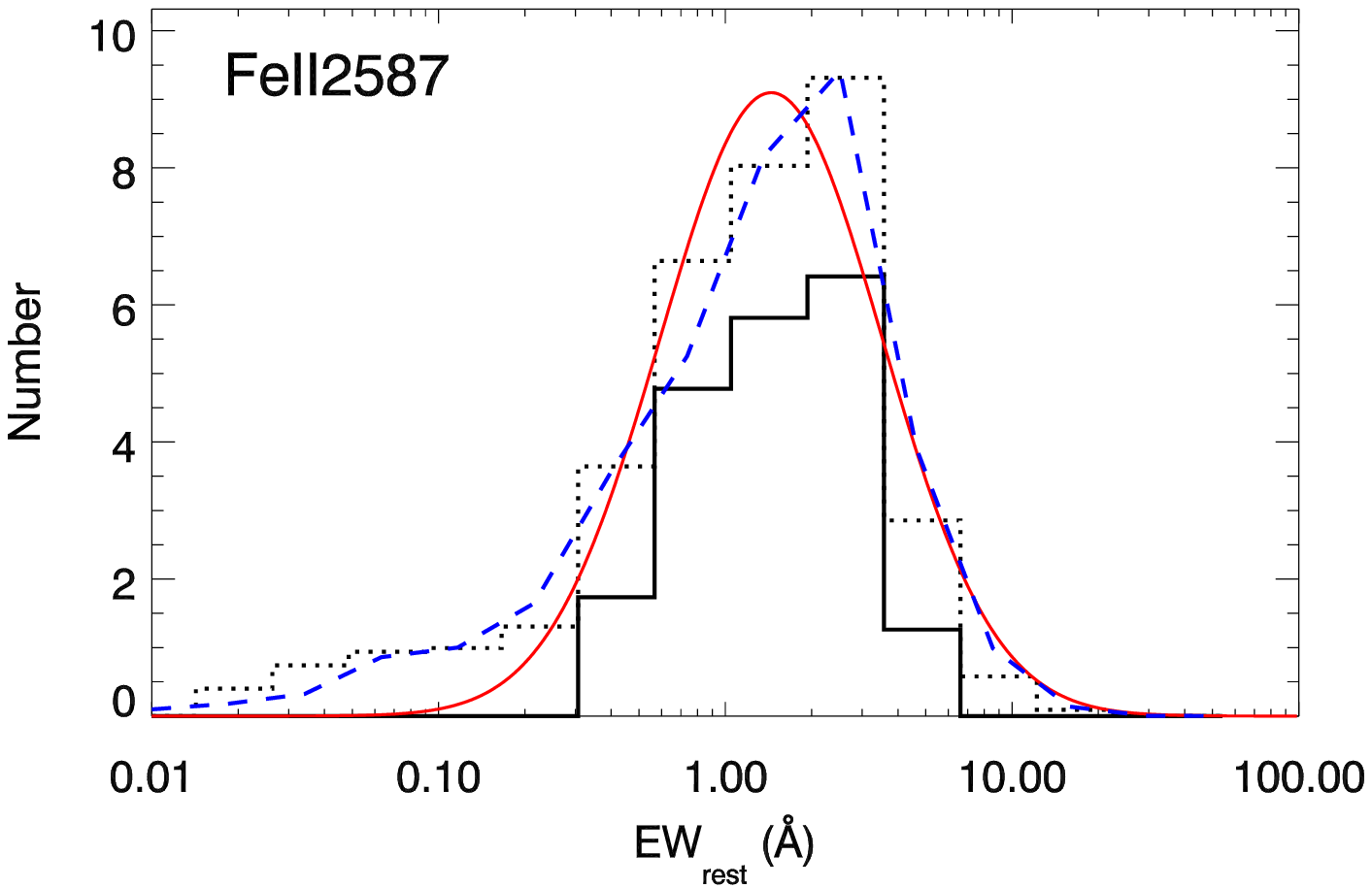} & \includegraphics[width=8cm,clip]{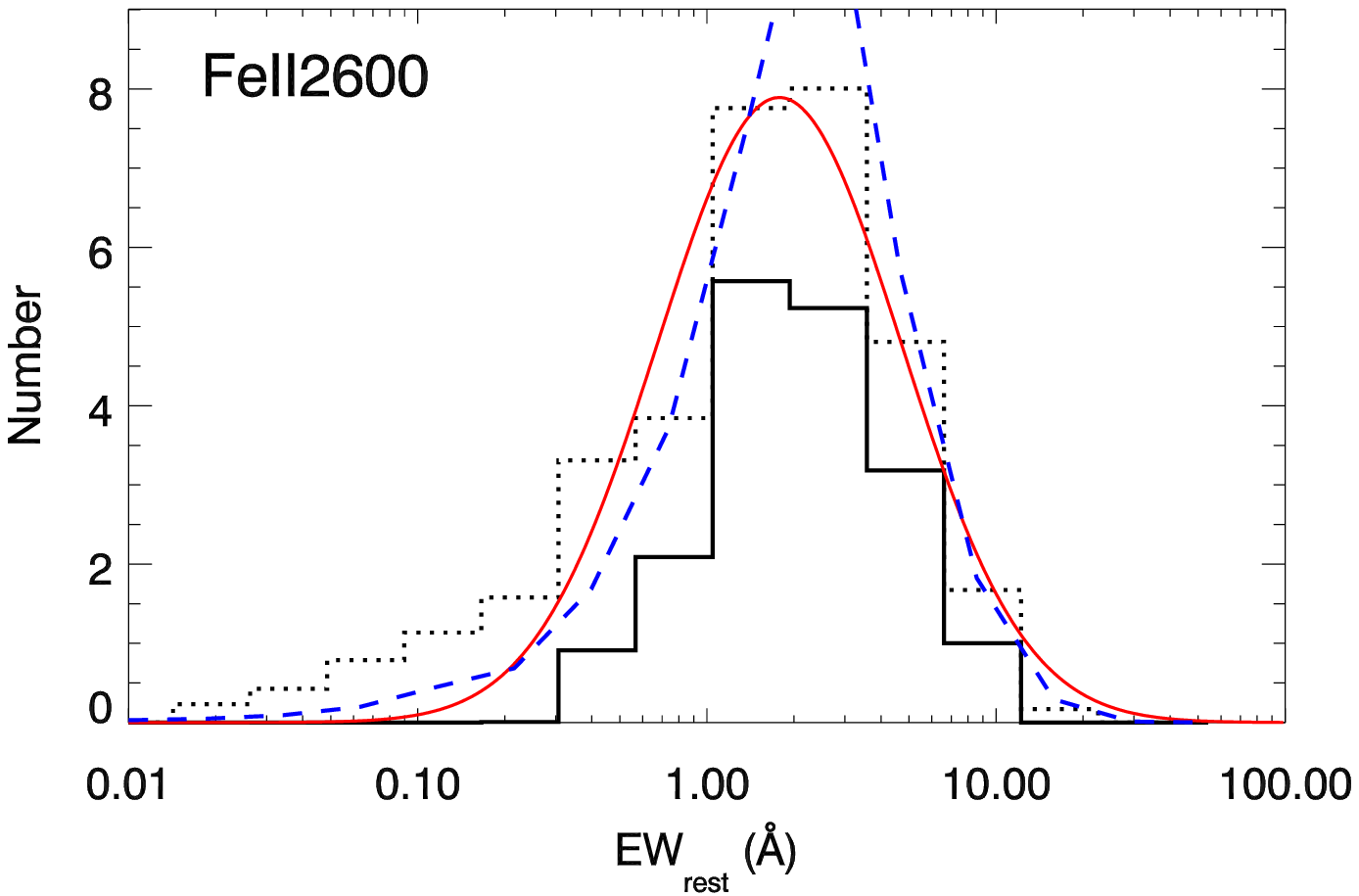} \\
\includegraphics[width=8cm,clip]{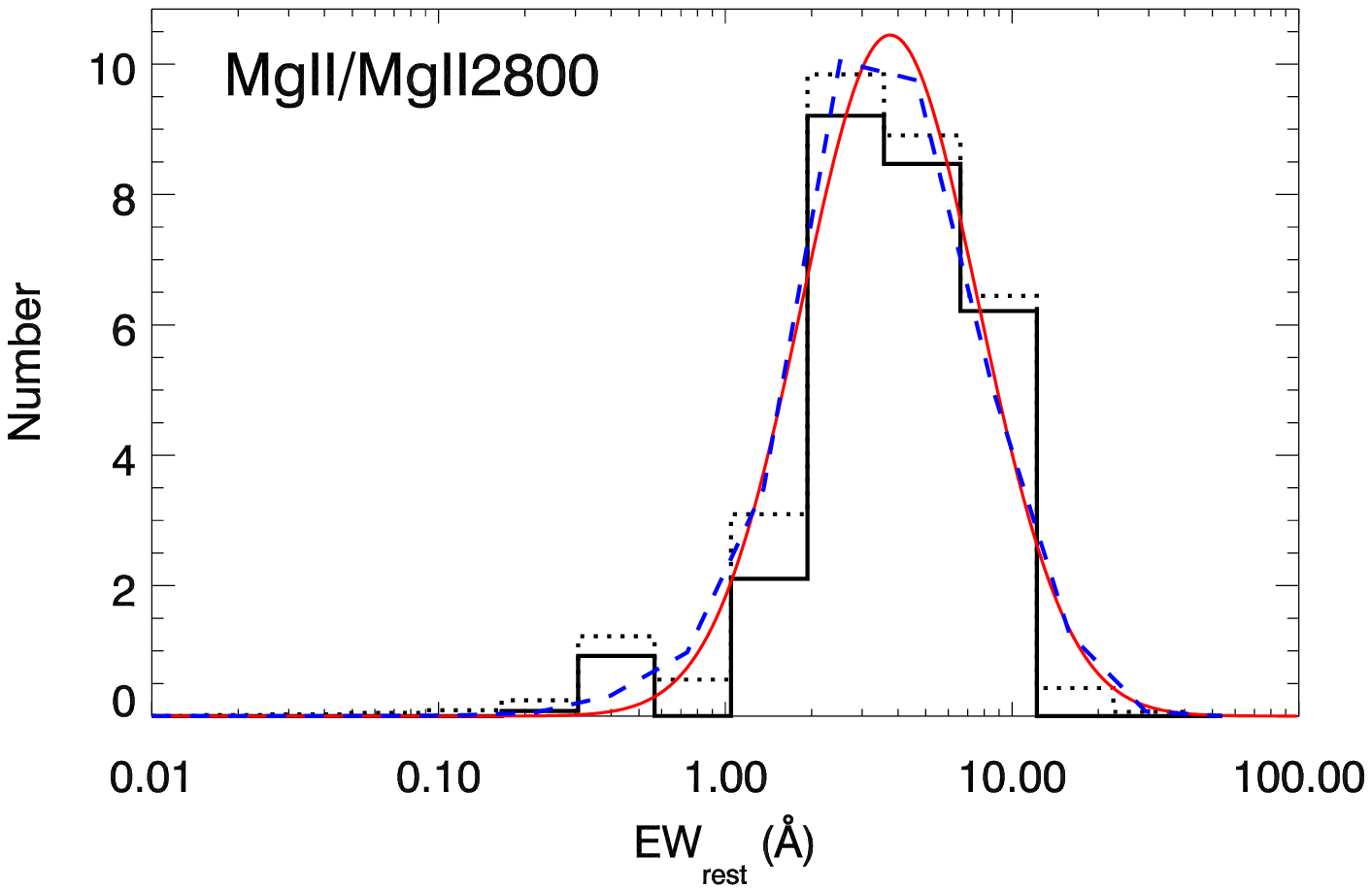} & \includegraphics[width=8cm,clip]{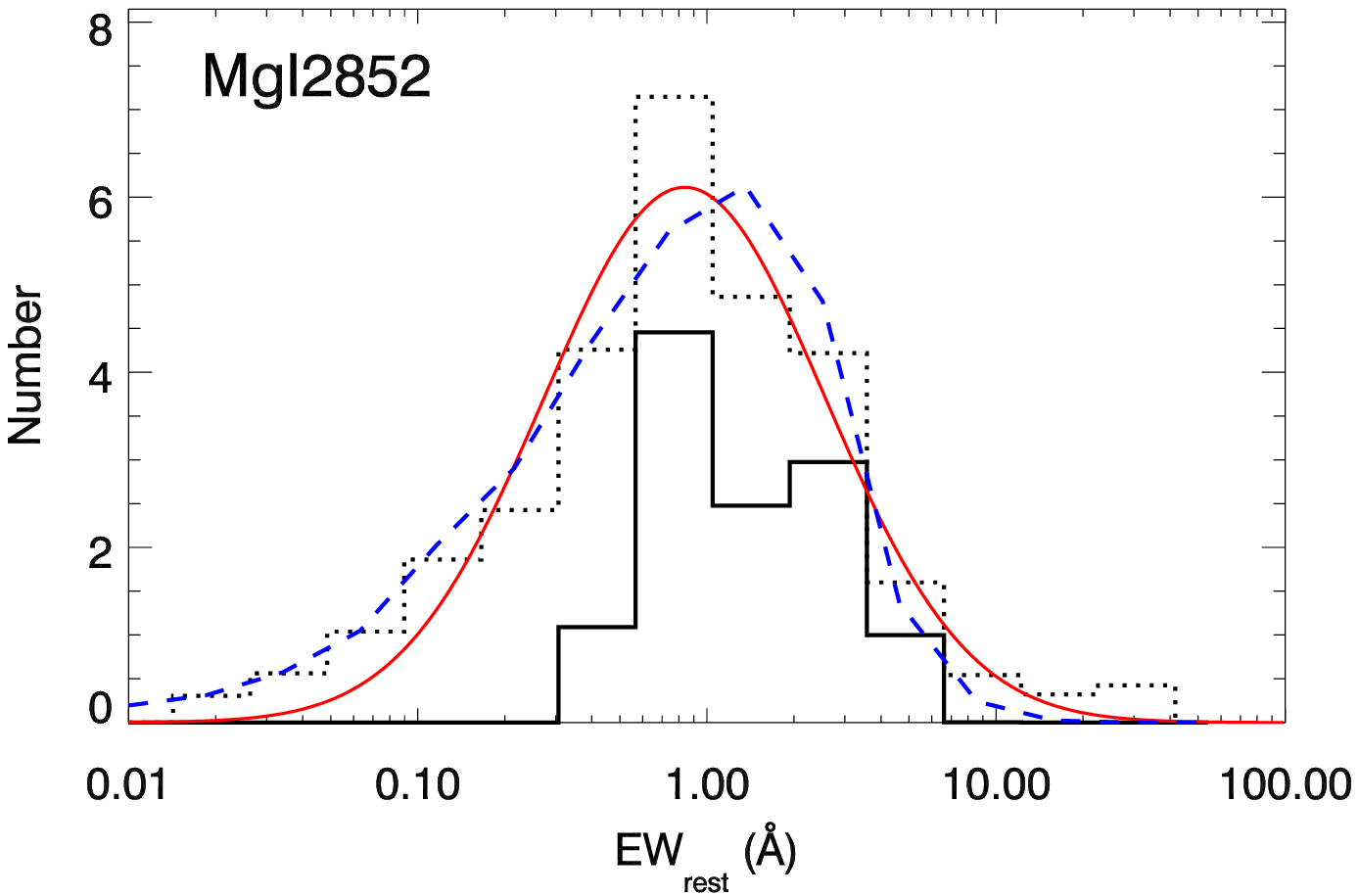} \\
\includegraphics[width=8cm,clip]{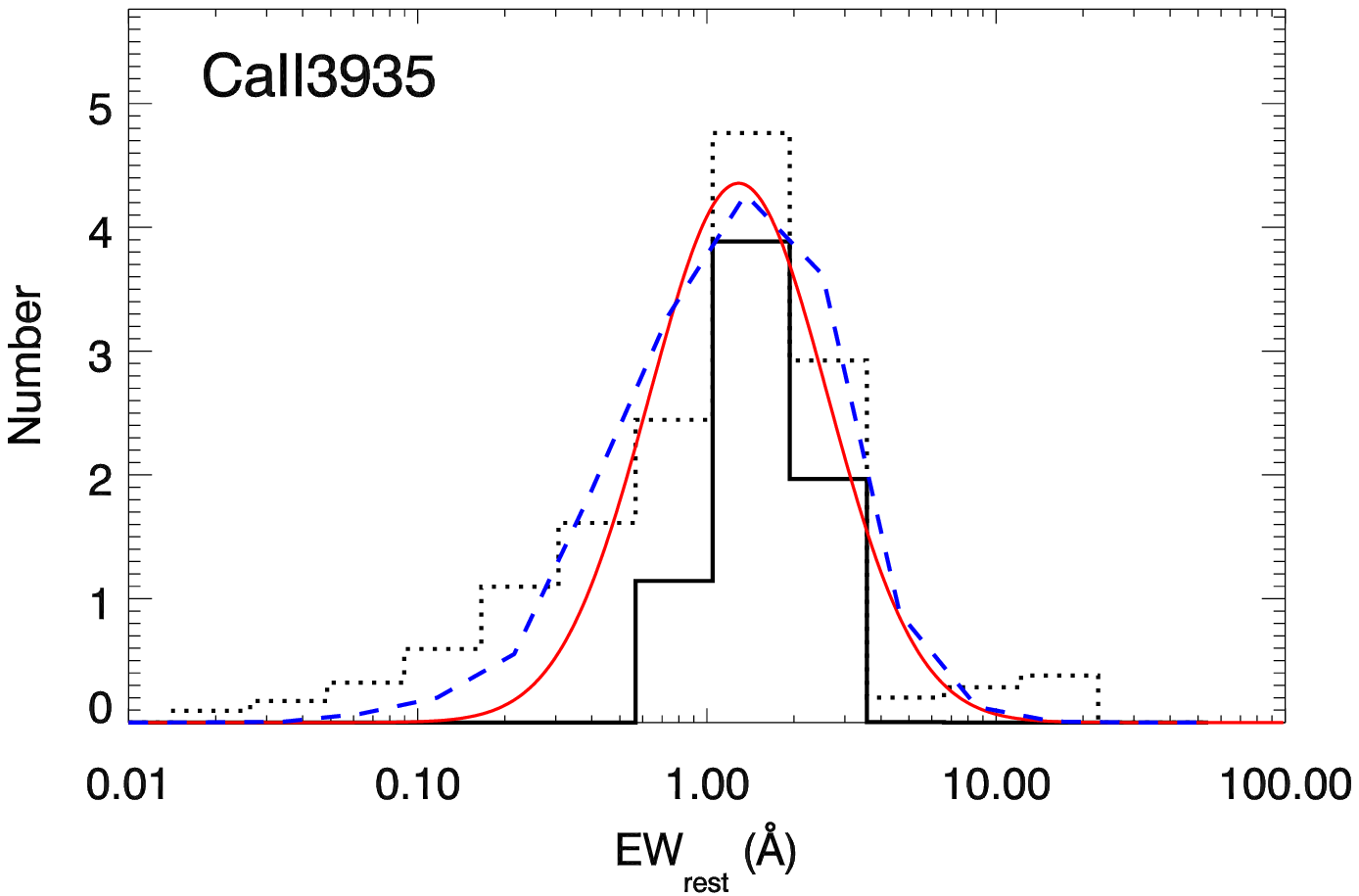} & \includegraphics[width=8cm,clip]{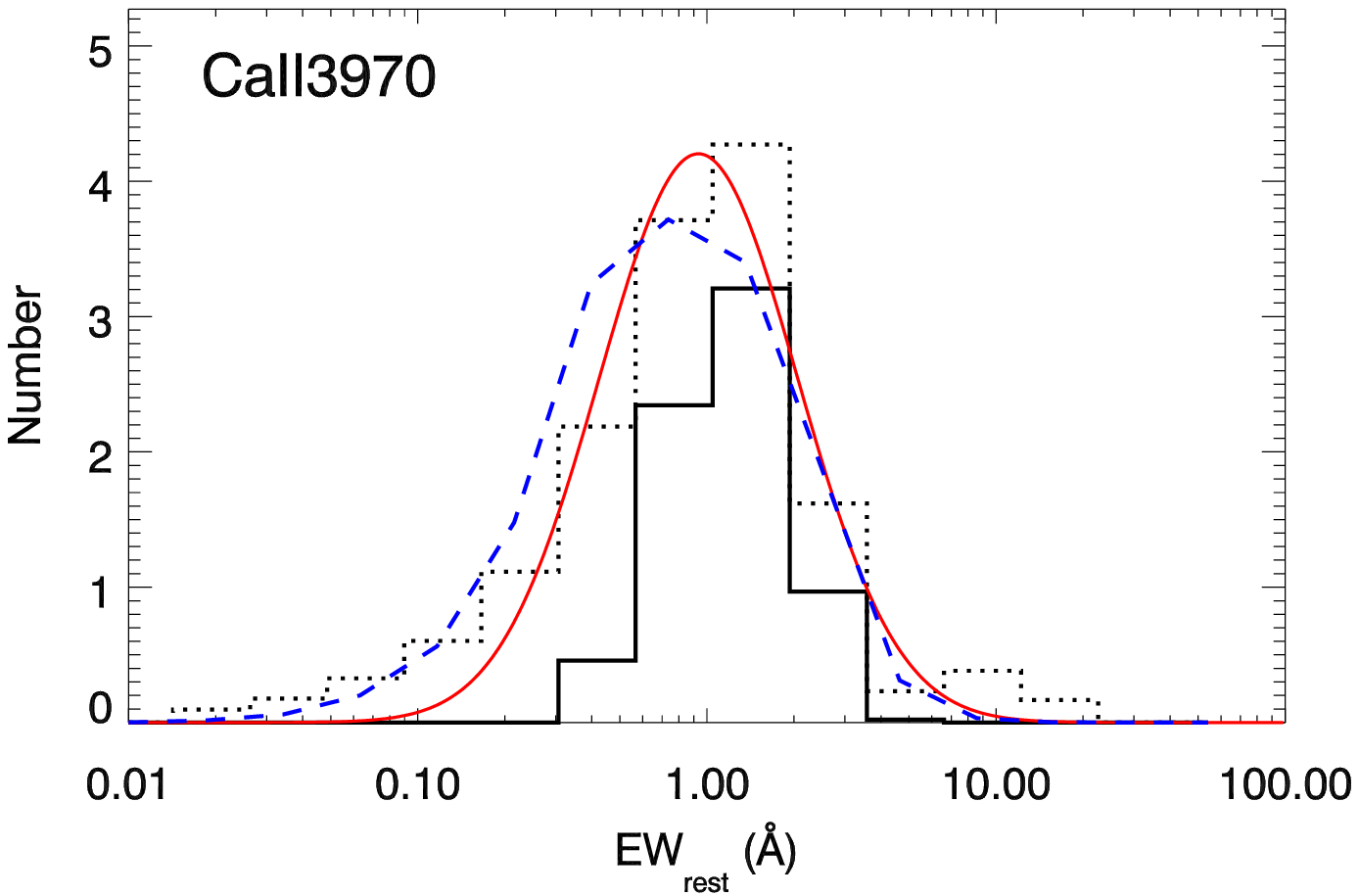} \\ 
\end{tabular}
\end{figure*}

\clearpage
   \newpage

\section{Catalogue of EW diagrams}

\begin{figure*}[]
\centering
\includegraphics[width=14cm,clip]{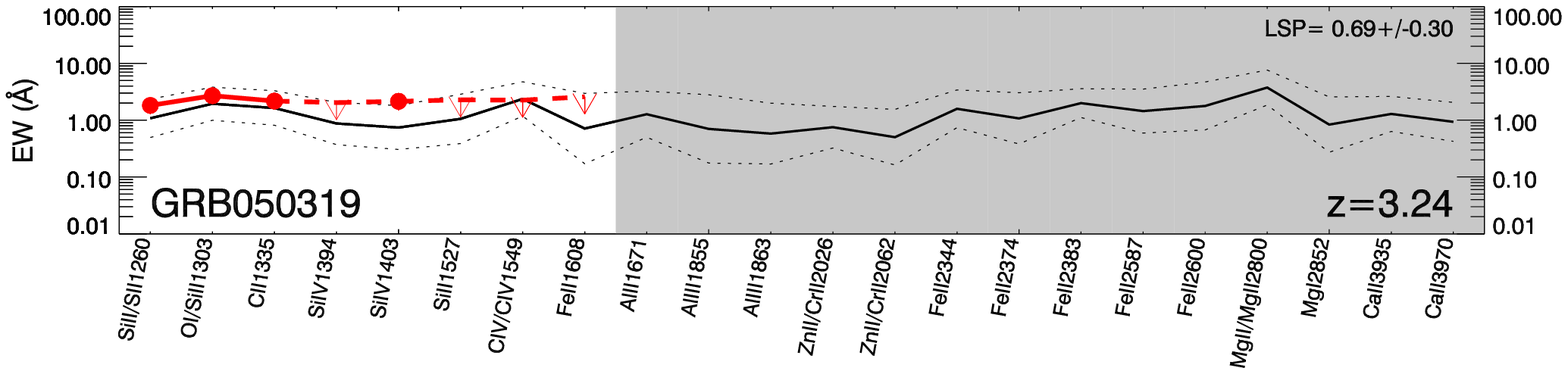}
\includegraphics[width=14cm,clip]{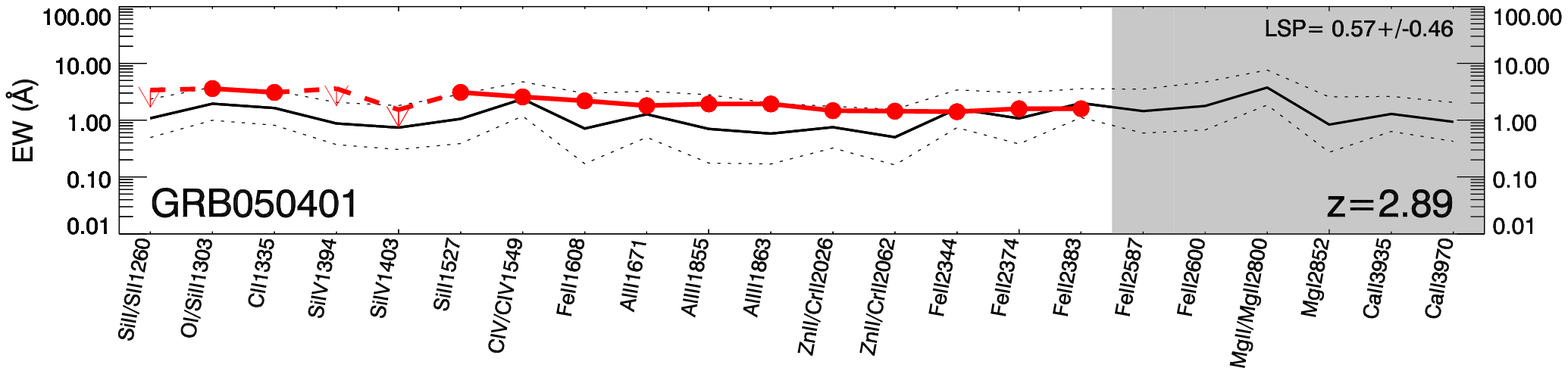}
\includegraphics[width=14cm,clip]{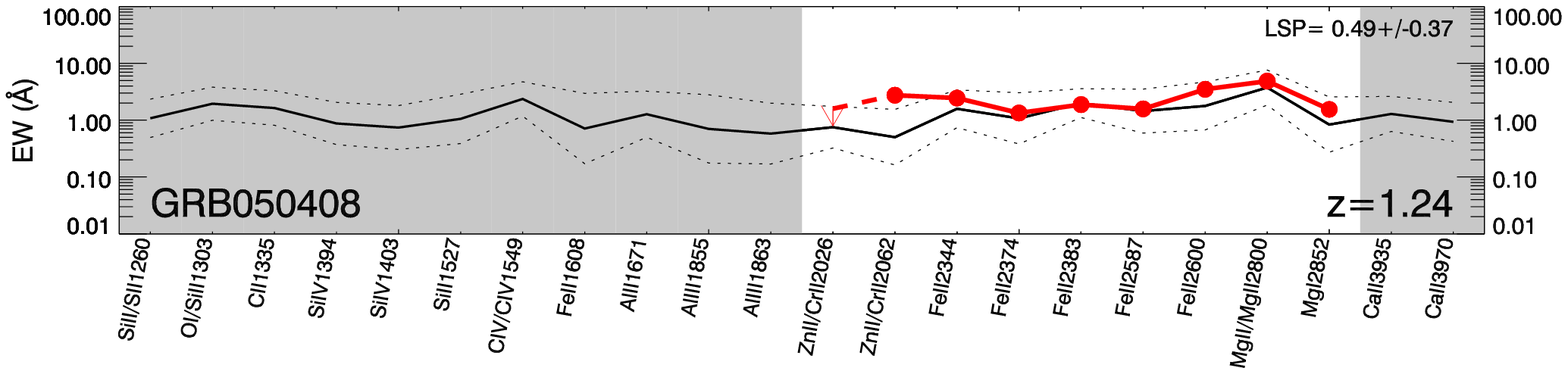}
\includegraphics[width=14cm,clip]{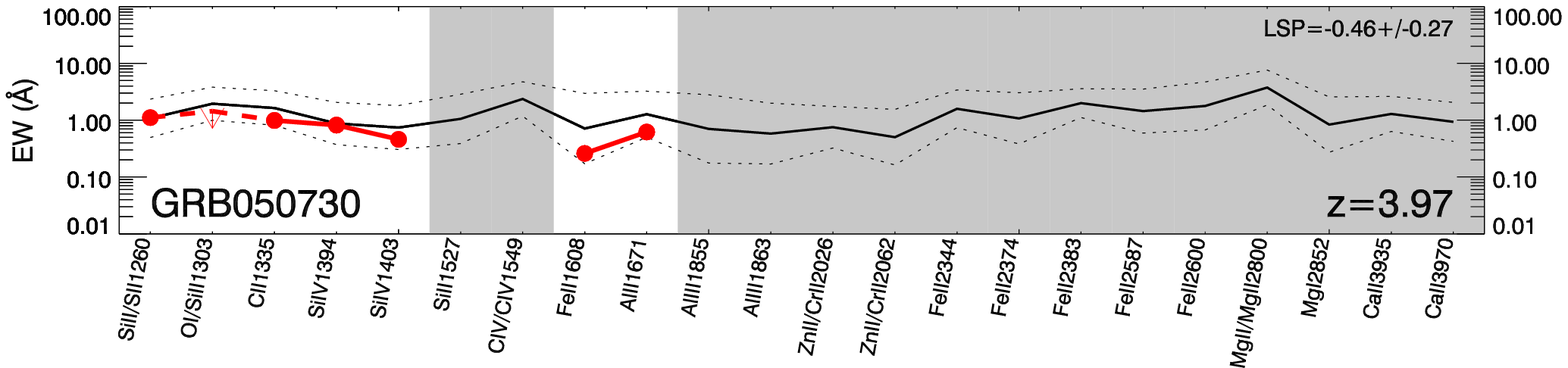}
\includegraphics[width=14cm,clip]{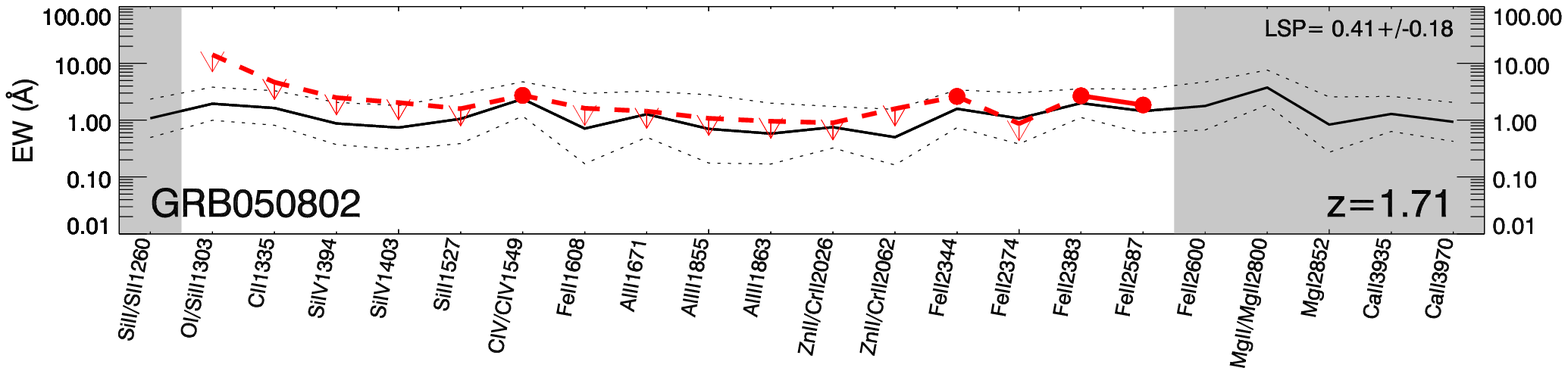}
\includegraphics[width=14cm,clip]{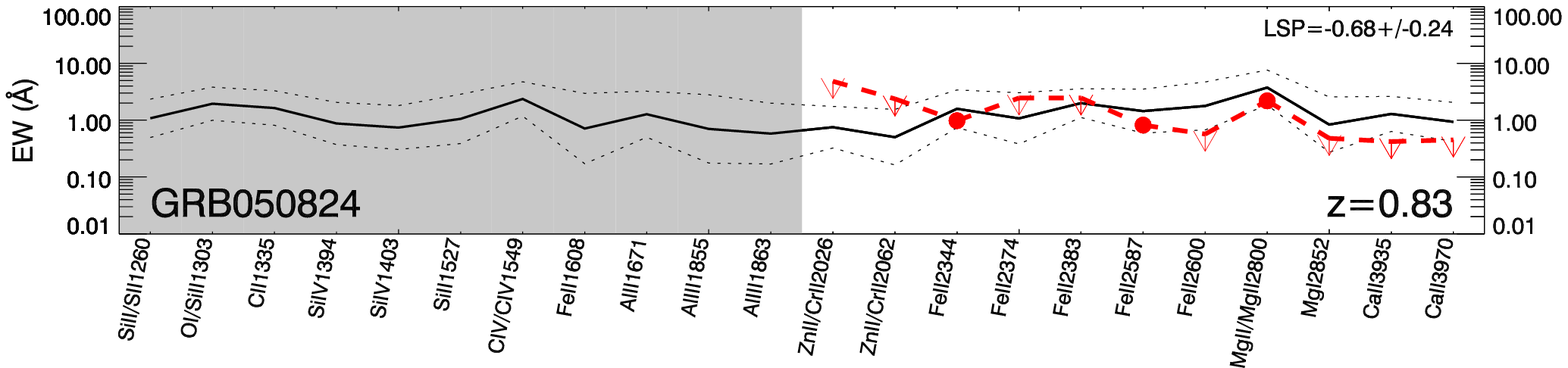}
\includegraphics[width=14cm,clip]{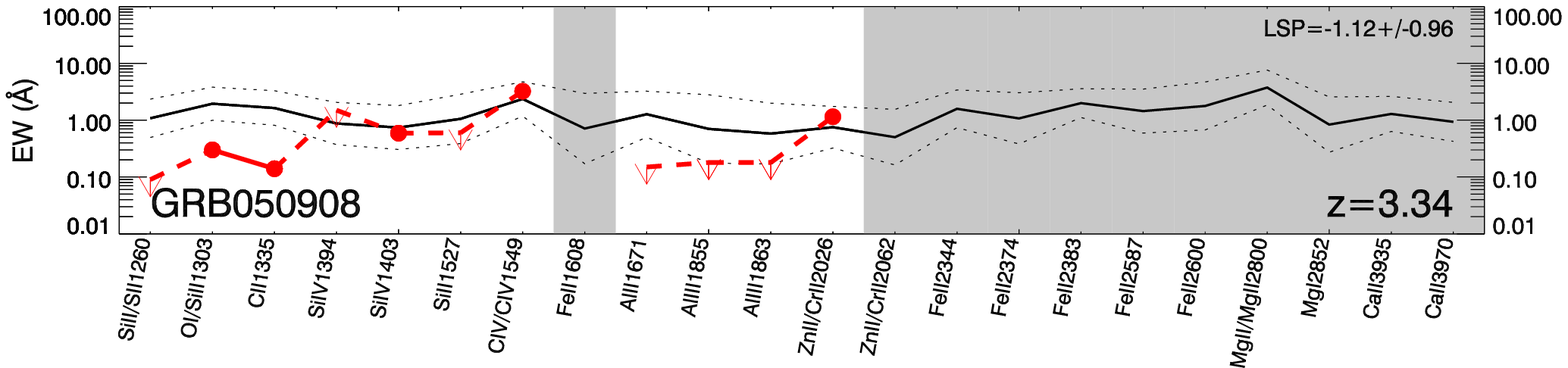}
\caption{Catalogue of GRBs and their spectral features. The GRB measurements are shown in red, while the average values and standard deviations of the fits for the complete sample are drawn in red. The plots also indicate the redshift of the GRB and the LSP. The shaded regions indicate regions of the spectra where there is no data.}
\label{Fig:GRBs}
\end{figure*}

Figure~\ref{Fig:GRBs} shows a catalogue of all the GRBs, where we show the strength of each of the features and compare them with the distributions calculated in this work.

\clearpage
\begin{figure*}[]
\centering
\includegraphics[width=15cm,clip]{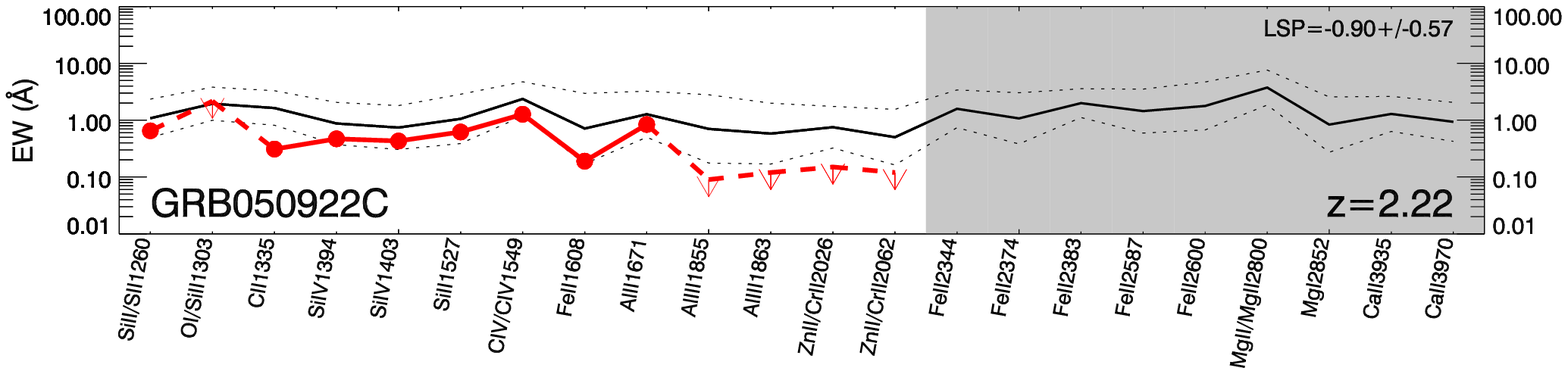}
\includegraphics[width=15cm,clip]{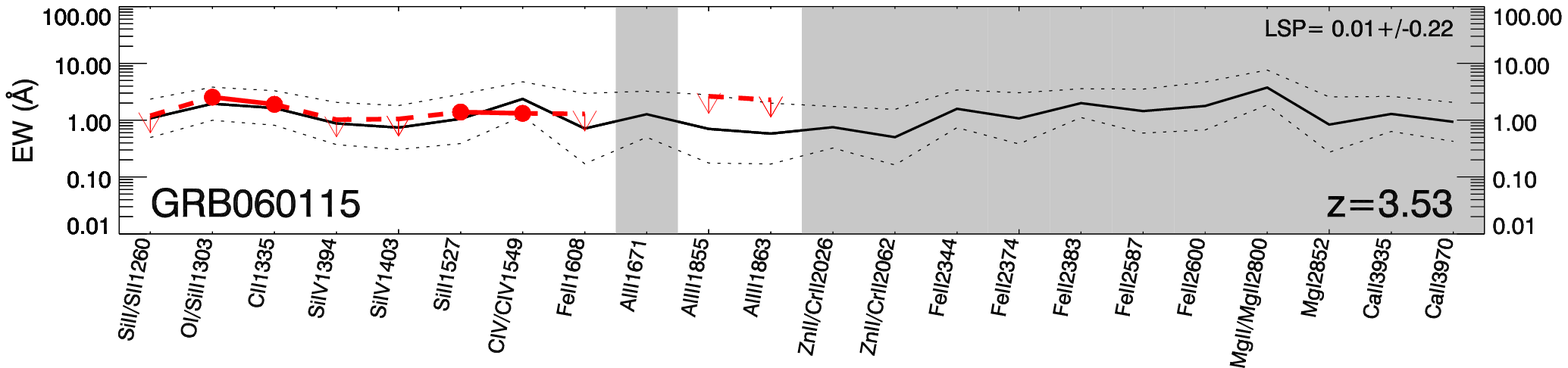}
\includegraphics[width=15cm,clip]{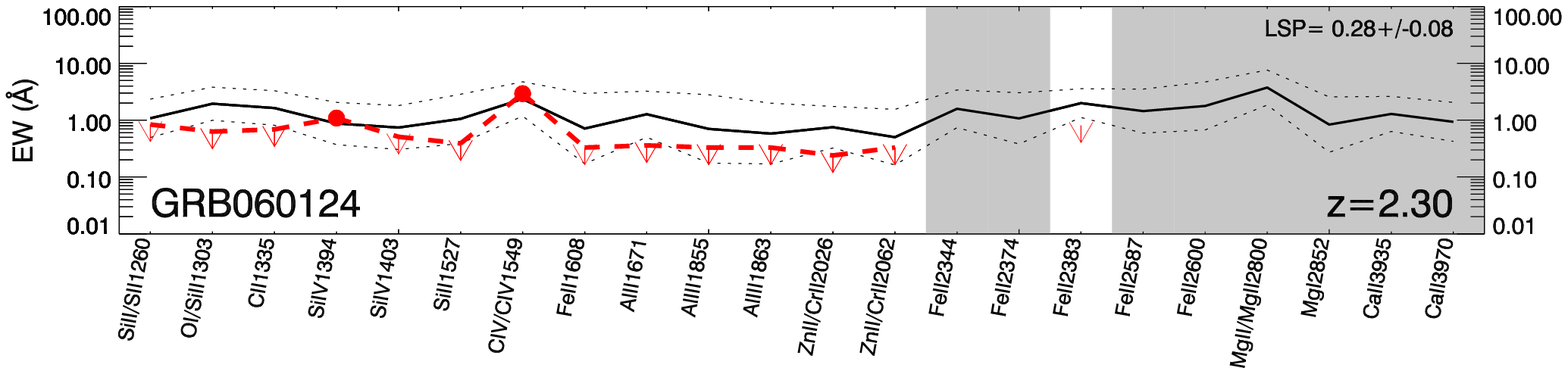}
\includegraphics[width=15cm,clip]{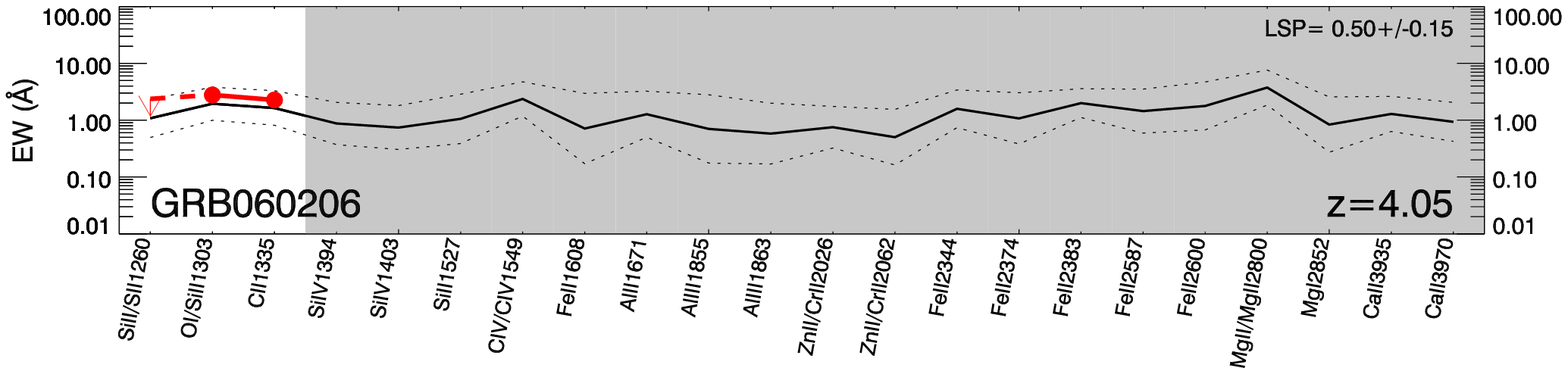}
\includegraphics[width=15cm,clip]{060210.eps}
\includegraphics[width=15cm,clip]{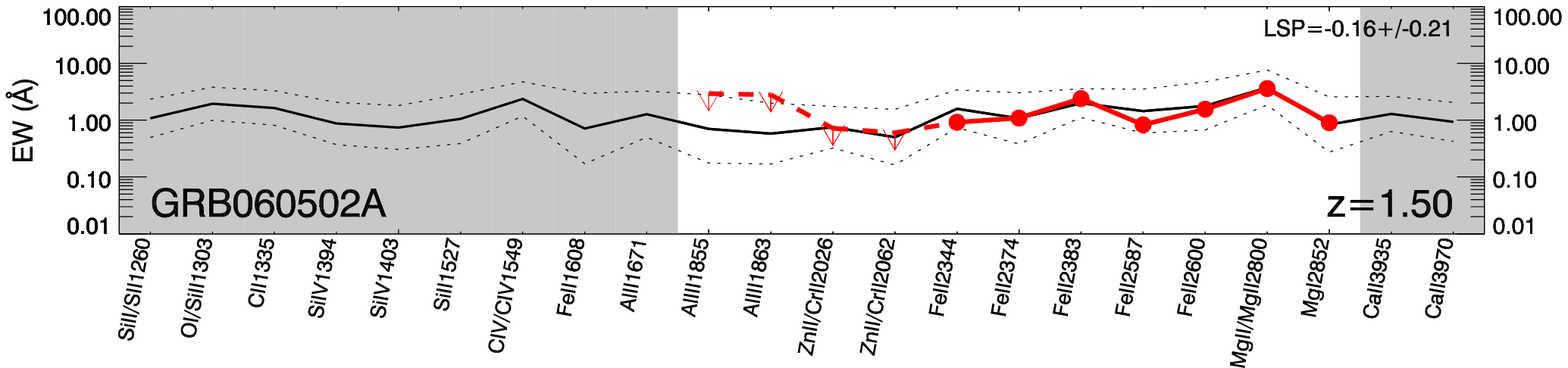}
\includegraphics[width=15cm,clip]{060526.eps}
\end{figure*}

\clearpage
\begin{figure*}[]
\centering
\includegraphics[width=15cm,clip]{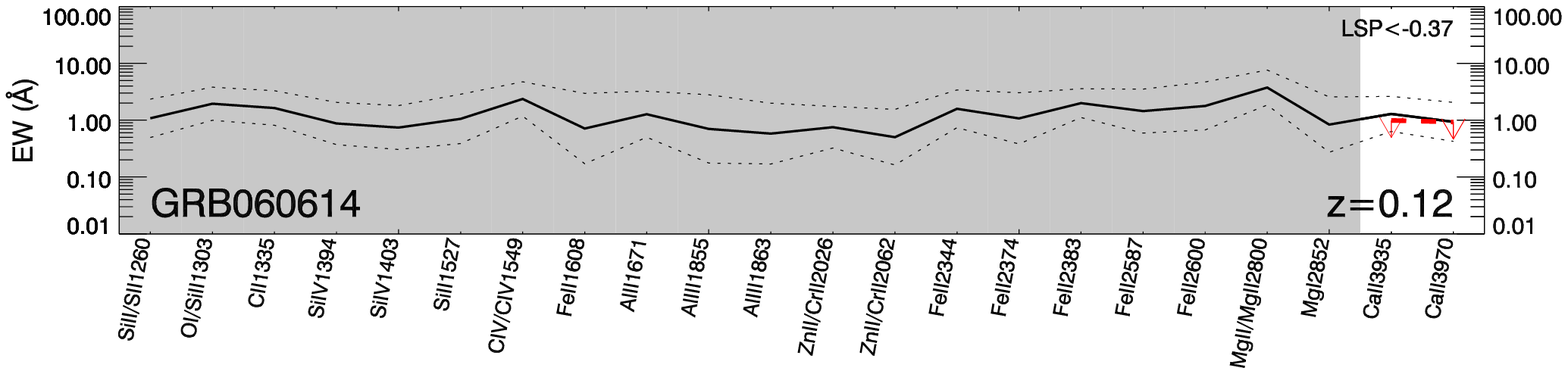}
\includegraphics[width=15cm,clip]{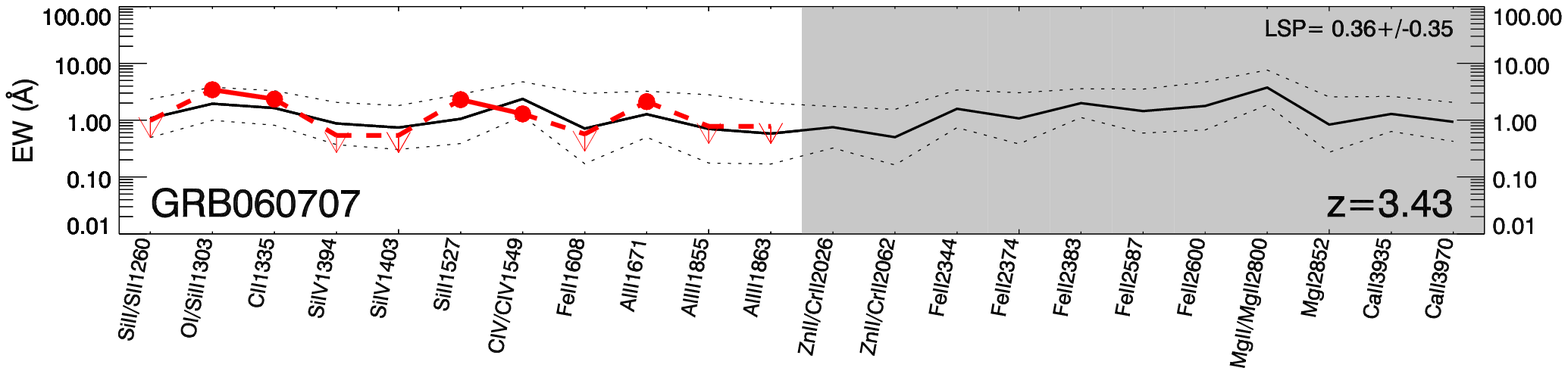}
\includegraphics[width=15cm,clip]{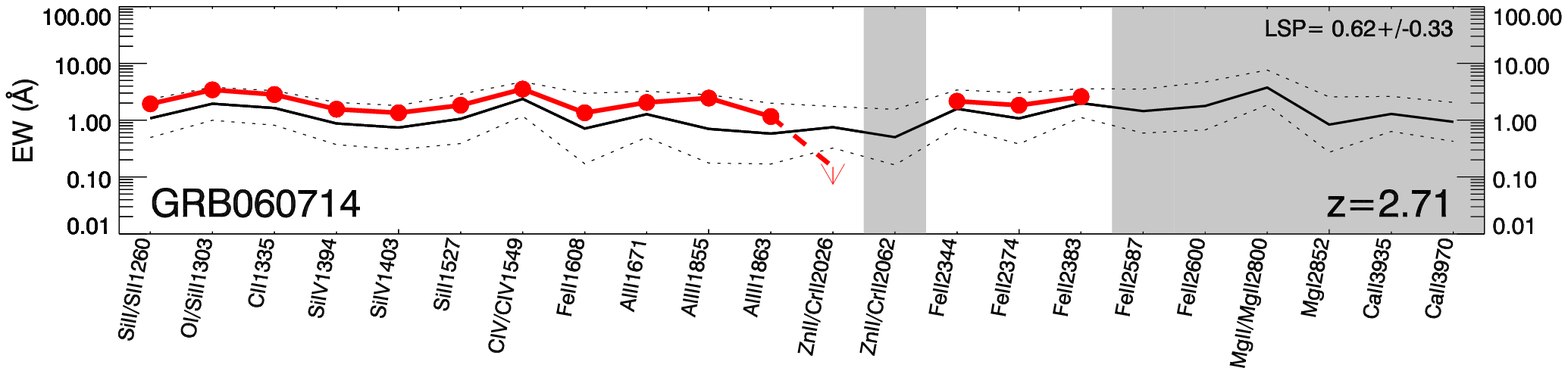}
\includegraphics[width=15cm,clip]{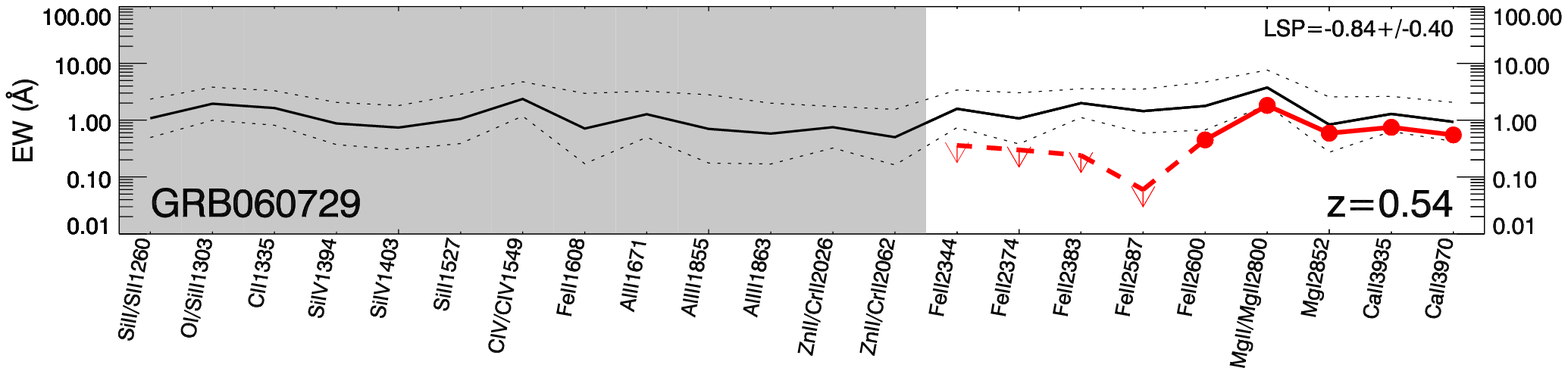}
\includegraphics[width=15cm,clip]{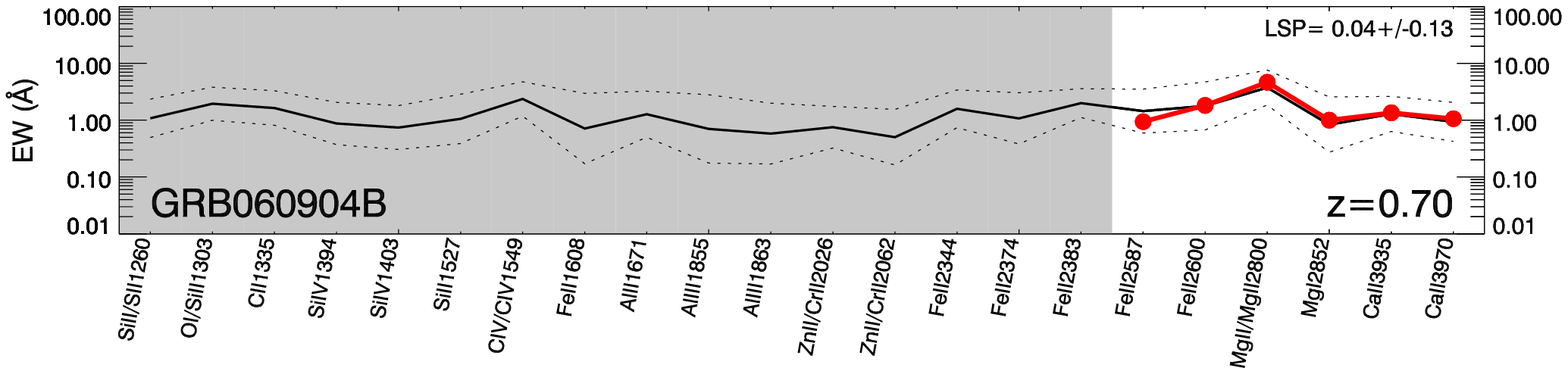}
\includegraphics[width=15cm,clip]{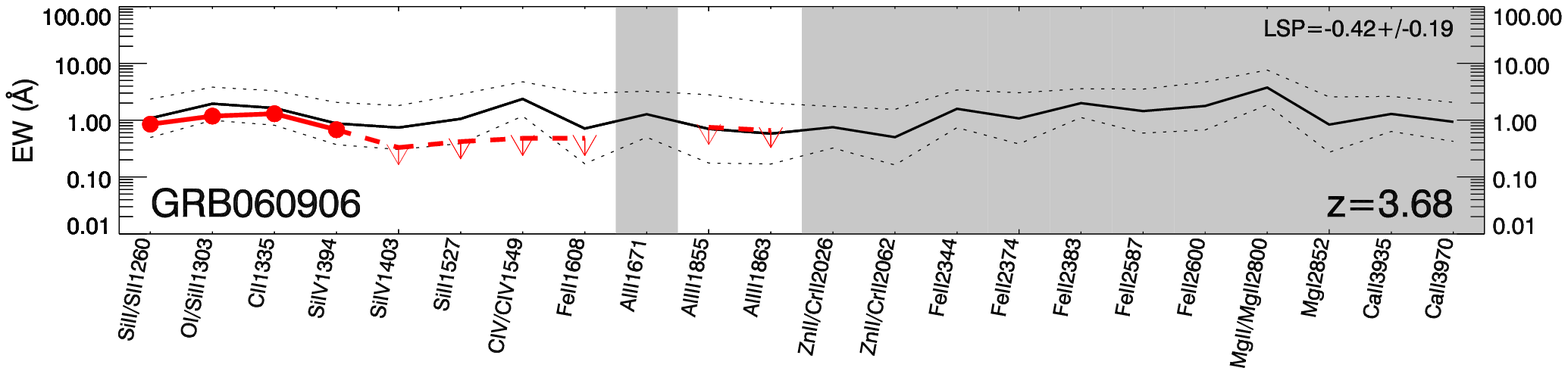}
\includegraphics[width=15cm,clip]{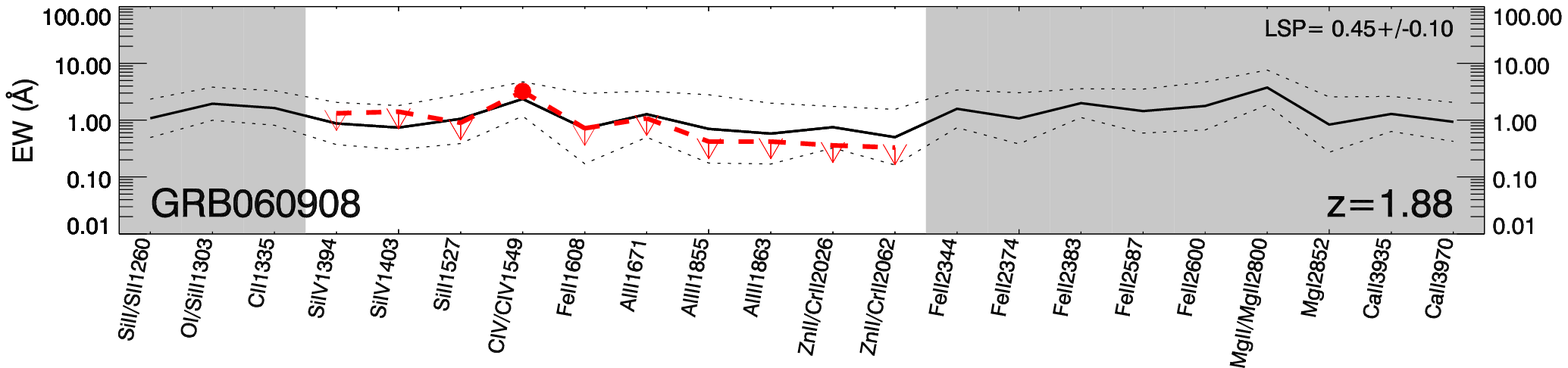}
\end{figure*}

\clearpage
\begin{figure*}[]
\centering
\includegraphics[width=15cm,clip]{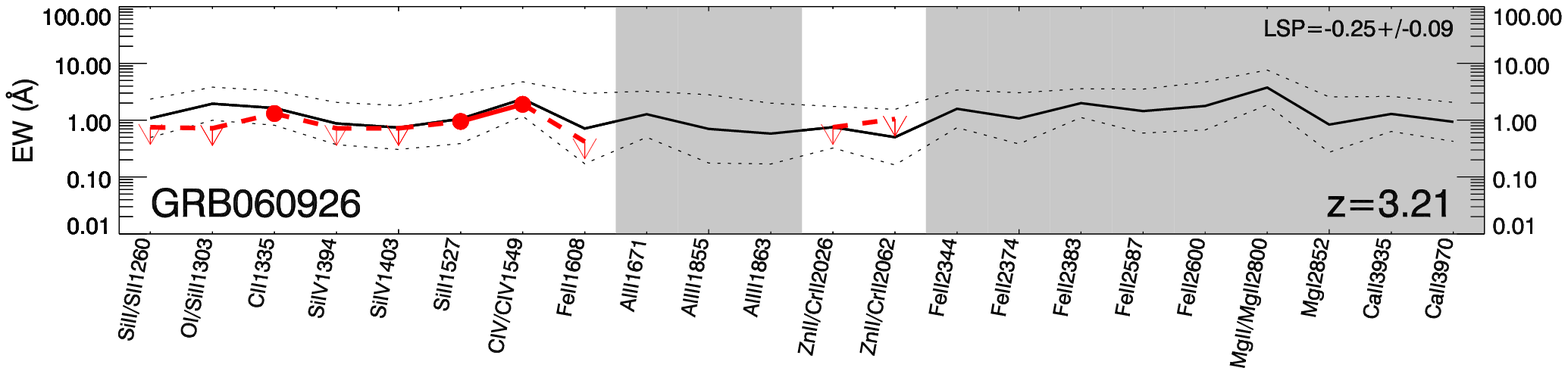}
\includegraphics[width=15cm,clip]{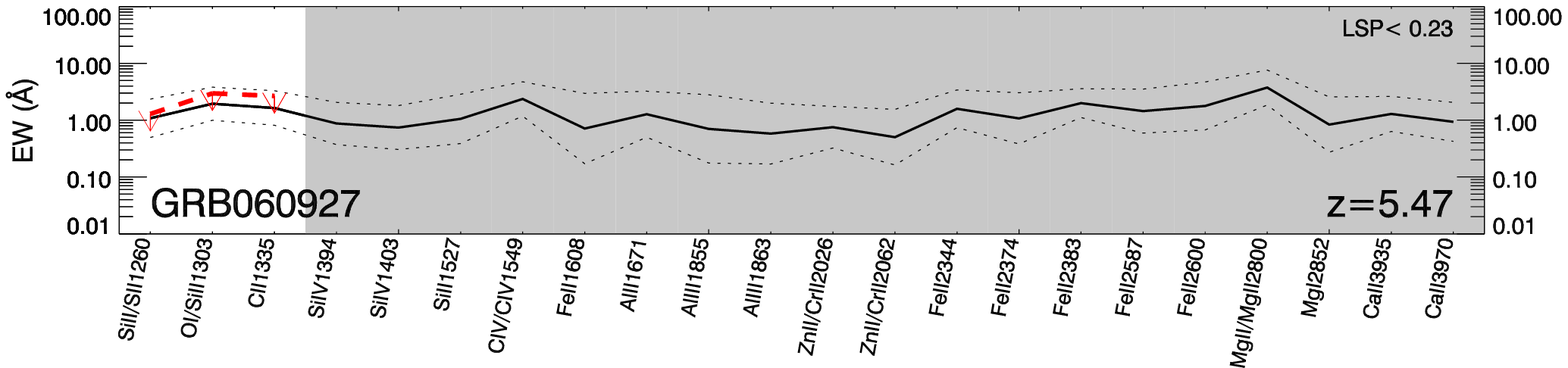}
\includegraphics[width=15cm,clip]{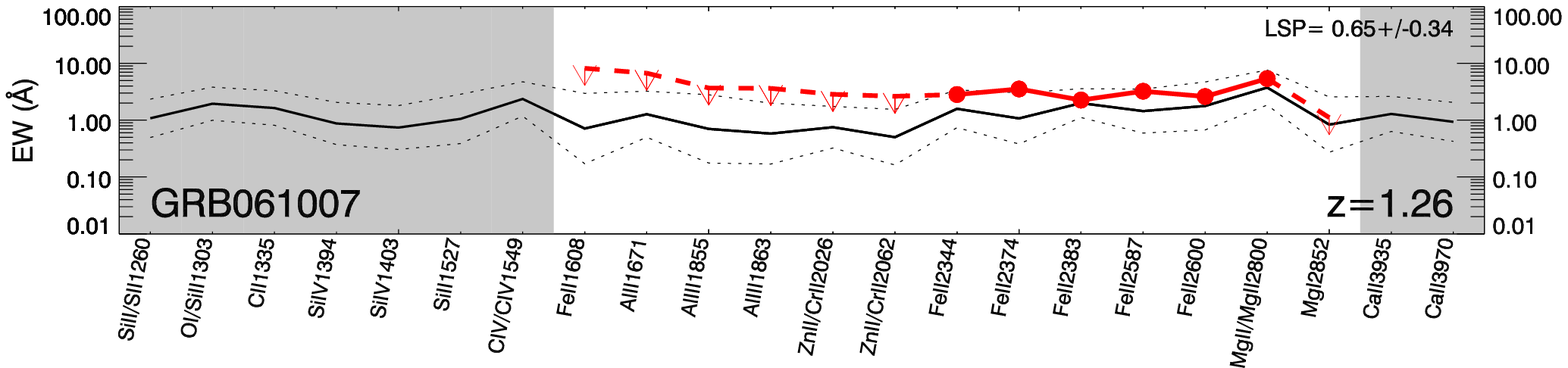}
\includegraphics[width=15cm,clip]{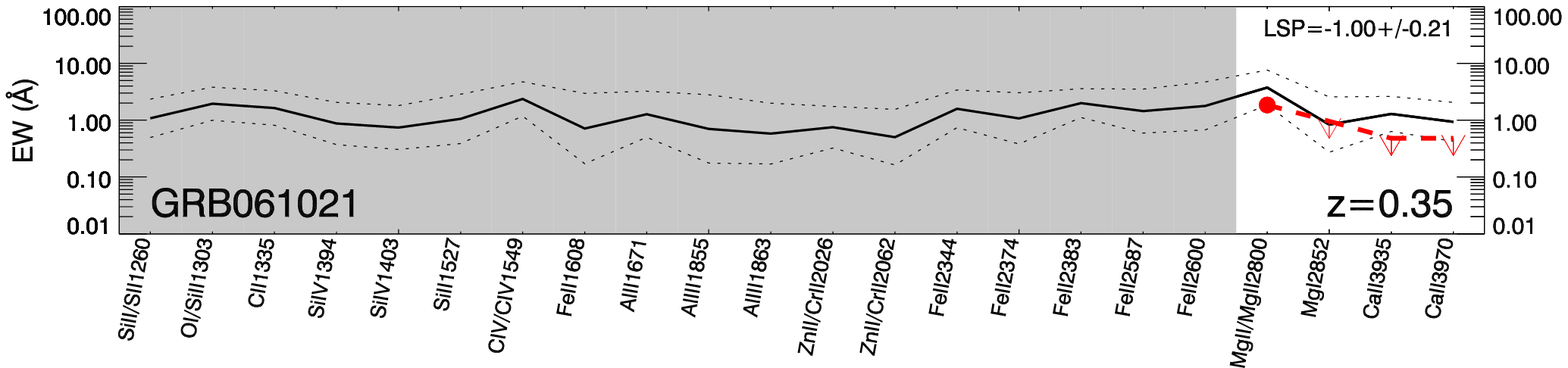}
\includegraphics[width=15cm,clip]{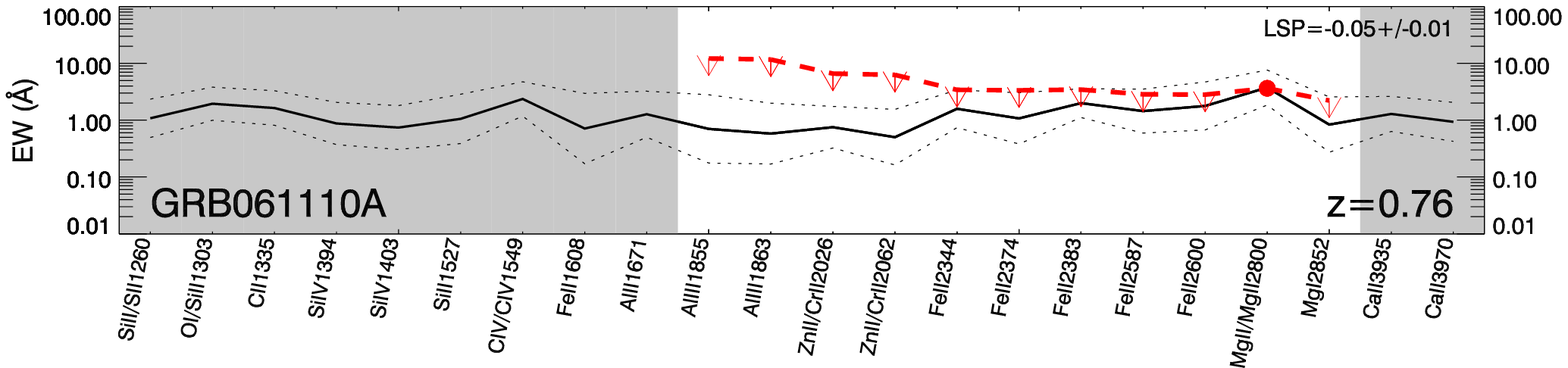}
\includegraphics[width=15cm,clip]{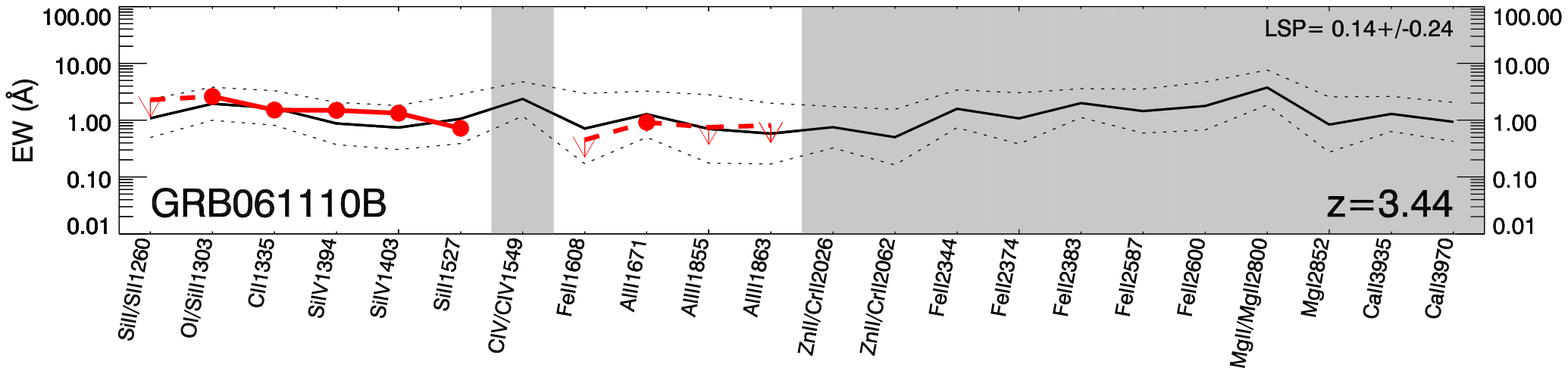}
\includegraphics[width=15cm,clip]{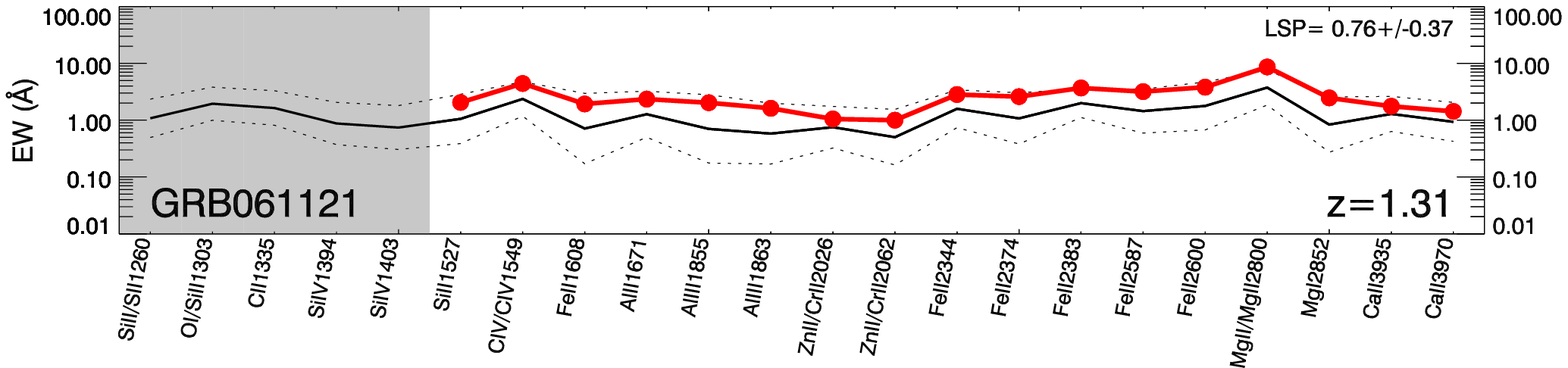}
\end{figure*}

\clearpage
\begin{figure*}[]
\centering
\includegraphics[width=15cm,clip]{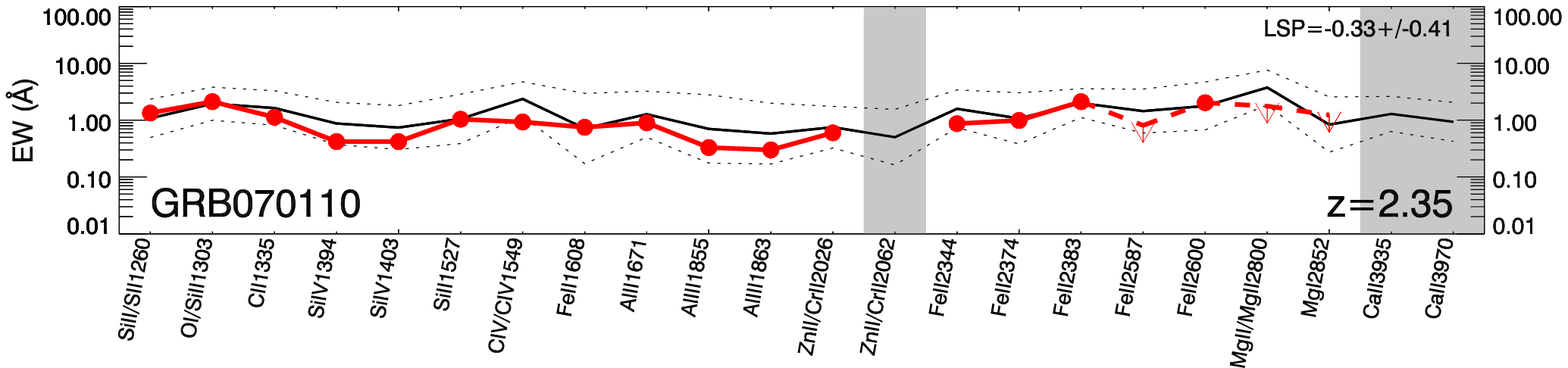}
\includegraphics[width=15cm,clip]{070125.eps}
\includegraphics[width=15cm,clip]{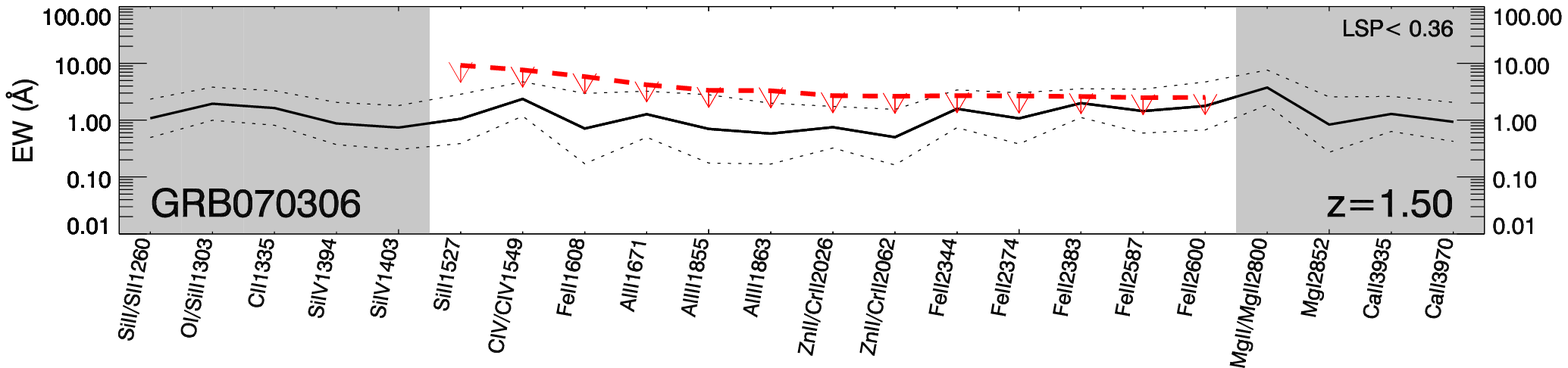}
\includegraphics[width=15cm,clip]{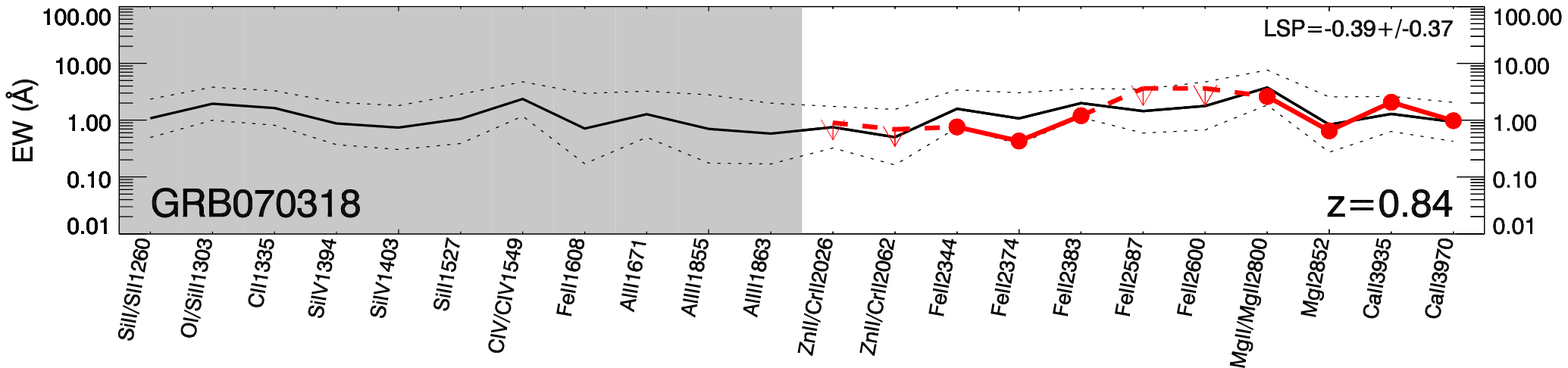}
\includegraphics[width=15cm,clip]{070411.eps}
\includegraphics[width=15cm,clip]{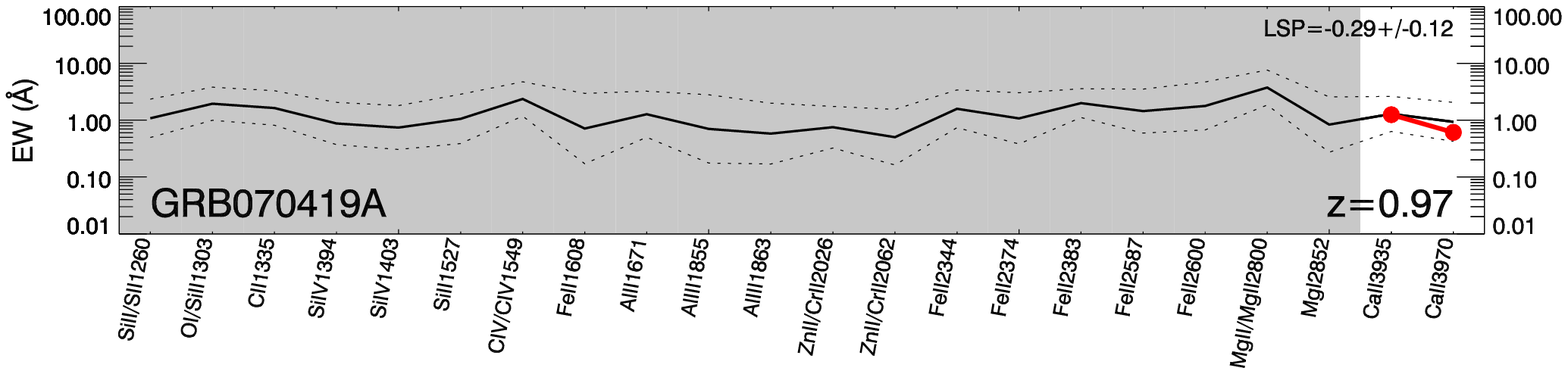}
\includegraphics[width=15cm,clip]{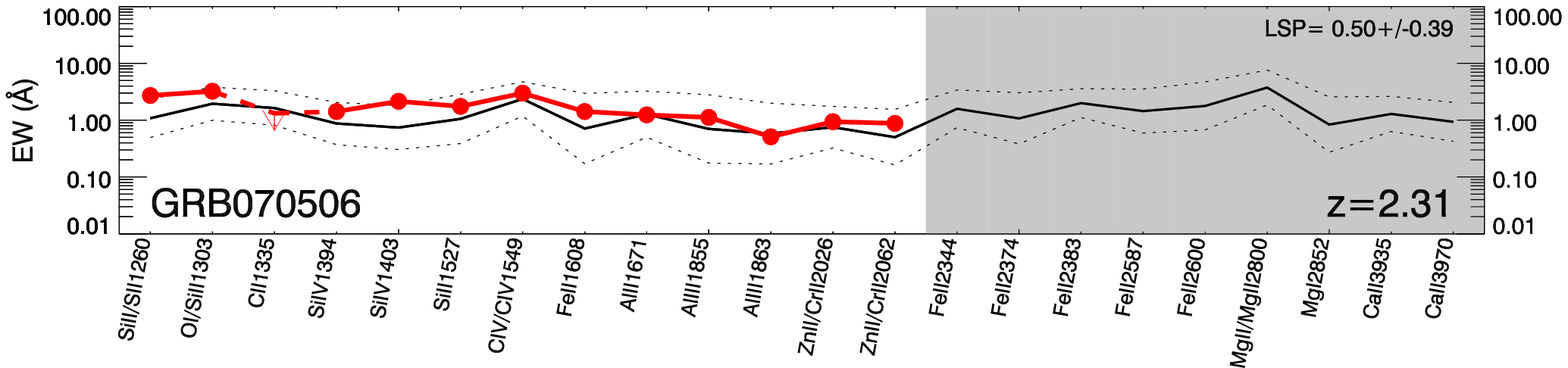}
\end{figure*}

\clearpage
\begin{figure*}[]
\centering
\includegraphics[width=15cm,clip]{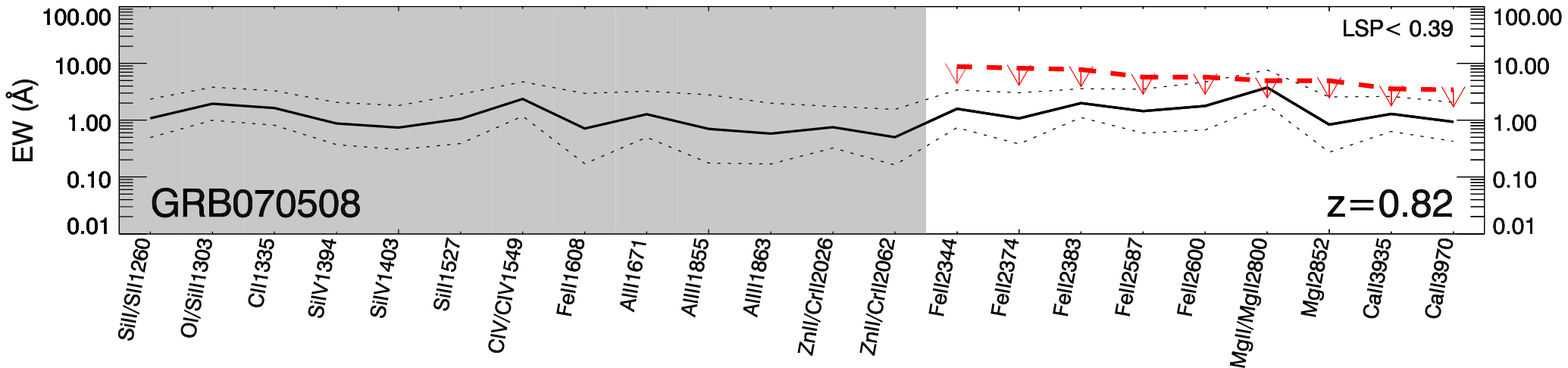}
\includegraphics[width=15cm,clip]{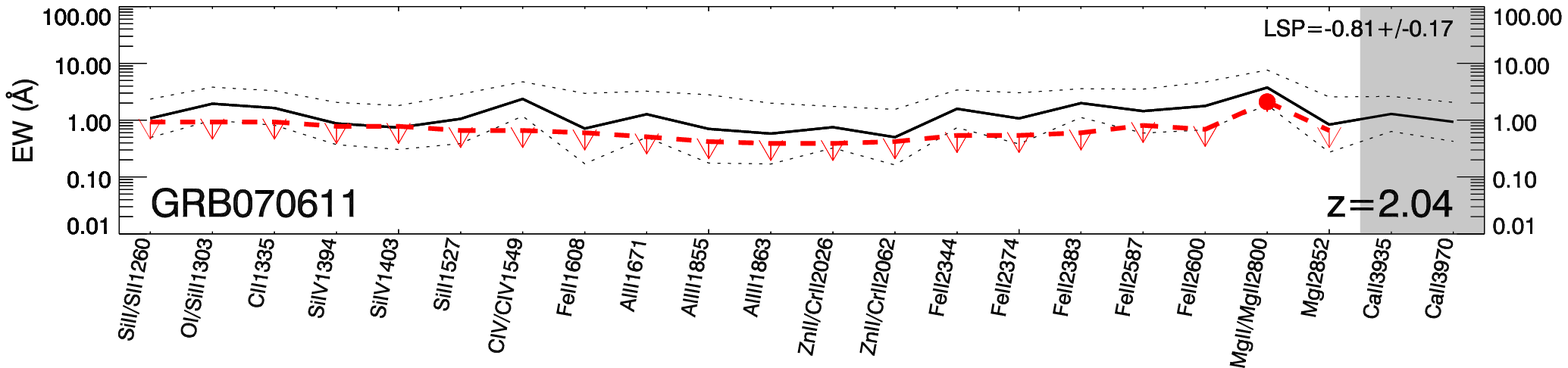}
\includegraphics[width=15cm,clip]{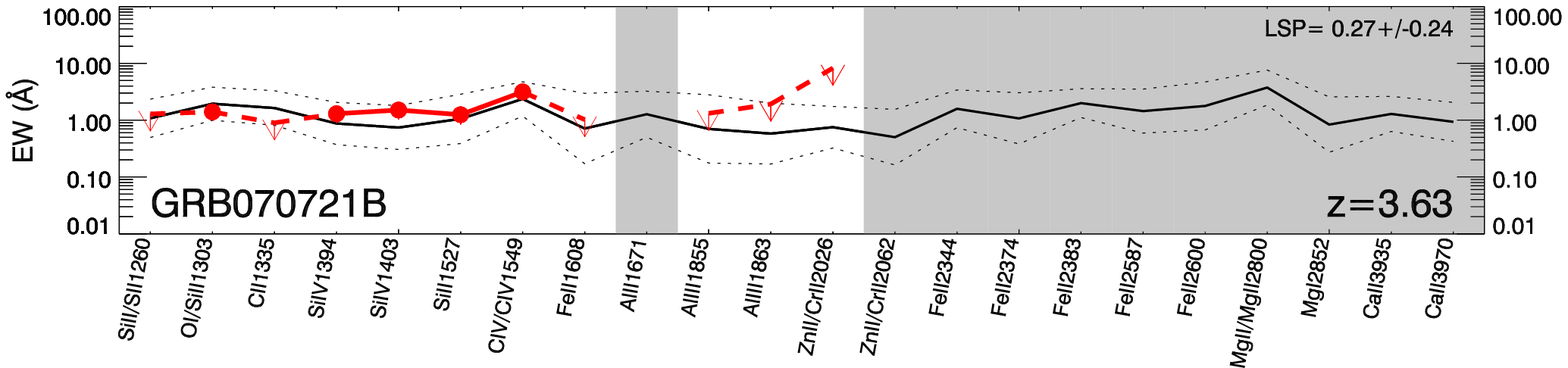}
\includegraphics[width=15cm,clip]{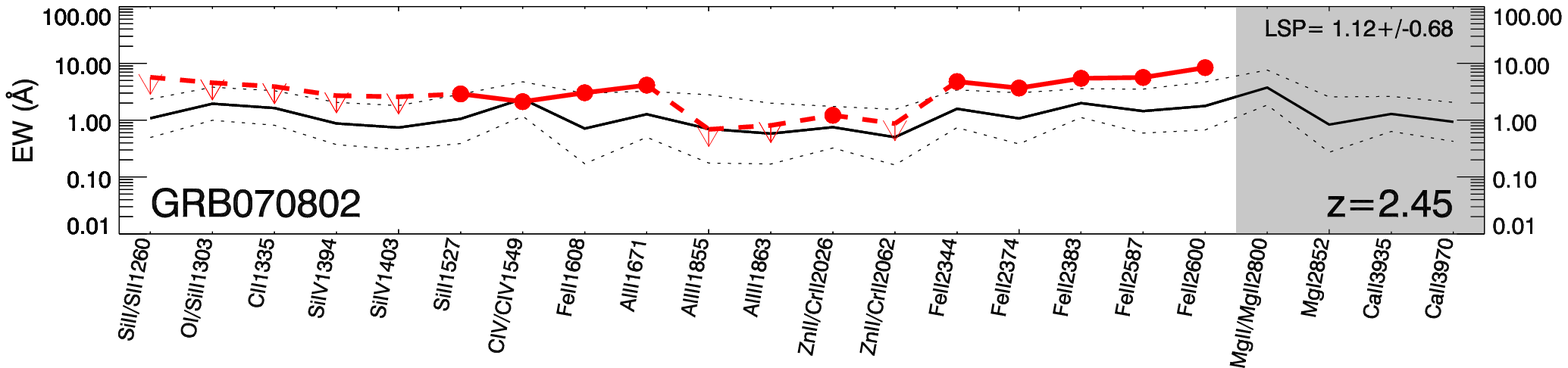}
\includegraphics[width=15cm,clip]{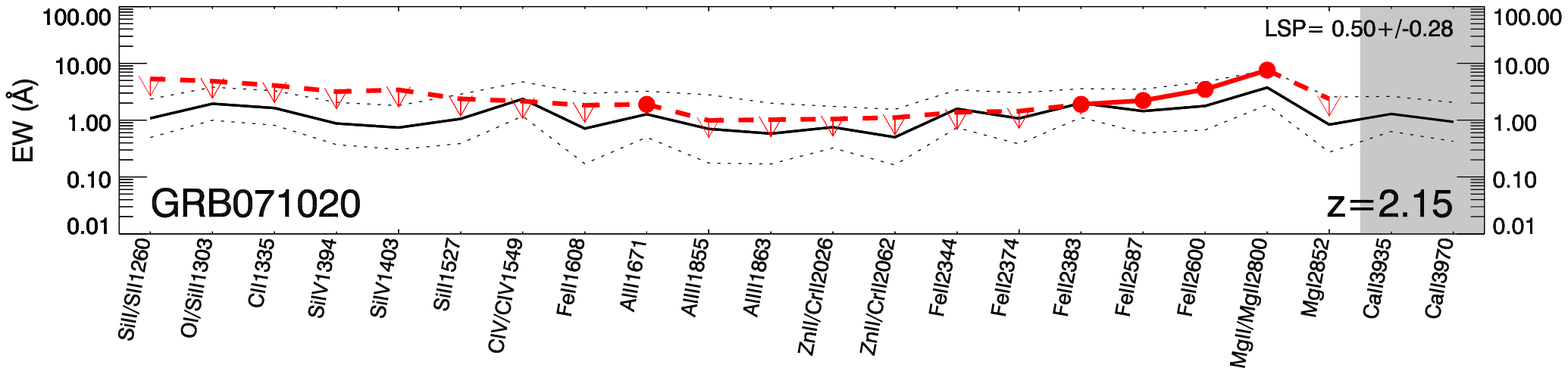}
\includegraphics[width=15cm,clip]{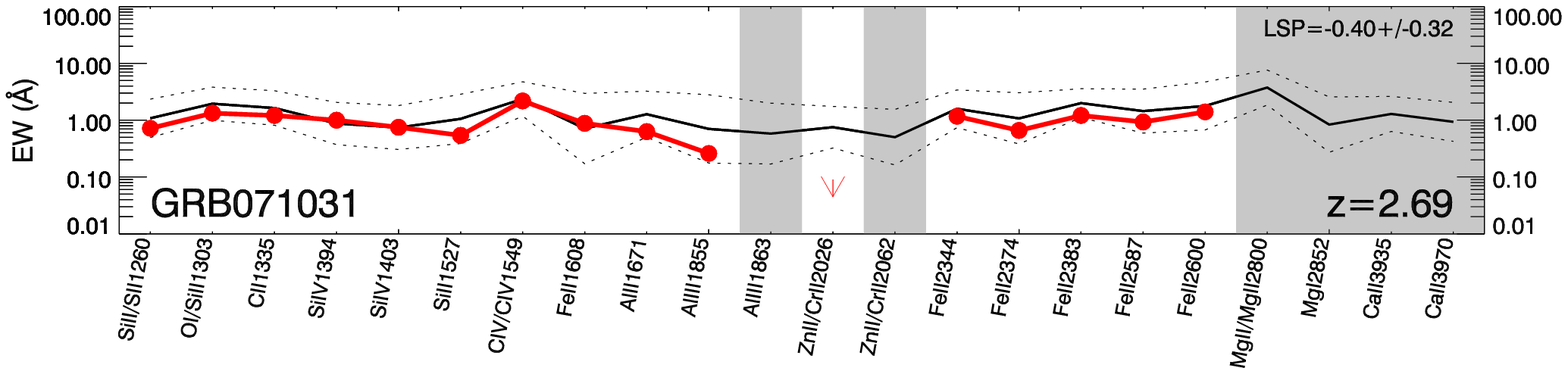}
\includegraphics[width=15cm,clip]{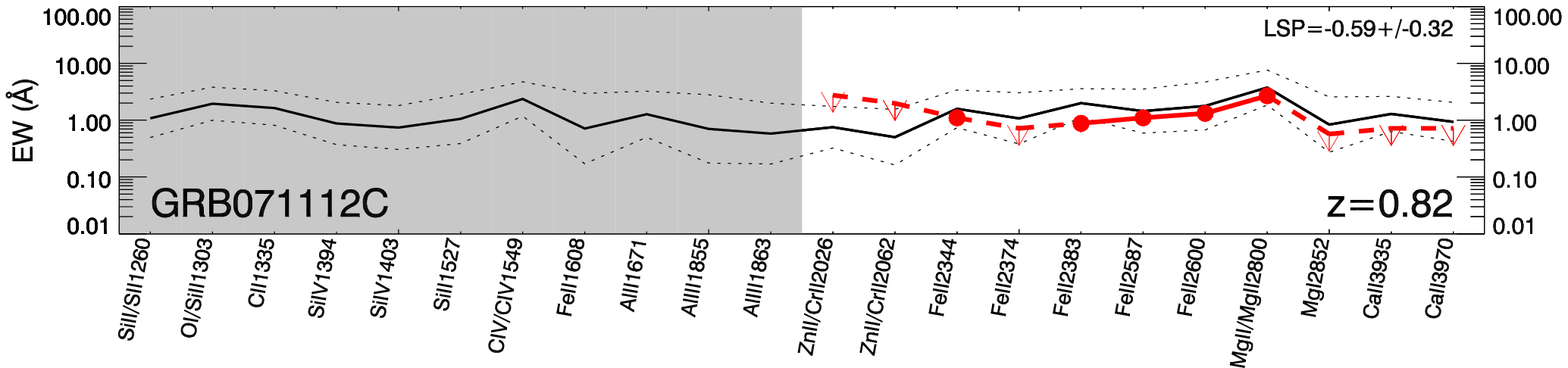}   
\end{figure*}

\clearpage
\begin{figure*}[]
\centering
\includegraphics[width=15cm,clip]{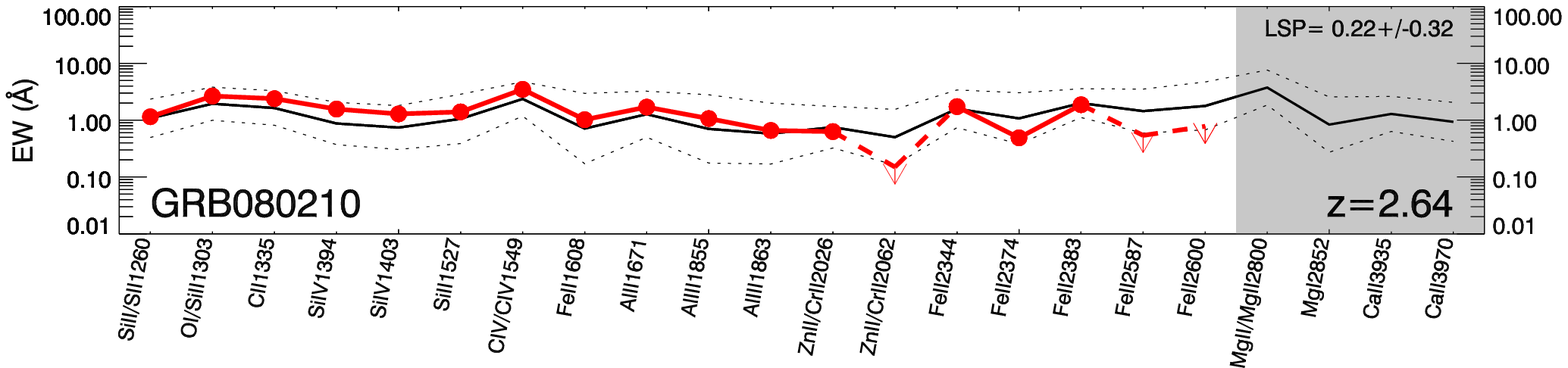}
\includegraphics[width=15cm,clip]{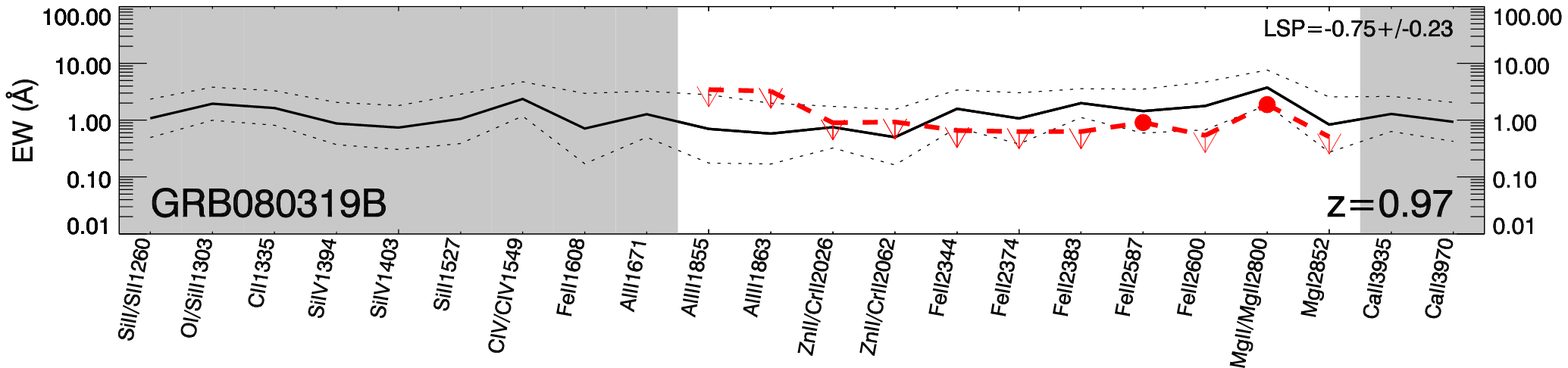}
\includegraphics[width=15cm,clip]{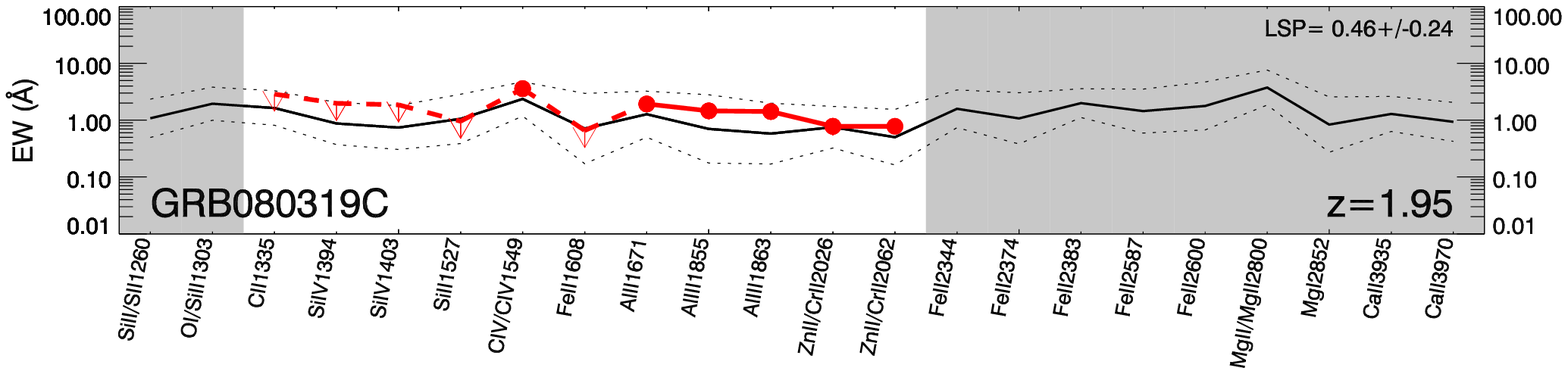}
\includegraphics[width=15cm,clip]{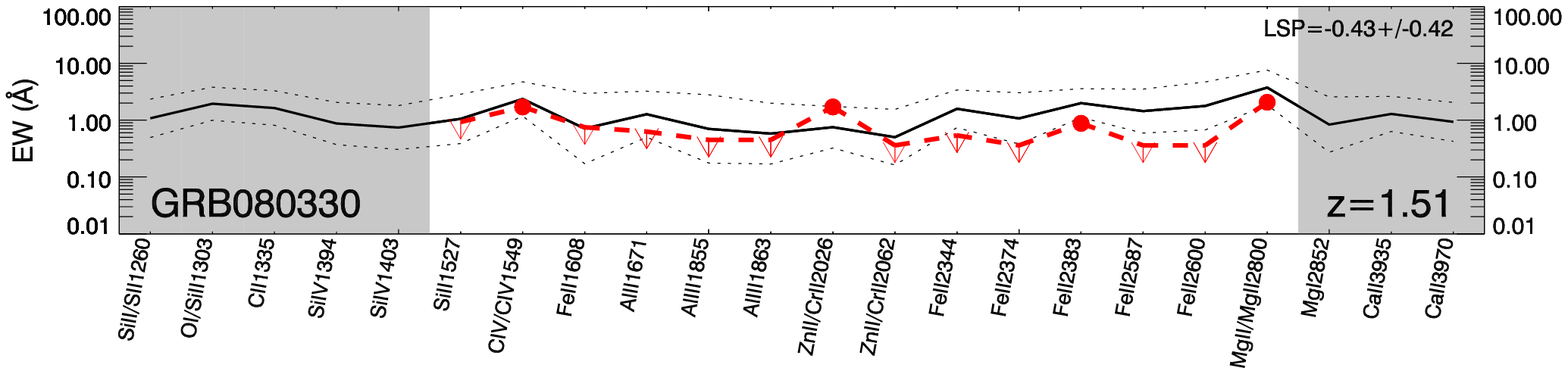}
\includegraphics[width=15cm,clip]{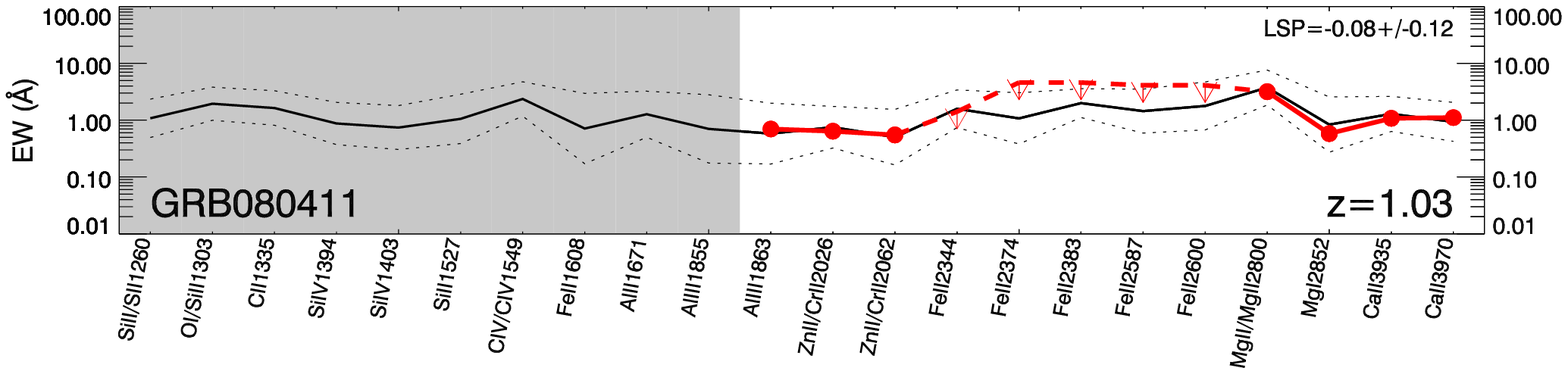}
\includegraphics[width=15cm,clip]{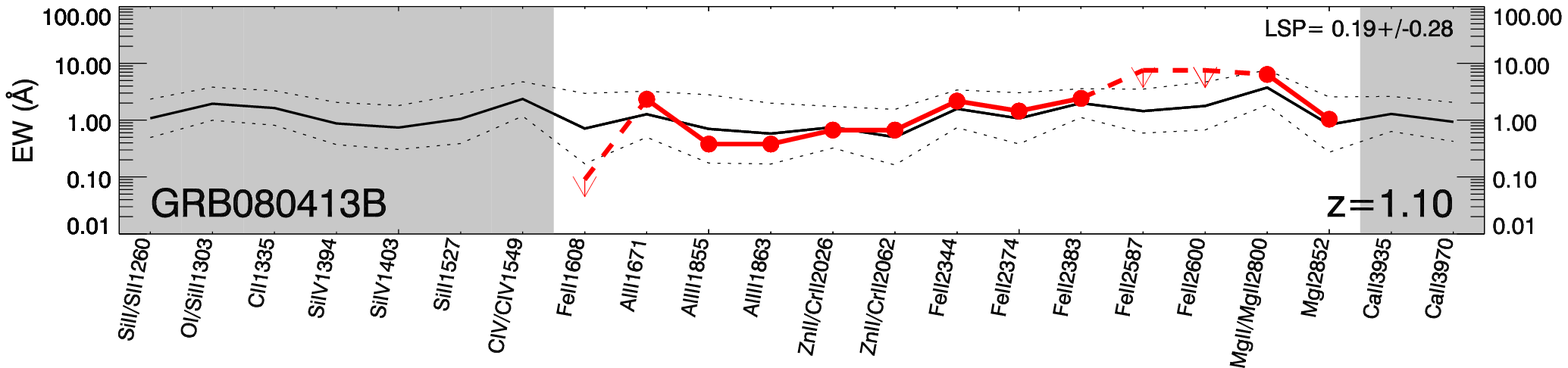}
\includegraphics[width=15cm,clip]{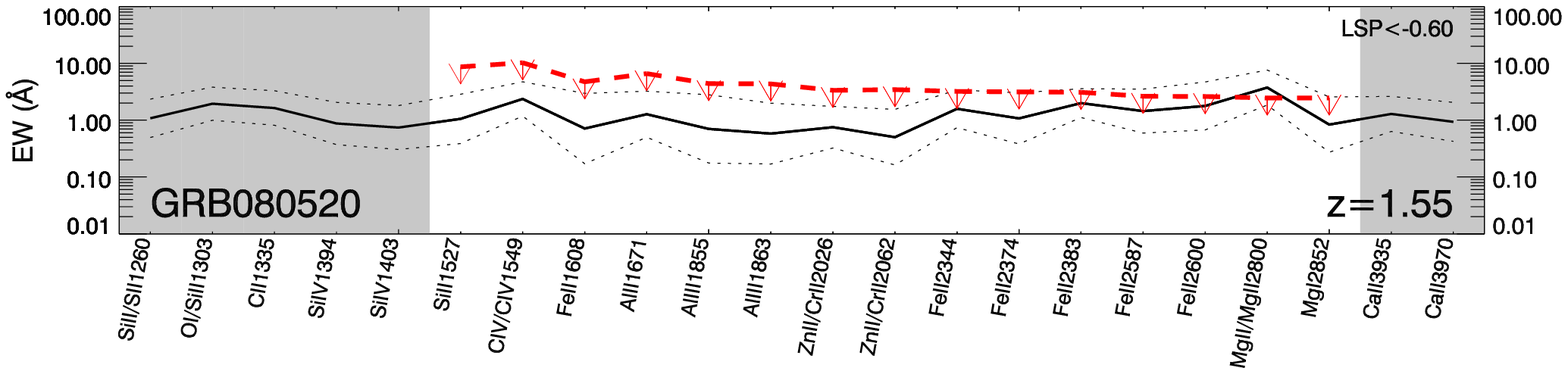}
\end{figure*}

\clearpage
\begin{figure*}[]
\centering
\includegraphics[width=15cm,clip]{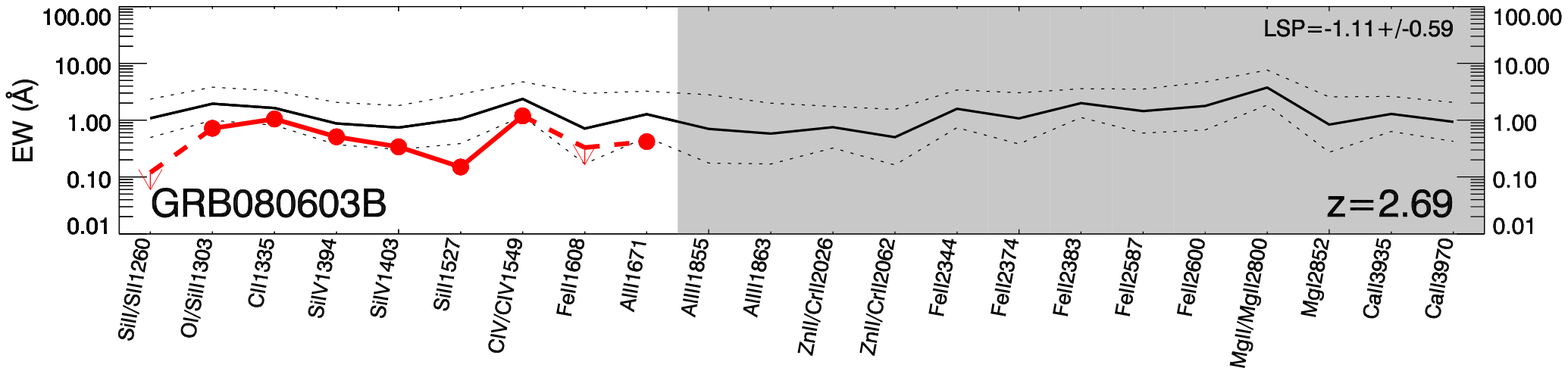}
\includegraphics[width=15cm,clip]{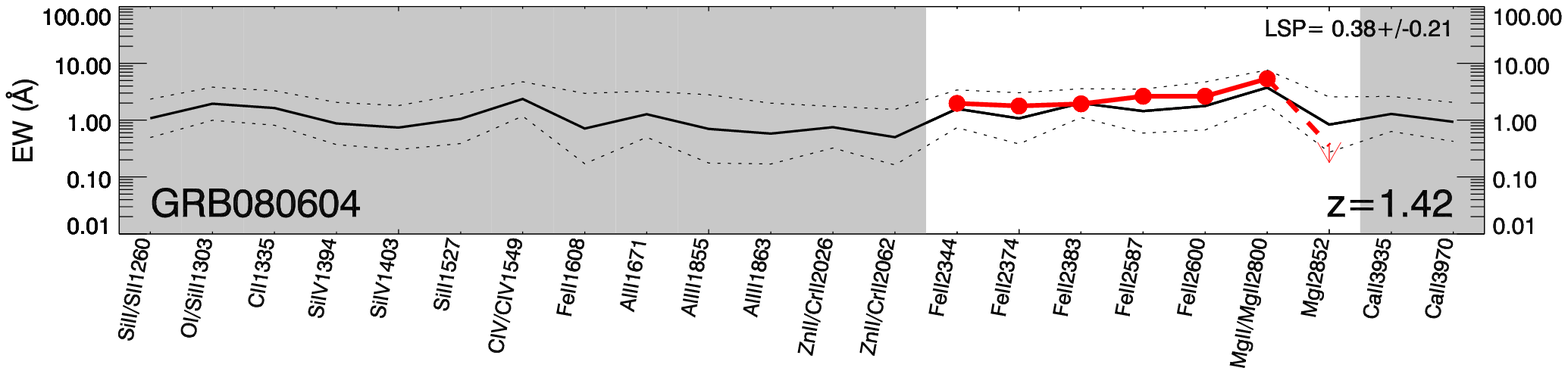}
\includegraphics[width=15cm,clip]{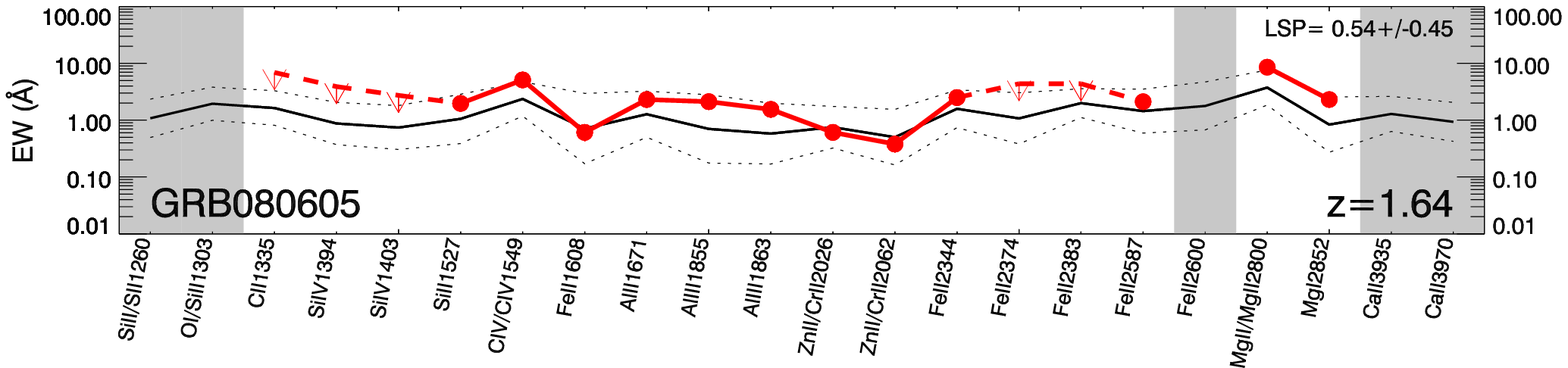}
\includegraphics[width=15cm,clip]{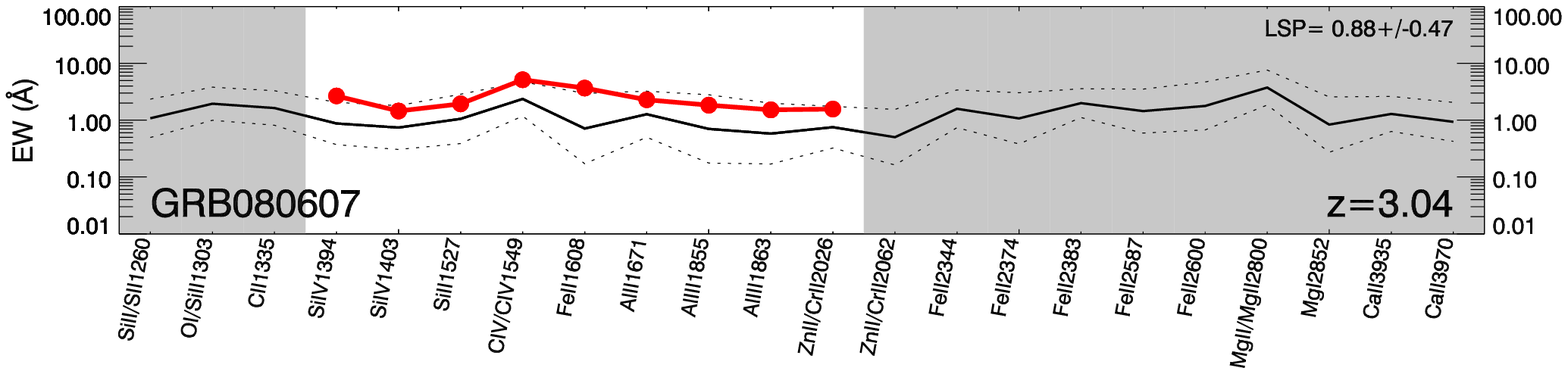}
\includegraphics[width=15cm,clip]{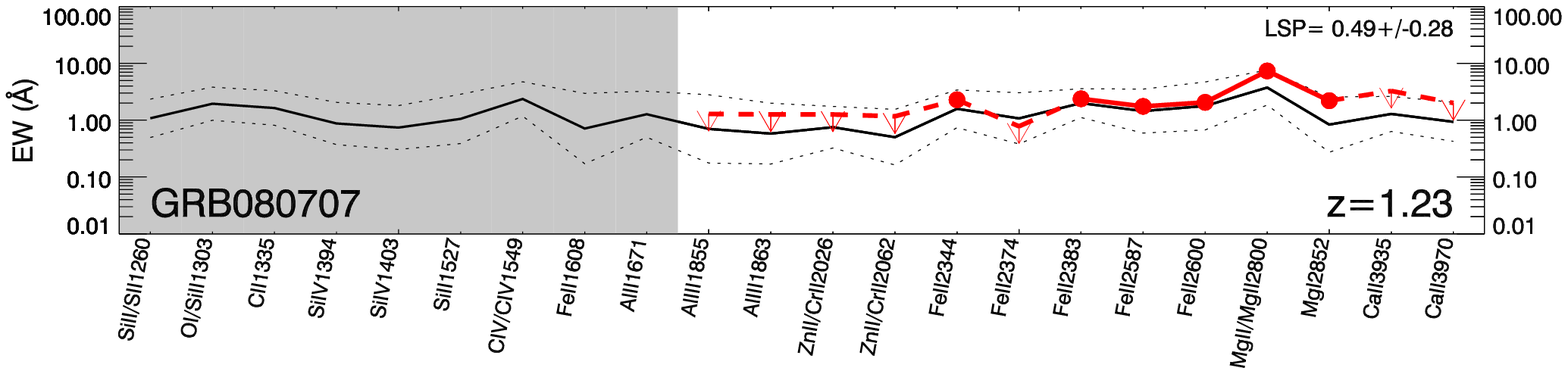}
\includegraphics[width=15cm,clip]{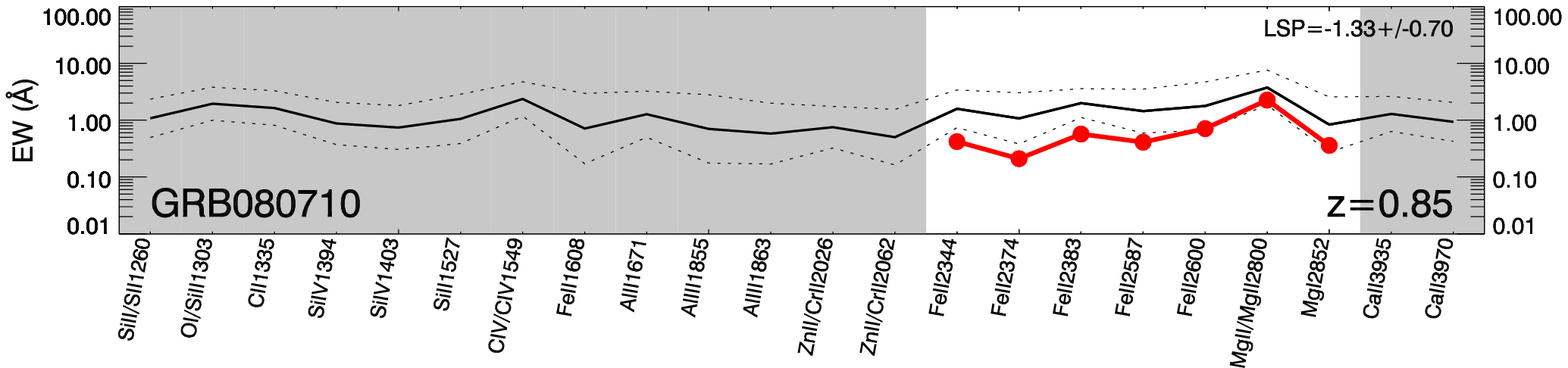}
\includegraphics[width=15cm,clip]{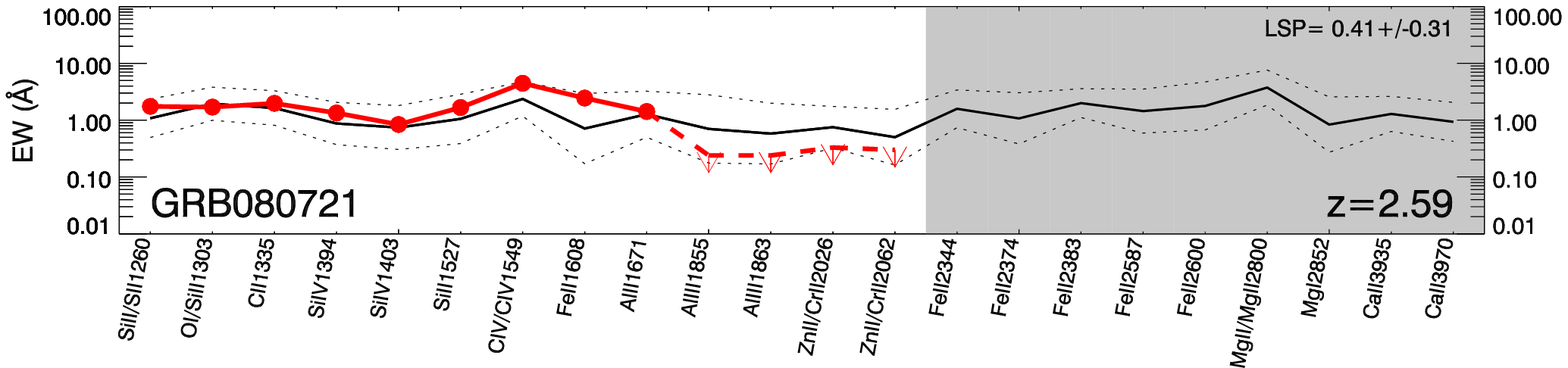}
\end{figure*}

\clearpage
\begin{figure*}[]
\centering
\includegraphics[width=15cm,clip]{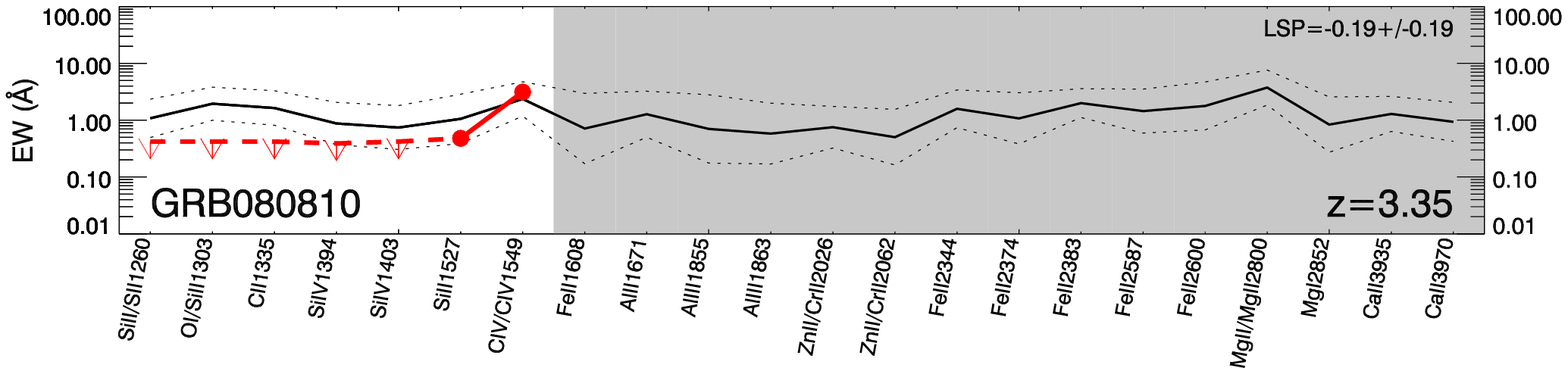}
\includegraphics[width=15cm,clip]{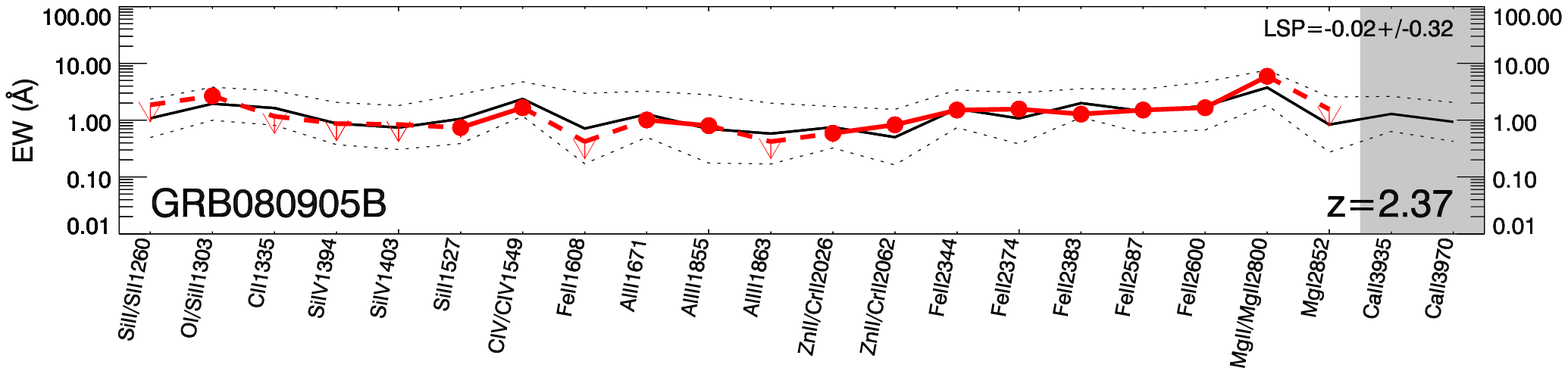}
\includegraphics[width=15cm,clip]{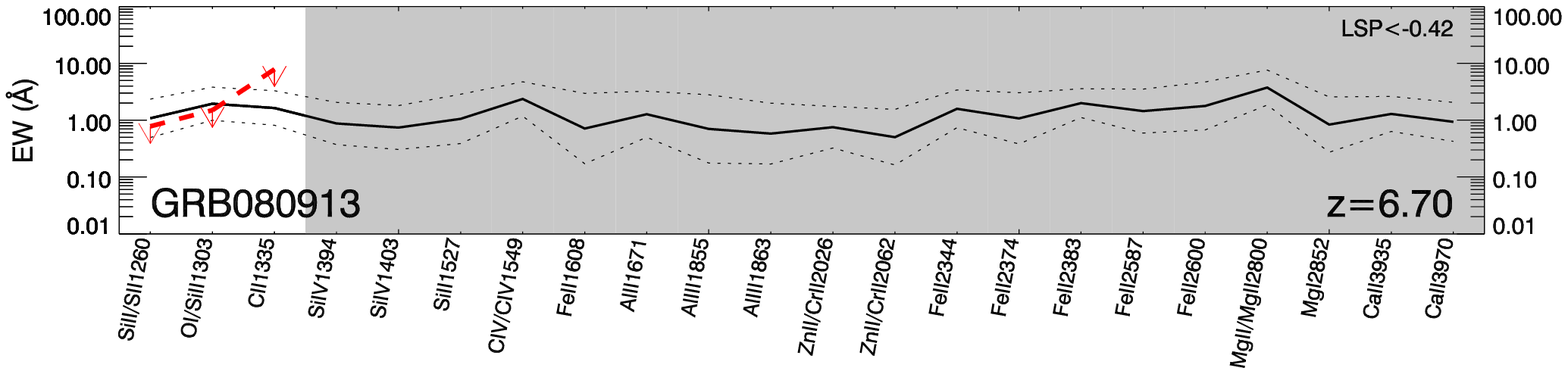}
\includegraphics[width=15cm,clip]{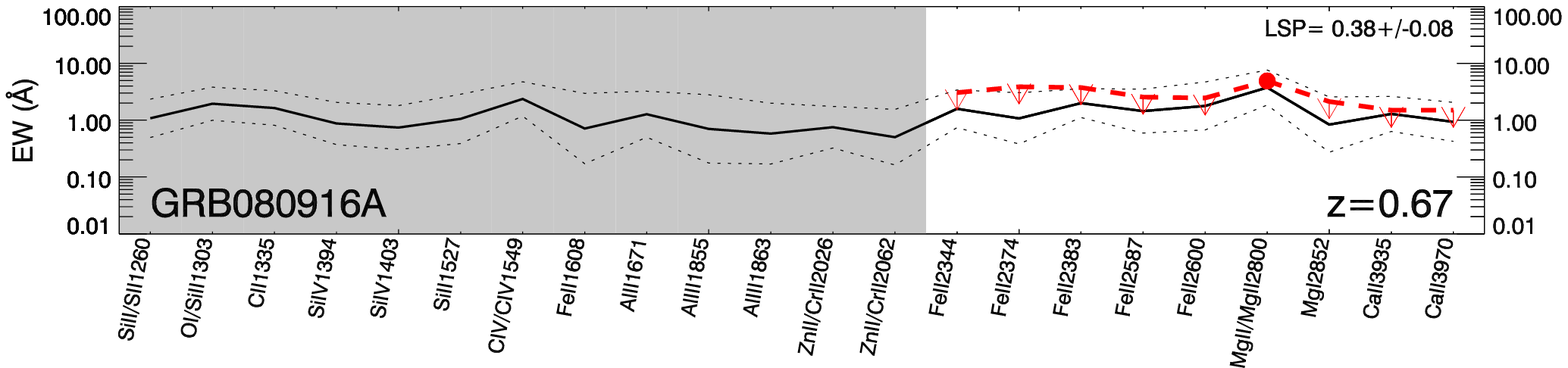}
\includegraphics[width=15cm,clip]{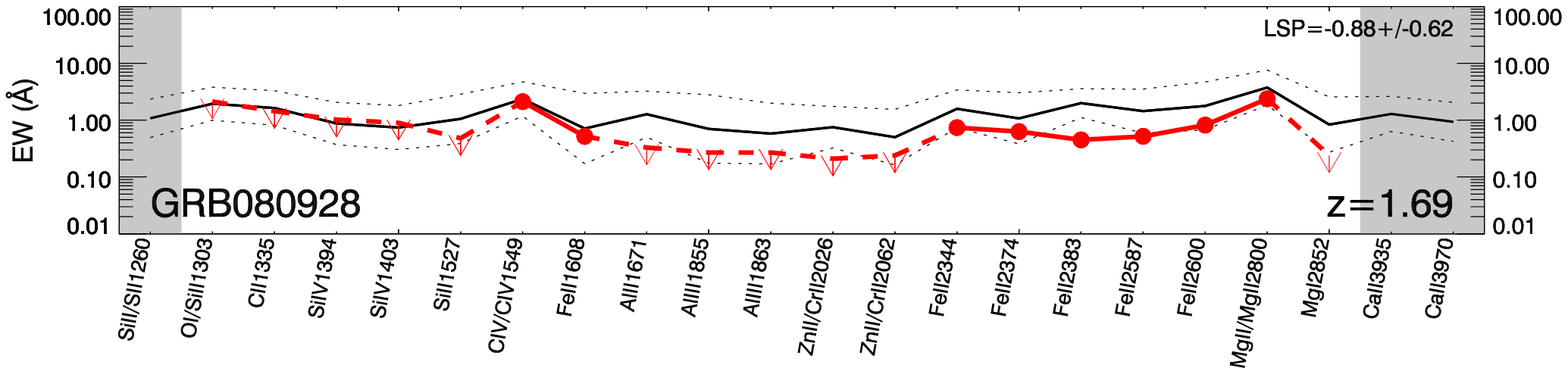}
\includegraphics[width=15cm,clip]{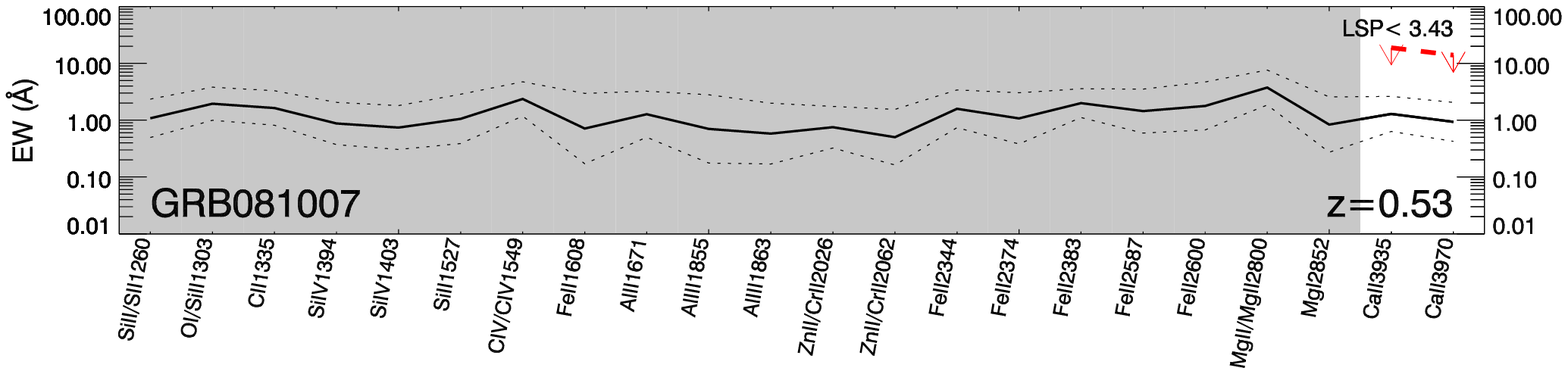}
\includegraphics[width=15cm,clip]{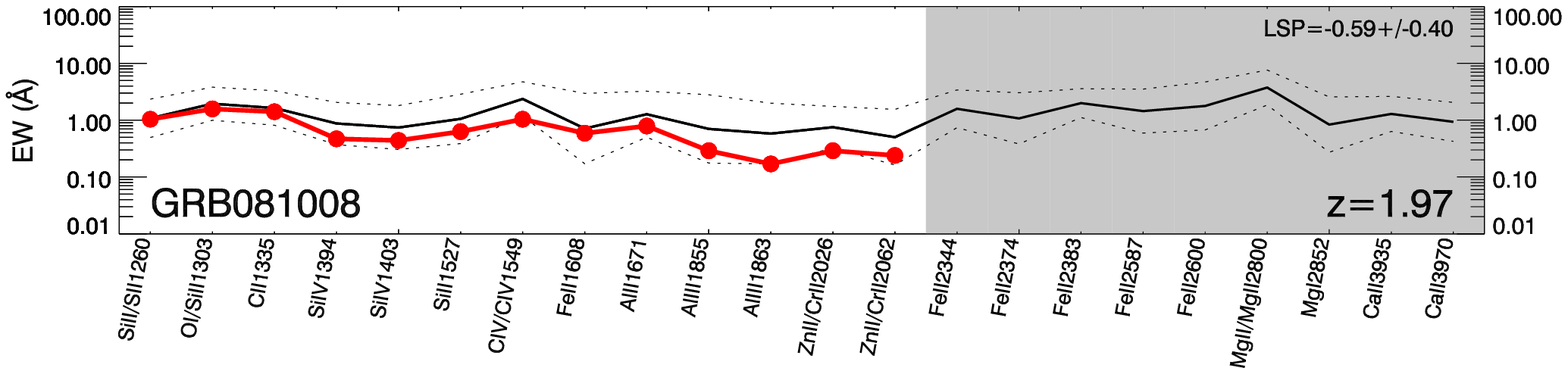}
\end{figure*}

\clearpage
\begin{figure*}[]
\centering
\includegraphics[width=15cm,clip]{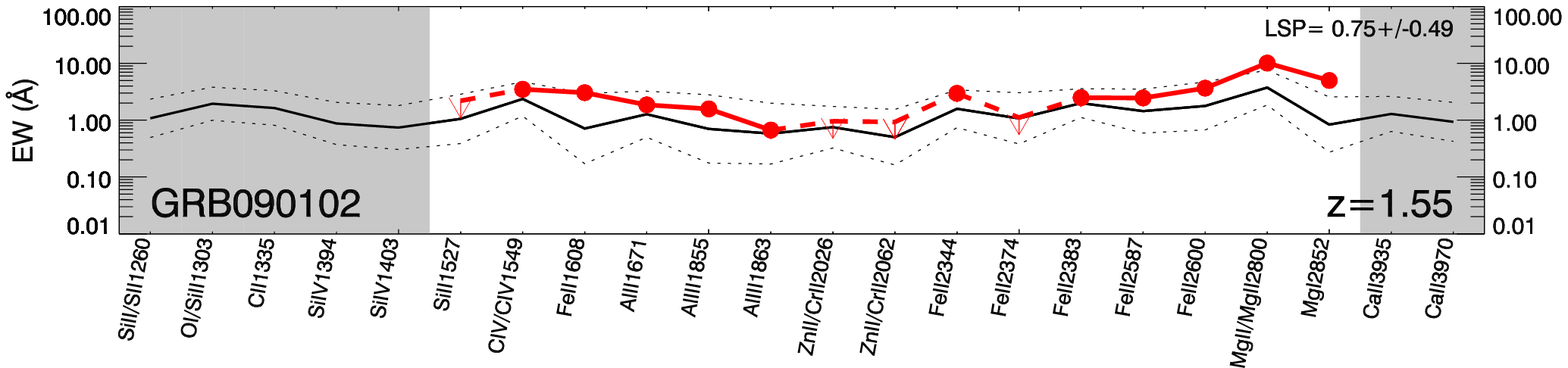}
\includegraphics[width=15cm,clip]{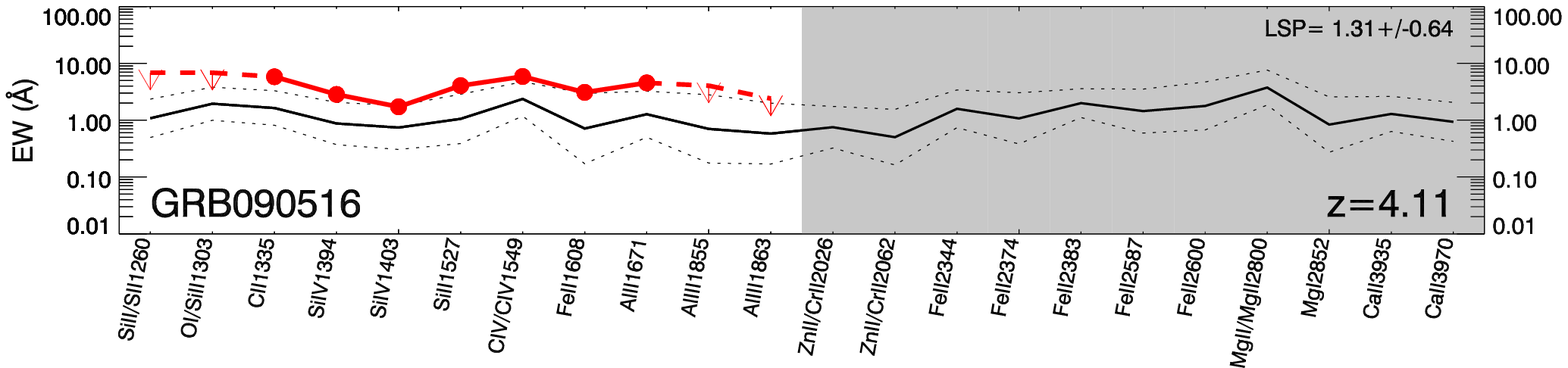}
\includegraphics[width=15cm,clip]{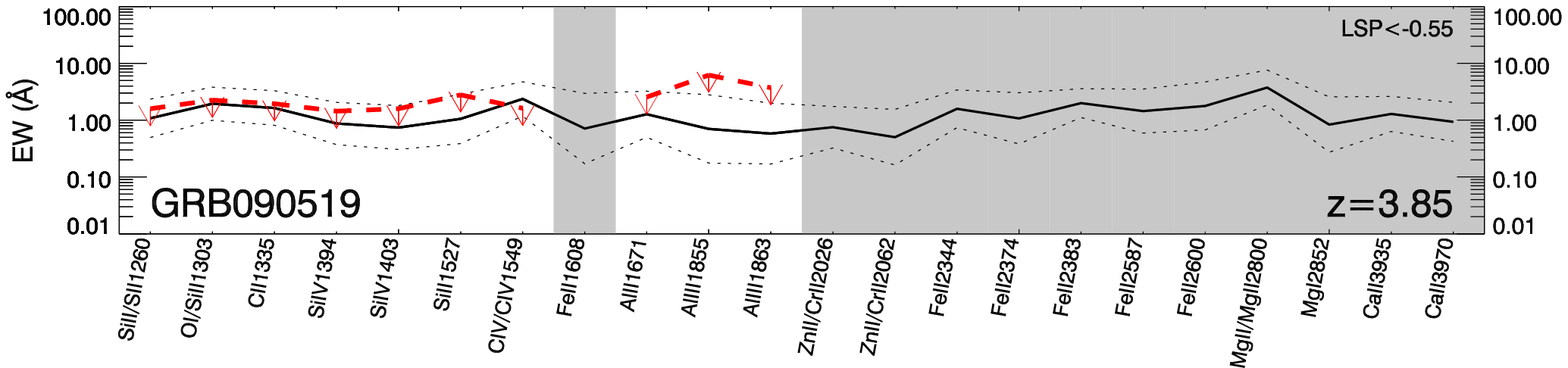}
\includegraphics[width=15cm,clip]{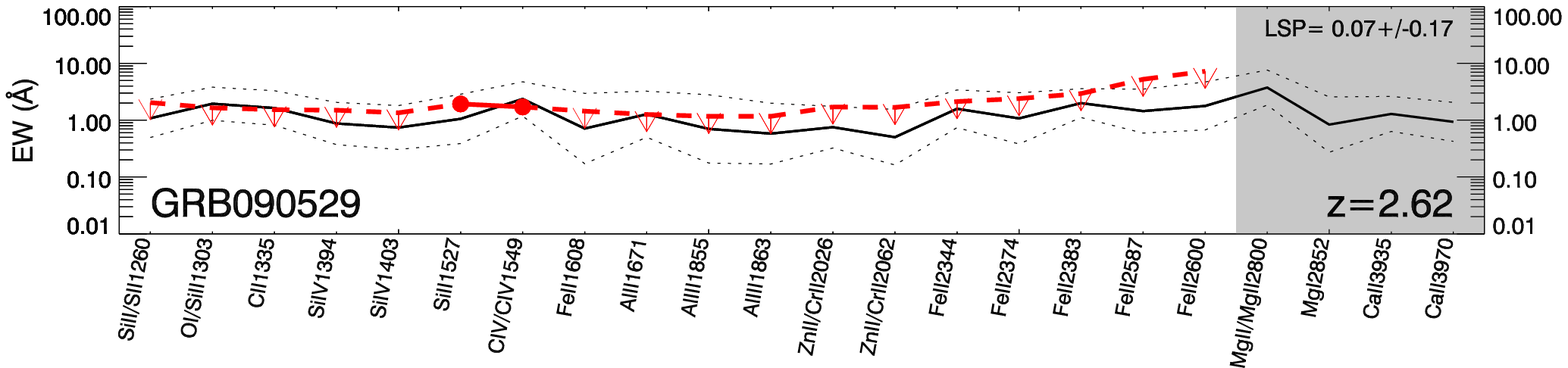}
\includegraphics[width=15cm,clip]{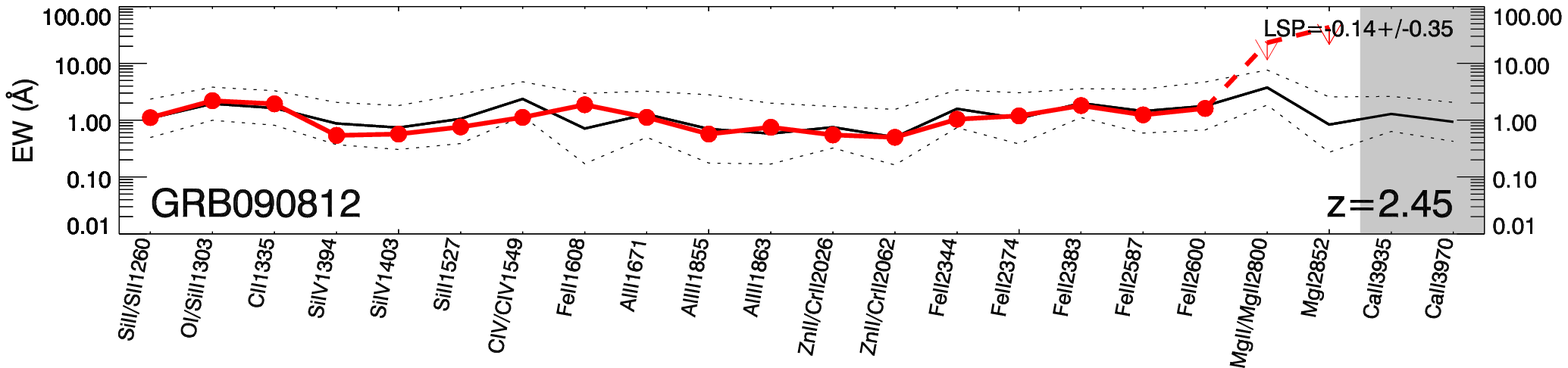}
\includegraphics[width=15cm,clip]{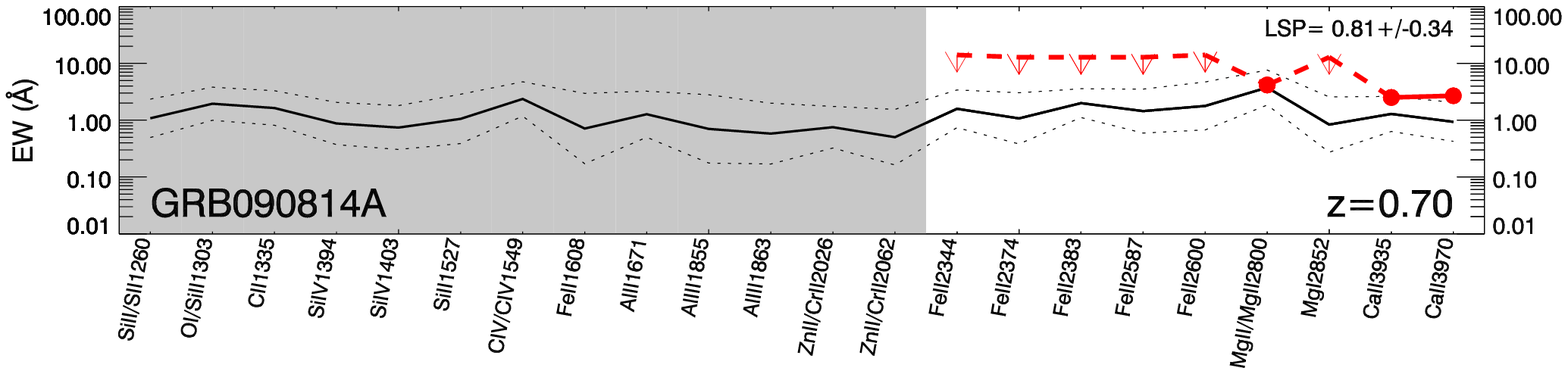}
\end{figure*}

\end{appendix}

\end{document}